\documentclass{article} 
\usepackage{iclr2026_conference,times}

\usepackage{amsmath,amsfonts,bm}

\def\eqref#1{equation~\ref{#1}}

\def\1{\bm{1}}

\def\vh{{\bm{h}}}

\def\mH{{\bm{H}}}

\def\mW{{\bm{W}}}

\def\mZ{{\bm{Z}}}

\DeclareMathAlphabet{\mathsfit}{\encodingdefault}{\sfdefault}{m}{sl}
\SetMathAlphabet{\mathsfit}{bold}{\encodingdefault}{\sfdefault}{bx}{n}

\def\gC{{\mathcal{C}}}

\def\gF{{\mathcal{F}}}

\def\gL{{\mathcal{L}}}
\def\gM{{\mathcal{M}}}

\def\gP{{\mathcal{P}}}

\def\gR{{\mathcal{R}}}

\def\sI{{\mathbb{I}}}

\newcommand{\R}{\mathbb{R}}

\usepackage{hyperref}
\usepackage{url}
\usepackage{graphicx}
\usepackage{wrapfig}
\usepackage{booktabs} 
\usepackage{multirow} 
\usepackage{rotating}
\usepackage{colortbl}
\usepackage{enumitem}
\usepackage{makecell}
\newcounter{bxincomm}
\usepackage{placeins}
\usepackage{textcomp}
\usepackage{subcaption}
\definecolor{aqua}{rgb}{0.00,0.67,0.80}
\definecolor{pink}{rgb}{0.8, 0.3,0.2}

\newcommand{\eg}{\textit{e}.\textit{g}., }
\newcommand{\ie}{\textit{i}.\textit{e}., }

\newcommand{\new}[1]{#1}

\title{Fast and Interpretable Protein Substructure Alignment via Optimal Transport}

\author{%
Zhiyu Wang$^{1,2}$ \thanks{Equal contribution first authors. }  \quad Bingxin Zhou$^{1\;*}$ \thanks{Corresponding authors (bingxin.zhou@sjtu.edu.cn; hongl3liang@sjtu.edu.cn).} \quad Jing Wang$^{1}$ \quad Yang Tan$^{1}$ \quad Weishu Zhao$^{1}$ \\ \textbf{Pietro Li\`o}$^{2}$ \quad \textbf{Liang Hong}$^{1 \dagger}$ \\
$^1$ Shanghai Jiao Tong University.\qquad
$^2$ University of Cambridge.}

\iclrfinalcopy

\begin{document}

\maketitle

\begin{abstract}
Proteins are essential biological macromolecules that execute life functions. \new{Local structural motifs}, such as active sites, are the most critical components for linking structure to function and are key to understanding protein evolution and enabling protein engineering. Existing computational methods struggle to identify and compare these local structures, which leaves a significant gap in understanding protein structures and harnessing their functions. This study presents PLASMA, \new{a deep-learning-based framework for efficient and interpretable residue-level local structural alignment}. We reformulate the problem as a regularized optimal transport task and leverage differentiable Sinkhorn iterations. For a pair of input protein structures, PLASMA outputs a clear alignment matrix with an interpretable overall similarity score. Through extensive quantitative evaluations and three biological case studies, we demonstrate that PLASMA achieves accurate, lightweight, and interpretable residue-level alignment. Additionally, we introduce PLASMA-PF, a training-free variant that provides a practical alternative when training data are unavailable. Our method addresses a critical gap in protein structure analysis tools and offers new opportunities for functional annotation, evolutionary studies, and structure-based drug design. Reproducibility is ensured via our official implementation at \url{https://github.com/ZW471/PLASMA-Protein-Local-Alignment.git}.
\end{abstract}

\section{Introduction}
Proteins are essential macromolecules responsible for life functions, from catalysis and signal transduction to structural support and transport. \new{Local structural motifs} (\eg catalytic residues, binding pockets, metal-binding sites) are critical for understanding mechanisms, designing therapeutics, and guiding protein engineering \citep{mills2018functional}. Structural conservation is three to ten times stronger than sequence conservation across evolution, suggesting that local structural comparison can reveal functional relationships invisible to sequence-based methods \citep{hvidsten2009comprehensive}.

Despite their importance, existing computational methods primarily emphasize global structure comparison or sequence alignment. The inability to detect \new{local structural motifs, \ie compact three-dimensional residue arrangements that often concentrate around catalytic pockets or interaction sites}, prevents researchers from understanding protein evolution, predicting functions of uncharacterized proteins, and rationally designing proteins with desired properties. While large-scale resources like AFDB \citep{jumperHighlyAccurateProtein2021,varadi2022afdb} open a unique opportunity to uncover conserved motifs across the protein universe, active sites often comprise spatially proximate residues that may be widely separated in sequence or embedded within different overall fold architectures \citep{liu2018learning}. Addressing this gap is key to advancing our understanding of protein function and evolution.

The development of robust local structure alignment methods specifically targeting \new{local structural motifs} is not merely a technical challenge but a fundamental requirement for advancing multiple areas of biological research and application. Existing methods for protein substructure alignment can be broadly divided into three categories. The first relies on template-based searches, where predefined motifs are used to identify similar substructures \citep{bittrichRealtimeStructuralMotif2020,kimStructuralMotifSearch2025}. These approaches are effective for detecting well-characterized patterns but cannot uncover novel similarities, making them \textbf{unsuitable for pairing novel structural motifs}. The second category estimates substructure similarity based on the global similarity of entire protein structures. Several studies leverage structural superposition \citep{zhangTMalignProteinStructure2005} or structural tokenization \citep{holmUsingDaliProtein2020} to produce residue-level matches with sequence alignment, but they are \textbf{computationally demanding and difficult to scale to large datasets}. More recent embedding-based methods \citep{hamamsyProteinRemoteHomology2024} are enabled by advances in protein representation learning, which make alignment faster and competitive for whole-protein comparison. However, they compress residue-level information into coarse embeddings, which causes \textbf{\new{problems} in producing interpretable local alignments}. The third category directly addresses substructure alignment by constructing pairwise similarity matrices and using dynamic programming to find matching regions. This approach captures local similarities more accurately than global methods and produces scores that reflect substructure correspondence \citep{kaminskiPLMBLASTDistantHomology2023, liuPLMSearchProteinLanguage2024, pantoliniEmbeddingbasedAlignmentCombining2024}. However, the results can be influenced by overall structural patterns, and \textbf{alignment matrices have limited interpretability} since they are optimized for algorithmic performance rather than clarity. Additionally, these methods are typically untrainable and cannot adapt to specific alignment tasks or incorporate domain knowledge, limiting their ability to improve through experience or be customized for particular biological contexts.

\begin{figure}[t]
    \centering
    \includegraphics[width=0.95\textwidth]{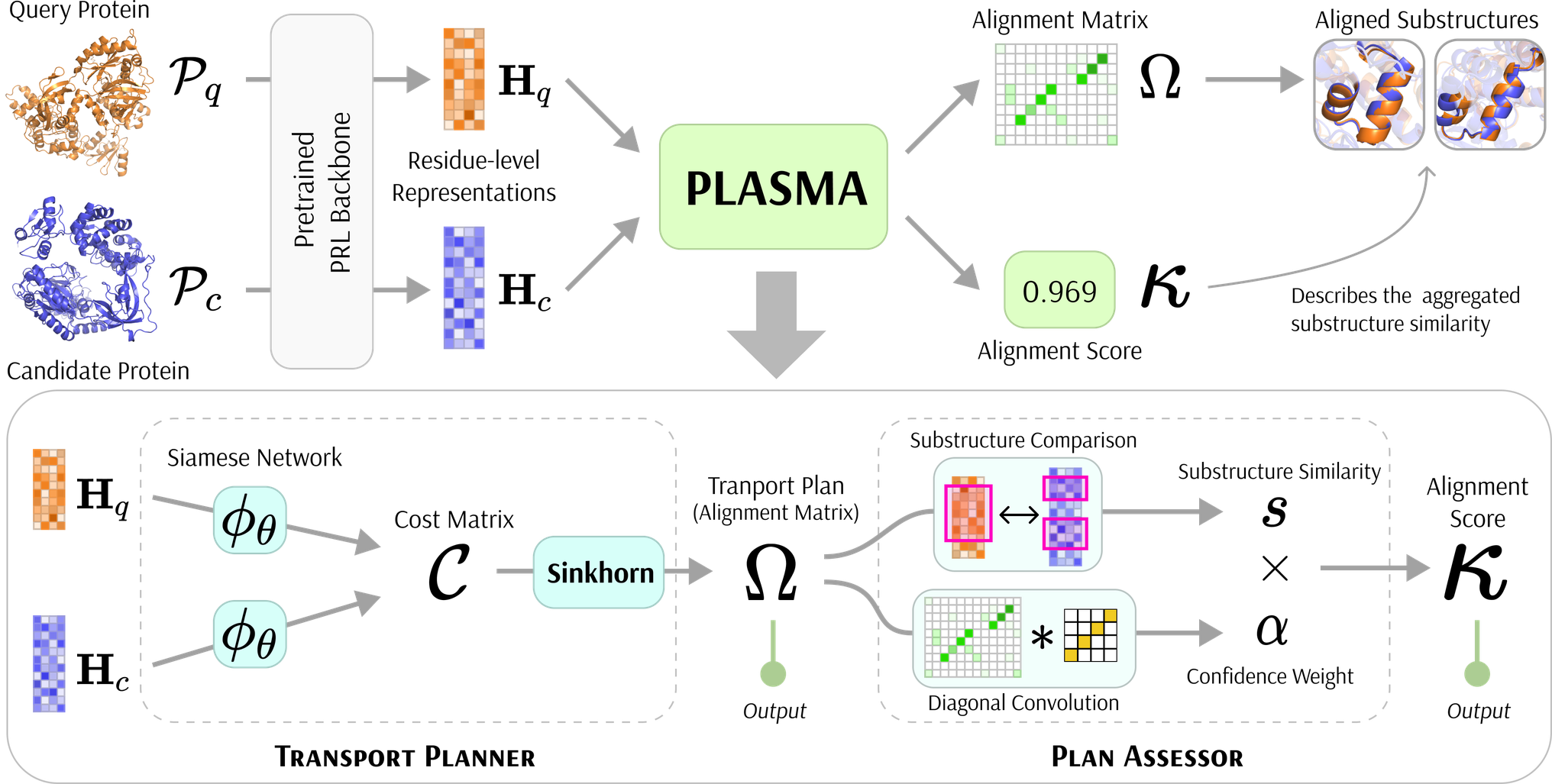}
    \caption{PLASMA Overview. PLASMA converts residue-level protein embeddings into substructure alignments using optimal transport. A \emph{Transport Planner} learns cost matrices with Sinkhorn iterations, and a \emph{Plan Assessor} produces similarity scores. The framework provides alignment matrices and quantitative scores without requiring model-specific designs.}
    \vspace{-5mm}
    \label{fig:visual_abstract}
\end{figure}

The challenges above point to the need for a novel protein substructure alignment method that combines accuracy, efficiency, and clarity. To this end, we explore optimal transport (OT), a mathematical framework proven effective in alignment problems \citep{menaLearningLatentPermutations2018}. In particular, the differentiable Sinkhorn algorithm \citep{sinkhornConcerningNonnegativeMatrices1967, cuturiSinkhornDistancesLightspeed2013} has shown strong ability to uncover meaningful correspondences in 3D shape analysis \citep{eisenbergerDeepShellsUnsupervised2020} and subgraph matching \citep{ramachandranIterativelyRefinedEarly2024}. Notably, these OT-based alignment methods assume strict one-to-one correspondences between all residues or that one set of residues is fully contained within the other. These constraints do not hold for protein substructure alignment, as functionally similar regions may only partially overlap and vary in length across proteins.

To address the aforementioned limitations, we reframe protein substructure alignment as an OT problem and introduce PLASMA (\textbf{P}luggable \textbf{L}ocal \textbf{A}lignment via \textbf{S}inkhorn \textbf{MA}trix). As illustrated in Figure~\ref{fig:visual_abstract}, PLASMA operates on residue-level embeddings from a pre-trained protein representation model and identifies the residue-level alignment between protein pairs. The \emph{Transport Planner} computes the pairwise matching using a learnable cost matrix and differentiable Sinkhorn iterations (Section~\ref{sec:transport-planner}), and the \emph{Plan Assessor} then summarizes the resulting alignment matrix into a single similarity score reflecting the overall similarity of the matched substructures (Section~\ref{sec:plan_assessor}). PLASMA functions as a lightweight, plug-and-play module for protein representation models. It is capable of efficiently aligning partial and variable-length matches between local structural regions.

\new{Our work addresses these limitations through three contributions. First, we introduce a formulation of residue-level local structural alignment based on regularized optimal transport with a learnable geometric cost, which provides a principled and flexible way to define correspondence and enables efficient, fully parallel implementation. Second, this formulation enables clear and interpretable residue–residue correspondences and naturally supports partial, variable-length, and non-sequential motif alignments, resolving the difficulty of obtaining reliable local alignments. Third, PLASMA produces a normalized and interpretable similarity score through its OT-based objective, overcoming the limitations of existing approaches whose alignment matrices or similarity measures lack a consistent probabilistic meaning. Our experiments show strong generalization to low-homology structures, and the case studies demonstrate the biological interpretability and practical utility of the resulting alignments.}

\section{Protein Substructure Alignment via Optimal Transport}
\label{sec:formulation}
\paragraph{Problem Formulation}
Consider a query protein $\gP_q = \{r_{q,1}, \dots, r_{q,N}\}$ of $N$ residues and a candidate protein $\gP_c = \{r_{c,1}, \dots, r_{c,M}\}$ of $M$ residues. Suppose the two proteins contain \new{local structural motifs} $\gF_q = \{f_{q,1},\dots,f_{q,n}\}\subseteq\gP_q$ and $\gF_c=\{f_{c,1},\dots,f_{c,m}\}\subseteq\gP_c$, where $n\leq N$ and $m \leq M$. The objective of protein substructure alignment is: (1) to identify the corresponding fragments $\gF_q$ and $\gF_c$ within $\gP_q$ and $\gP_c$, and (2) to score their level of similarity. 

The task is challenging for several reasons: the overall structures of $\gP_q$ and $\gP_c$ may differ substantially, the fragments $\gF_q$ and $\gF_c$ may vary in sequence length or composition, and alignments require remaining meaningful in a biological context. In particular, biologically relevant alignments should capture functional similarities, such as common enzymatic activities or conserved structural roles.

\paragraph{Optimal Transport Reformulation}
To address the protein substructure alignment problem, we reformulate it as an entropy-regularized OT problem between the residues of two proteins $\gP_q$ and $\gP_c$. Each protein is represented as a set of residue embeddings that capture local biochemical and structural context. The OT solver then computes a soft alignment matrix $\Omega\in\R^{N\times M}$ by assigning weights between residues so as to minimize the overall transport cost $\gC$. This formulation bypasses explicit fragment enumeration, naturally accommodates partial and variable-length matches, and produces interpretable alignment matrices that highlight the underlying substructures (Appendix~\ref{sec:ot-formulation}).

\paragraph{Overview of PLASMA}
We implement entropy-regularized OT and propose \textbf{PLASMA}, a module that transforms $\mH_q\in\R^{N\times d}$ and $\mH_c\in\R^{M\times d}$, residue-level $d$-dimensional hidden representations of $\gP_q$ and $\gP_c$ (\eg from pre-trained protein language models), into a soft alignment matrix $\Omega\in\R^{N\times M}$ and a similarity score $\kappa\in [0,1]$. \new{In our experiments, we instantiate $H_q$ and $H_c$ with seven diverse protein 
representation backbones (Section~\ref{sec:experiments}), and observe consistent alignment 
behavior across them, indicating that PLASMA is not tied to a particular choice of encoder.} Formally,
\begin{equation}
    (\Omega,\kappa)={\rm PLASMA}(\mH_q, \mH_c).
\end{equation}
PLASMA consists of two complementary components (visualized in Figure~\ref{fig:visual_abstract}, with details introduced in the next two sections). The first component, the \emph{Transport Planner}, produces $\Omega$ to highlight local correspondences between $\gP_q$ and $\gP_c$. The second component, the \emph{Plan Assessor}, summarizes this alignment matrix into a similarity score $\kappa \in [0,1]$, providing a quantitative measure of alignment quality. The framework achieves a computational complexity of $O(N^2)$ (Appendix~\ref{sec:complexity-analysis}).

\section{Transport Planner}
\label{sec:transport-planner}
The Transport Planner module handles the core OT computation. It defines cost functions between residue pairs and solves the regularized OT problem to produce an $\Omega$ that captures residue-level matching between query and candidate proteins $(\gP_q,\gP_c)$.

\paragraph{Cost Matrix}
\label{para:cost-matrix}
We formulate a learnable cost matrix with a siamese network architecture to capture complex residue-level similarities. This approach enables PLASMA to learn task-specific representations that optimize alignment quality through end-to-end training. The cost from $r_{q,i}$ to $r_{c,j}$ is denoted by $\gC_{ij}$ in the learnable cost matrix, defined as
\begin{equation}
\label{eq:cost-matrix}
    \gC_{ij} = \Bigl\| \bigl[\phi_\theta({\rm LN}(\vh_{q, i}))- \phi_\theta({\rm LN}(\vh_{c, j}))\bigr]_+ \Bigr\|_1. 
\end{equation}
Here $\vh_{q,i}$ and $\vh_{c,j}$ denote the hidden representations of residues $r_{q,i}$ and $r_{c,j}$, respectively. The operator $[\cdot]_+$ applies a hinge non-linearity, shown to outperform dot-product similarity in subgraph matching tasks \citep{rajChartingDesignSpace2025}. The layer normalization ${\rm LN(\cdot)}$ facilitates robust optimization dynamics with numerical stability and scale-invariant representations. The siamese network $\phi_\theta(\cdot)$ processes query and candidate residues using a twin architecture with shared parameters $\theta$. 

\paragraph{Learnable and Parameter-Free Implementations}
The siamese network architecture can be chosen flexibly, ranging from Transformer-based \citep{hamamsyProteinRemoteHomology2024} models to graph neural networks \citep{jamasbEvaluatingRepresentationLearning2024}, depending on the inductive bias of the input data and the computational budget. Here we also provide a simple implementation using fully connected layers:
\begin{equation}
    \phi_\theta(\vh) = {\rm ReLU}(\vh \cdot \mW_1)\cdot \mW_2,
\end{equation}
where $\mW_1 \in \R^{d \times d^{\prime}}$ and $\mW_2 \in \R^{d^{\prime} \times d^{\prime}}$ are learnable transformation matrices with $d^{\prime}$ hidden dimension. For simplicity, we omit the subscript of $\mH$ as the siamese network applies the same set of parameters to both the query and candidate proteins. This lightweight design serves as an effective default while allowing more sophisticated architectures to be substituted without modifying the overall PLASMA architecture. In addition, for scenarios with a lack of labeled data, we introduce a parameter-free variant, \textbf{PLASMA-PF}, which bypasses the siamese network and \new{operates directly on residue embeddings. The cost used in the OT objective follows (\ref{eq:cost-matrix}) with no architectural components removed other than the encoder}. PLASMA-PF preserves the fundamental alignment functionality and offers a fast baseline for substructure similarity evaluation. Notably, the learnable version remains preferable for improved stability and extrapolation (See Section~\ref{subsec:alignment-quality-analysis} and Figure~\ref{fig:lms-comparison}).

\paragraph{Sinkhorn Alignment Matrix}
\label{para:sinkhorn-alignment-matrix}
Based on the cost matrix $\gC$ defined in (\ref{eq:cost-matrix}), we formulate the corresponding OT problem (Appendix~\ref{sec:ot-formulation}) and solve it using the Sinkhorn algorithm \citep{cuturiSinkhornDistancesLightspeed2013}. The algorithm approximates the OT plan by iteratively scaling the matrix to satisfy the marginal constraints with row and column normalizations, ensuring that the total alignment weights of each residue are properly distributed across residues of the other protein:
\begin{align}
\Omega^{(t+1)}_{ij} = \frac{\mZ^{(t)}_{ij}}{\sum_{v=1}^{M} \mZ^{(t)}_{iv}}, \quad {\rm where}\;\; \mZ^{(t)}_{ij} = \frac{\Omega^{(t)}_{ij}}{\sum_{u=1}^{N} \Omega^{(t)}_{uj}}.
\end{align}
The iteration is initialized as $\Omega^{(0)} = \exp(-\gC/\tau)$, where $\tau$ is a temperature parameter controlling the alignment sharpness (Appendix~\ref{sec:temperature-analysis}). The optimal $\Omega^{\star} = \Omega^{(T)}$ after $T$ iterations serves as the Sinkhorn alignment matrix. For simplicity, we denote it as $\Omega$ in the subsequent discussions. 

The original Sinkhorn algorithm converges to a fully doubly stochastic matrix, forcing each query residue to distribute across all candidate residues (and vice versa). This strict matching is often biologically meaningless, as most residues lack relevant counterparts. PLASMA achieves implicit partial alignments via two mechanisms. First, \emph{early termination} preserves sparsity by limiting Sinkhorn iterations, letting poorly matching residues retain low weights. Second, the \emph{temperature parameter} $\tau$ controls alignment mass, with lower values producing sparser, focused alignments. Together, these mechanisms emphasize biologically relevant correspondences while avoiding forced matches, without hard constraints on the transport budget \citep{caffarelliFreeBoundariesOptimal2010, figalliOptimalPartialTransport2010}. Representative alignment matrices demonstrating these patterns are shown in Appendix~\ref{sec:alignment-matrix-visualizations}.

\section{Plan Assessor}
\label{sec:plan_assessor}
The Plan Assessor receives the alignment matrix $\Omega$ from the Transport Planner and transforms it into an interpretable single similarity score $\kappa \in [0,1]$ that quantifies the existence and degree of similarity of the aligned substructures. This is computed by first calculating a substructure similarity score for the aligned regions, then adjusting it with a confidence weight to correct potential bias.

\paragraph{Substructure Similarity}
We calculate the alignment score on \emph{matched substructure}. With a threshold $\rho$, a residue pair $r_{q,i} \in \gP_q$ and $r_{c,j} \in \gP_c$ is treated as matched if $\Omega_{ij} > \rho$. The matched residues then form two sets, $\gR_q = \{r_{q,i} \mid \forall j, \Omega_{ij} >\rho\}$ and $\gR_c = \{r_{c,j} \mid \forall i, \Omega_{ij} > \rho \}$. A matched substructure is a subset of these residues. The representation of the matched substructure can be approximated by summing the embeddings of residues from $\gR_q$ and $\gR_c$. Therefore, the \emph{substructure similarity score} $s \in [-1, 1]$ is defined as the cosine similarity between the summed representations:
\begin{align}
    s = \frac{\sum_{i \in \gR_q} \vh_{q,i} \cdot \sum_{j \in \gR_c} \vh_{c,j}}{\|\sum_{i \in \gR_q} \vh_{q,i}\| \cdot \|\sum_{j \in \gR_c} \vh_{c,j}\|}.
\end{align}
This substructure similarity score is effective when a sufficient number of residues are matched between the two proteins. However, it becomes less reliable when only a few residues are aligned or when the matched residues are dispersed along the sequence rather than forming a continuous region. In such cases, the score reduces to a residue-level similarity measure, which may appear deceptively high even though the aligned residues do not cluster into a structurally interpretable substructure. We thus introduce a \emph{confidence weight} to adjust the initial similarity score.

\paragraph{Alignment Score with Confidence Weight Correction}
The \emph{confidence weight} $\alpha \in [0, 1]$ is derived from $\Omega$ using a 2D convolution with an identity kernel $K = \sI_k \in \R^{k \times k}$ of size $k$:
\begin{align}
\alpha_{ij} = \sum_{u=0}^{k-1} \sum_{v=0}^{k-1} \Omega_{i+u,j+v} \cdot K_{uv}
= \sum_{u=0}^{k-1} \Omega_{i+u,j+u}.
\end{align}
This convolution operation \new{highlights continuous diagonal segments} in $\Omega$ and emphasizes core regions where consecutive residues in the query align with consecutive residues in the candidate. A max-pooling layer then produces a scalar confidence weight $\alpha = \max_{i,j} \alpha_{ij}$, summarizing the strongest local alignment signal used to weight the similarity score and obtain the final  \emph{alignment score} $\kappa = \alpha \cdot s_+ \in [0, 1]$. Here $s_+$ is the non-negative substructure similarity score. This formulation provides an intuitive and interpretable measure: $\kappa = 0$ indicates no residue matches and $\kappa = 1$ represents perfect substructure alignment. We follow the convention of established alignment methods (\eg TM-align \citep{zhangTMalignProteinStructure2005}) and exclude negative similarity values, since matched substructures with opposite orientations in the representation space lack meaningful biological interpretation. See Appendix~\ref{sec:alignment-matrix-visualizations} for visual examples of alignment matrices with different similarity scores.

\section{Model Optimization}
\label{subsec:module-integration-and-training}
PLASMA is trained with two complementary objectives: predicting the presence of aligned substructures via the alignment score $\kappa$ and recovering precise residue-level matches via the alignment matrix $\Omega$. Training data consists of protein pairs $(\gP_q,\gP_c)$, where a subset of pairs contains matched substructures with shared functions. For each input protein pair, two mask vectors $\gM_q \in \{0,1\}^N$ and $\gM_c \in \{0,1\}^M$ are respectively defined to indicate the position of target substructures $\gF_q$ and $\gF_c$, where $1$ marks the residues that belong to the substructure of interest.

\paragraph{Alignment Score Optimization} 
The alignment score $\kappa$ serves as the model’s prediction on whether the input protein pair contains aligned substructures. We define the ground truth $y = 1$ if the pair contains matched substructures and $y = 0$ otherwise. The prediction is optimized by $\gL_{{\rm BCE}} = -y \log(\sigma(\kappa)) - (1-y) \log(1-\sigma(\kappa))$, where $\sigma(\cdot)$ is the sigmoid function.

\paragraph{Alignment Matrix Optimization}
Unlike the alignment score, optimizing the alignment matrix is challenging because unlabeled residues may correspond to valid but unannotated matches. Treating these residues as negative examples would impose inappropriate penalties on the model. To address this, we propose the \emph{Label Match Loss} (LML) to focus exclusively on the labeled substructures. Specifically, when $\|\gM_c\|_1>0$ and $\|\gM_q\|_1>0$, the LML for protein pairs is defined as
\begin{equation}
\label{eq:lml}
    \gL_{\rm LML} = {\|[\gM_c-\Omega^{\top}\gM_q]_{+}\|_1}/{\|\gM_c\|_1},
\end{equation}
where $[\cdot]_+$ retains only non-negative elements, and $\|\cdot\|_1$ denotes the $\ell_1$ norm. This loss evaluates how well the constructed alignment matrix $\Omega$ aligns the labeled substructures $(\gF_q, \gF_c)$ in $(\gP_q, \gP_c)$. For each residue $r_j\in\gP_c$, $(\Omega^{\top}\gM_q)_j$ gives the alignment weight with respect to labeled residues in $\gP_q$. The non-negative contributions by $[\gM_c-\Omega^{\top}\gM_q]_+$ are normalized by $\|\gM_c\|_1$ across all labeled residues. When no labeled substructures exist, $\gL_{\rm LML}=0$, which allows the model to focus on known substructures without penalizing unlabeled but potentially valid matches. \new{This loss provides an optional bias toward annotated local structural motifs when such labels exist. These regions are typically small and structurally meaningful (\eg catalytic or binding motifs), and emphasizing them helps the model avoid being dominated by background alignments}.

The final $\gL = \gL_{\rm BCE} + \gL_{\rm LML}$ jointly detects substructure existence by $\kappa$ and localizes known substructures by $\Omega$, while staying robust to missing or incomplete labels in the training data.

\section{Empirical Analysis}
\label{sec:experiments}
We conduct extensive quantitative and qualitative evaluations to comprehensively assess the validity and advancement of PLASMA in \new{local structural motif} alignment tasks. All experiments are programmed with PyTorch v2.5.1 and run on NVIDIA RTX 4090 32 GB GPU.

\subsection{Experimental Setup}
\label{sec:experimentSetup}
\paragraph{Prediction Tasks and Benchmark Datasets} Our experiments are based on a residue-level functional alignment benchmark, \textbf{VenusX} \citep{tanVenusXUnlockingFinegrained2025}. We consider three common classes of functional substructures: activation sites, binding sites, and motifs. Across all test sets, the sequence identity between training and test proteins is kept below $50\%$. For quantitative evaluation, we design two levels of difficulty: (i) interpolation (\texttt{test\_inter}), where the test set contains proteins from InterPro families already present in training; and (ii) extrapolation (\texttt{test\_extra}), where the test set only includes novel substructures from unseen families. Further details are in Appendix~\ref{sec:app:dataset}.

\paragraph{Baseline Methods} 
We compare PLASMA with popular baselines in protein structure alignment, including structure-based methods (\textsc{Foldseek} \citep{vankempenFastAccurateProtein2024}, \textsc{TM-Align} \citep{zhangTMalignProteinStructure2005}, and \textsc{TM-vec} \citep{hamamsyProteinRemoteHomology2024}) and embedding-based methods (\textsc{EBA} \citep{pantoliniEmbeddingbasedAlignmentCombining2024} and \textsc{CosineSim}, a cosine similarity over protein embeddings).  
For all embedding-based methods, we implement seven popular pre-trained models to extract residue-level sequence and structure representations, including \textsc{ProtT5} \citep{elnaggarProtTransCrackingLanguage2020}, \textsc{ProstT5} \citep{heinzingerBilingualLanguageModel2024}, \textsc{Ankh} \citep{elnaggarAnkhOptimizedProtein2023}, \textsc{ESM2} \citep{linEvolutionaryscalePredictionAtomiclevel2023}, \textsc{ProtBERT} \citep{brandesProteinBERTUniversalDeeplearning2022}, \textsc{TM-Vec} \citep{hamamsyProteinRemoteHomology2024}, and \textsc{ProtSSN} \citep{tanSemanticalGeometricalProtein2025}. All baselines use the authors’ official code and checkpoints (see Appendices~\ref{sec:app:baselines} for details).

\paragraph{Evaluation Metrics} 
To assess the ability to detect the existence of \new{local structural motifs}, we use standard binary classification metrics, including ROC-AUC, PR-AUC, and F1-Max. Additionally, to evaluate alignment quality, we introduce the Label Match Score (LMS) by (\ref{eq:lml}) with ${\rm LMS}=1-{\rm LML}$ to measure correspondence between predicted alignments and annotated functional regions. 

\begin{table}[tb]
    \centering
    \caption{Model performance on \texttt{test\_extra} (mean $\pm$ std over three independent seeds). Colors indicate relative performance versus \textsc{TM-Align}.
    }
    \renewcommand{\arraystretch}{1.2}
    \resizebox{\textwidth}{!}{
    \begin{tabular}{cc ccc ccc ccc}
        \toprule
        \midrule
        \multirow{2}[4]{*}{\textbf{Metrics}} & \multirow{2}[4]{*}{\textbf{Methods}} & \multicolumn{3}{c}{\textbf{Motif}} & \multicolumn{3}{c}{\textbf{Binding Site}} & \multicolumn{3}{c}{\textbf{Active Site}} \\
        \cmidrule(lr){3-5}\cmidrule(lr){6-8}\cmidrule(lr){9-11}          &       & \textsc{Ankh}  & \textsc{ESM2}  & \textsc{ProtSSN} & \textsc{Ankh}  & \textsc{ESM2}  & \textsc{ProtSSN} & \textsc{Ankh}  & \textsc{ESM2}  & \textsc{ProtSSN} \\
        \toprule
        \multirow{6}[4]{*}{\begin{sideways}ROC-AUC\end{sideways}} & PLASMA & \cellcolor[RGB]{255,201,92}$\mathbf{.98_{\pm.008}}$ & \cellcolor[RGB]{255,207,112}$\mathbf{.97_{\pm.013}}$ & \cellcolor[RGB]{255,215,135}$\mathbf{.96_{\pm.016}}$ & \cellcolor[RGB]{255,206,107}$\mathbf{.99_{\pm.008}}$ & \cellcolor[RGB]{255,213,131}$\mathbf{.98_{\pm.013}}$ & \cellcolor[RGB]{255,218,143}$\mathbf{.98_{\pm.014}}$ & \cellcolor[RGB]{255,206,108}$\mathbf{.98_{\pm.012}}$ & \cellcolor[RGB]{255,201,92}$\mathbf{.98_{\pm.010}}$ & \cellcolor[RGB]{255,210,121}$\mathbf{.97_{\pm.011}}$ \\
        & PLASMA-PF & \cellcolor[RGB]{255,201,94}$.98_{\pm.009}$ & \cellcolor[RGB]{255,228,175}$.93_{\pm.004}$ & \cellcolor[RGB]{255,238,205}$.90_{\pm.005}$ & \cellcolor[RGB]{255,201,93}$.99_{\pm.006}$ & \cellcolor[RGB]{254,252,247}$.92_{\pm.052}$ & \cellcolor[RGB]{255,230,180}$.96_{\pm.012}$ & \cellcolor[RGB]{255,211,124}$.97_{\pm.015}$ & \cellcolor[RGB]{255,237,200}$.96_{\pm.006}$ & \cellcolor[RGB]{255,222,158}$.97_{\pm.008}$ \\
        & EBA   & \cellcolor[RGB]{255,240,211}$.90_{\pm.033}$ & \cellcolor[RGB]{255,231,184}$.92_{\pm.021}$ & \cellcolor[RGB]{180,180,255}$.32_{\pm.043}$ & \cellcolor[RGB]{255,201,92}$.99_{\pm.007}$ & \cellcolor[RGB]{255,221,153}$.97_{\pm.021}$ & \cellcolor[RGB]{180,180,255}$.30_{\pm.060}$ & \cellcolor[RGB]{255,212,127}$.97_{\pm.013}$ & \cellcolor[RGB]{255,213,129}$.97_{\pm.012}$ & \cellcolor[RGB]{180,180,255}$.43_{\pm.066}$ \\
        & Backbone & \cellcolor[RGB]{254,250,240}$.85_{\pm.019}$ & \cellcolor[RGB]{243,243,255}$.74_{\pm.033}$ & \cellcolor[RGB]{252,252,255}$.79_{\pm.018}$ & \cellcolor[RGB]{255,217,141}$.98_{\pm.010}$ & \cellcolor[RGB]{231,231,255}$.72_{\pm.060}$ & \cellcolor[RGB]{229,229,255}$.70_{\pm.070}$ & \cellcolor[RGB]{255,229,176}$.96_{\pm.012}$ & \cellcolor[RGB]{233,233,255}$.79_{\pm.068}$ & \cellcolor[RGB]{229,229,255}$.76_{\pm.033}$ \\
        \cmidrule{2-11}          & Foldseek & \multicolumn{3}{c}{\cellcolor[RGB]{255,242,217}$.89_{\pm.033}$} & \multicolumn{3}{c}{\cellcolor[RGB]{254,254,255}$.90_{\pm.013}$} & \multicolumn{3}{c}{\cellcolor[RGB]{245,245,255}$.87_{\pm.022}$} \\
        & TM-Align & \multicolumn{3}{c}{\cellcolor[RGB]{255,255,255}$.81_{\pm.014}$} & \multicolumn{3}{c}{\cellcolor[RGB]{255,255,255}$.91_{\pm.040}$} & \multicolumn{3}{c}{\cellcolor[RGB]{255,255,255}$.93_{\pm.009}$} \\
        \toprule
        \multirow{6}[4]{*}{\begin{sideways}PR-AUC\end{sideways}} & PLASMA & \cellcolor[RGB]{255,201,92}$\mathbf{.98_{\pm.011}}$ & \cellcolor[RGB]{255,206,110}$\mathbf{.97_{\pm.014}}$ & \cellcolor[RGB]{255,215,135}$\mathbf{.96_{\pm.017}}$ & \cellcolor[RGB]{255,204,101}$\mathbf{.98_{\pm.011}}$ & \cellcolor[RGB]{255,213,130}$\mathbf{.97_{\pm.019}}$ & \cellcolor[RGB]{255,215,136}$\mathbf{.97_{\pm.019}}$ & \cellcolor[RGB]{255,215,134}$\mathbf{.97_{\pm.014}}$ & \cellcolor[RGB]{255,201,92}$\mathbf{.98_{\pm.011}}$ & \cellcolor[RGB]{255,211,123}$\mathbf{.97_{\pm.012}}$ \\
        & PLASMA-PF & \cellcolor[RGB]{255,201,93}$.98_{\pm.010}$ & \cellcolor[RGB]{255,224,163}$.95_{\pm.005}$ & \cellcolor[RGB]{255,239,209}$.92_{\pm.007}$ & \cellcolor[RGB]{255,206,108}$.98_{\pm.012}$ & \cellcolor[RGB]{255,254,253}$.90_{\pm.079}$ & \cellcolor[RGB]{254,233,188}$.95_{\pm.026}$ & \cellcolor[RGB]{255,219,148}$.97_{\pm.015}$ & \cellcolor[RGB]{255,234,193}$.96_{\pm.006}$ & \cellcolor[RGB]{255,221,155}$.97_{\pm.009}$ \\
        & EBA   & \cellcolor[RGB]{255,242,217}$.91_{\pm.035}$ & \cellcolor[RGB]{255,233,189}$.93_{\pm.019}$ & \cellcolor[RGB]{180,180,255}$.38_{\pm.014}$ & \cellcolor[RGB]{255,201,92}$.98_{\pm.012}$ & \cellcolor[RGB]{255,227,171}$.96_{\pm.035}$ & \cellcolor[RGB]{180,180,255}$.28_{\pm.063}$ & \cellcolor[RGB]{255,208,115}$.97_{\pm.012}$ & \cellcolor[RGB]{255,218,144}$.97_{\pm.012}$ & \cellcolor[RGB]{180,180,255}$.43_{\pm.032}$ \\
        & Backbone & \cellcolor[RGB]{255,254,254}$.86_{\pm.023}$ & \cellcolor[RGB]{241,241,255}$.77_{\pm.041}$ & \cellcolor[RGB]{249,249,255}$.82_{\pm.027}$ & \cellcolor[RGB]{255,225,164}$.96_{\pm.023}$ & \cellcolor[RGB]{227,227,255}$.67_{\pm.093}$ & \cellcolor[RGB]{225,225,255}$.65_{\pm.118}$ & \cellcolor[RGB]{255,237,202}$.96_{\pm.016}$ & \cellcolor[RGB]{239,239,255}$.84_{\pm.059}$ & \cellcolor[RGB]{234,234,255}$.80_{\pm.038}$ \\
        \cmidrule{2-11}          & Foldseek & \multicolumn{3}{c}{\cellcolor[RGB]{252,252,255}$.84_{\pm.031}$} & \multicolumn{3}{c}{\cellcolor[RGB]{239,239,255}$.76_{\pm.065}$} & \multicolumn{3}{c}{\cellcolor[RGB]{235,235,255}$.81_{\pm.026}$} \\
        & TM-Align & \multicolumn{3}{c}{\cellcolor[RGB]{255,255,255}$.86_{\pm.020}$} & \multicolumn{3}{c}{\cellcolor[RGB]{255,255,255}$.89_{\pm.064}$} & \multicolumn{3}{c}{\cellcolor[RGB]{255,255,255}$.94_{\pm.012}$} \\
        \toprule
        \multirow{6}[4]{*}{\begin{sideways}F1-MAX\end{sideways}} & PLASMA & \cellcolor[RGB]{255,201,92}$\mathbf{.97_{\pm.009}}$ & \cellcolor[RGB]{255,212,125}$\mathbf{.95_{\pm.018}}$ & \cellcolor[RGB]{255,223,158}$\mathbf{.92_{\pm.022}}$ & \cellcolor[RGB]{255,204,103}$.96_{\pm.022}$ & \cellcolor[RGB]{255,219,147}$.95_{\pm.030}$ & \cellcolor[RGB]{255,230,181}$.93_{\pm.026}$ & \cellcolor[RGB]{255,201,92}$\mathbf{.98_{\pm.013}}$ & \cellcolor[RGB]{255,207,111}$\mathbf{.97_{\pm.011}}$ & \cellcolor[RGB]{255,211,124}$\mathbf{.97_{\pm.011}}$ \\
        & PLASMA-PF & \cellcolor[RGB]{255,207,110}$.96_{\pm.013}$ & \cellcolor[RGB]{255,231,183}$.90_{\pm.006}$ & \cellcolor[RGB]{255,245,226}$.84_{\pm.008}$ & \cellcolor[RGB]{255,211,124}$.96_{\pm.027}$ & \cellcolor[RGB]{253,253,255}$.85_{\pm.082}$ & \cellcolor[RGB]{255,247,232}$.90_{\pm.031}$ & \cellcolor[RGB]{255,215,136}$.97_{\pm.018}$ & \cellcolor[RGB]{255,242,217}$.94_{\pm.016}$ & \cellcolor[RGB]{255,228,175}$.95_{\pm.012}$ \\
        & EBA   & \cellcolor[RGB]{255,241,213}$.86_{\pm.035}$ & \cellcolor[RGB]{255,238,206}$.87_{\pm.024}$ & \cellcolor[RGB]{180,180,255}$.00_{\pm.000}$ & \cellcolor[RGB]{255,201,92}$\mathbf{.97_{\pm.021}}$ & \cellcolor[RGB]{255,232,187}$.93_{\pm.049}$ & \cellcolor[RGB]{180,180,255}$.00_{\pm.000}$ & \cellcolor[RGB]{255,212,126}$.97_{\pm.013}$ & \cellcolor[RGB]{255,211,124}$.97_{\pm.008}$ & \cellcolor[RGB]{180,180,255}$.00_{\pm.000}$ \\
        & Backbone & \cellcolor[RGB]{255,252,246}$.79_{\pm.008}$ & \cellcolor[RGB]{249,249,255}$.70_{\pm.014}$ & \cellcolor[RGB]{252,252,255}$.73_{\pm.013}$ & \cellcolor[RGB]{255,241,215}$.91_{\pm.034}$ & \cellcolor[RGB]{233,233,255}$.62_{\pm.087}$ & \cellcolor[RGB]{231,231,255}$.60_{\pm.107}$ & \cellcolor[RGB]{255,248,234}$.92_{\pm.020}$ & \cellcolor[RGB]{242,242,255}$.75_{\pm.044}$ & \cellcolor[RGB]{239,239,255}$.71_{\pm.018}$ \\
        \cmidrule{2-11}          & Foldseek & \multicolumn{3}{c}{\cellcolor[RGB]{255,228,175}$.91_{\pm.046}$} & \multicolumn{3}{c}{\cellcolor[RGB]{255,201,92}$.97_{\pm.014}$} & \multicolumn{3}{c}{\cellcolor[RGB]{255,221,153}$.96_{\pm.015}$} \\
        & TM-Align & \multicolumn{3}{c}{\cellcolor[RGB]{255,255,255}$.76_{\pm.015}$} & \multicolumn{3}{c}{\cellcolor[RGB]{255,255,255}$.87_{\pm.063}$} & \multicolumn{3}{c}{\cellcolor[RGB]{255,255,255}$.90_{\pm.014}$} \\
        \toprule
        \multirow{2}[2]{*}{\begin{sideways}LMS\end{sideways} \vspace{3pt}} & PLASMA & \cellcolor[RGB]{255,255,255}$.75_{\pm.045}$ & \cellcolor[RGB]{255,201,92}$\mathbf{.69_{\pm.019}}$ & \cellcolor[RGB]{255,201,92}$\mathbf{.52_{\pm.046}}$ & \cellcolor[RGB]{255,255,255}$.82_{\pm.062}$ & \cellcolor[RGB]{255,201,92}$\mathbf{.77_{\pm.105}}$ & \cellcolor[RGB]{255,201,92}$\mathbf{.65_{\pm.088}}$ & \cellcolor[RGB]{255,255,255}$.90_{\pm.034}$ & \cellcolor[RGB]{255,201,92}$\mathbf{.87_{\pm.038}}$ & \cellcolor[RGB]{255,201,92}$\mathbf{.67_{\pm.044}}$ \\
        & PLASMA-PF & \cellcolor[RGB]{255,201,92}$\mathbf{.78_{\pm.055}}$ & \cellcolor[RGB]{255,255,255}$.48_{\pm.074}$ & \cellcolor[RGB]{255,255,255}$.23_{\pm.021}$ & \cellcolor[RGB]{255,201,92}$\mathbf{.85_{\pm.058}}$ & \cellcolor[RGB]{255,255,255}$.49_{\pm.082}$ & \cellcolor[RGB]{255,255,255}$.36_{\pm.055}$ & \cellcolor[RGB]{255,201,92}$\mathbf{.94_{\pm.029}}$ & \cellcolor[RGB]{255,255,255}$.68_{\pm.067}$ & \cellcolor[RGB]{255,255,255}$.43_{\pm.032}$ \\
        
        \midrule
        \bottomrule
    \end{tabular}%
}
    \label{tab:main_results_test_extra}%
    \hspace{1pt}\includegraphics[width=0.99\textwidth]{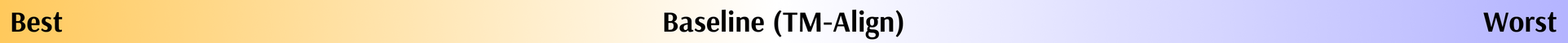}
\end{table}%


\subsection{Quantitative Performance Evaluation}
\label{subsec:performance-evaluation}
\new{Table~\ref{tab:main_results_test_extra} reports performance on \texttt{test\_extra}, which contains functional substructures from protein families not seen during training. This setting evaluates the generalizability of the alignment framework, which is essential in practice because new functional substructures are continuously discovered. Full results on seven backbone models are provided in Appendix~\ref{sec:full-extrapolation-performance-comparison}, and all hyperparameter and dataset details are summarized in Appendix~\ref{sec:app:hyperparameters}. Corresponding interpolation results on \texttt{test\_inter} are reported in Appendix~\ref{sec:full-performance-comparision}.}

\new{Across all three substructure detection tasks and all evaluation metrics, PLASMA achieves consistent top performance, highlighting its robustness in capturing fundamental local structural similarities for novel substructures beyond the training distribution. PLASMA-PF also performs strongly and remains competitive without task-specific training. However, unlike in the interpolation setting, PLASMA-PF does not surpass the learnable PLASMA variant on \texttt{test\_extra}; this emphasizes the value of supervised examples in improving alignment accuracy for entirely new functional substructures. In contrast, baseline methods show large performance variation across backbone models. EBA performs reasonably well with sequence-based \textsc{Ankh} and \textsc{ESM2} yet drops substantially with structure-based \textsc{ProtSSN}, especially under the extrapolation split. \textsc{Foldseek} and \textsc{TM-Align} remain consistently below PLASMA across nearly all conditions, reflecting the limited usefulness of global structural similarity for residue-level motif detection.}

\begin{figure}[!t]
    \centering
    \includegraphics[width=\linewidth]{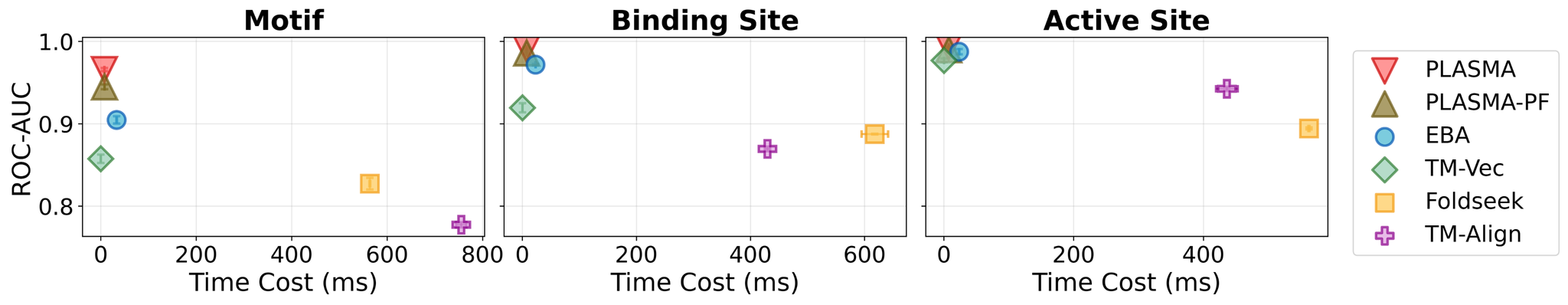}
    \vspace{-5mm}
    \caption{Performance versus computational efficiency comparison. ROC-AUC scores plotted against inference time (milliseconds) for motif and binding/active site detection using \textsc{ProstT5} embeddings. Points represent averages across three splits with standard error bars on both axes.}
    \label{fig:performance-vs-time-inter}
    \vspace{1mm}

    \centering
    \includegraphics[width=\linewidth]{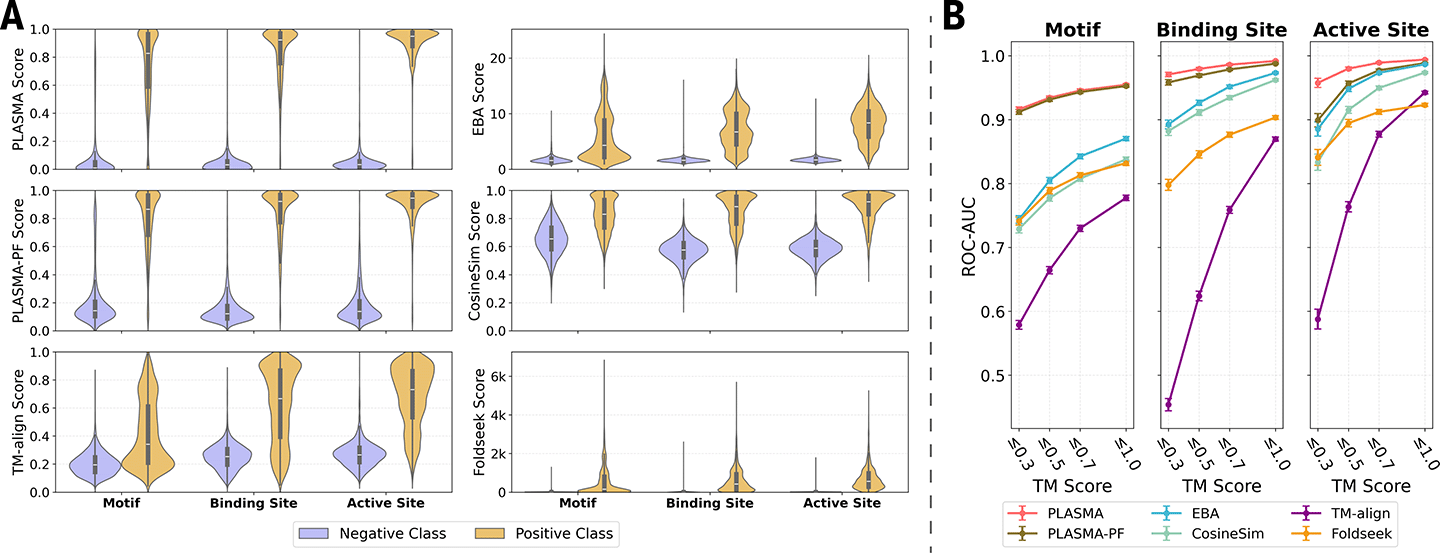}
    \vspace{-5mm}
    \caption{Alignment quality analysis across different approaches. \textbf{A}. Distribution of alignment scores for positive and negative protein pairs. \textbf{B}. ROC-AUC score trend at different global structural similarity levels.}
    \label{fig:score-comparison}
    \vspace{1mm}

    \centering
    \includegraphics[width=\linewidth]{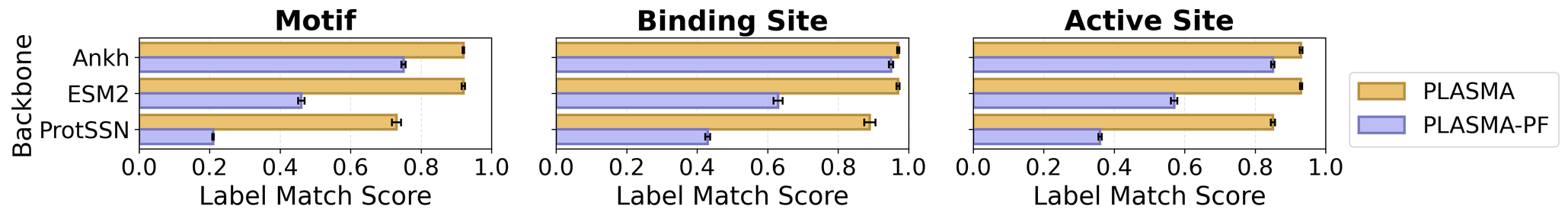}
    \vspace{-5mm}
    \caption{Label Match Score comparison between PLASMA and PLASMA-PF across different substructure types, demonstrating the improved alignment quality achieved through training.}
    \label{fig:lms-comparison}
\end{figure}

Beyond accuracy, PLASMA demonstrates exceptional computational efficiency. As shown in Figure~\ref{fig:performance-vs-time-inter}, PLASMA achieves the best performance while requiring minimal time per protein pair—approximately 10ms for PLASMA and 7ms for PLASMA-PF. This represents a roughly $50$ times speedup over global structure alignment methods like TM-Align and Foldseek, which require costly structural superposition, and about $3$ times faster than EBA due to PLASMA’s fully differentiable OT formulation that is efficiently accelerated on GPUs, compared to EBA’s inherently sequential dynamic programming approach.

\subsection{Quality of Predicted Alignments}
\label{subsec:alignment-quality-analysis}
Beyond quantitative metrics, we assess PLASMA’s robustness in identifying biologically meaningful substructures by examining both alignment scores and alignment matrices.

PLASMA effectively distinguishes proteins with shared local functional substructures even when overall structural similarity is low. Figure~\ref{fig:score-comparison} provides evidence from two perspectives, with all embedding-based methods obtaining protein representations from \textsc{Ankh}. Figure~\ref{fig:score-comparison}A compares similarity score distributions for protein pairs from \texttt{test\_inter}, where PLASMA and PLASMA-PF clearly separate positive and negative pairs. This advantage comes from the OT framework, which emphasizes local correspondences independent of overall similarity. In contrast, EBA and \textsc{CosineSim} show substantial overlap between positive and negative distributions. EBA in particular lacks an upper bound on its scores, making them difficult to interpret and subject to calibration problems (\ie scores cannot be directly used as probabilities and lead to unstable thresholds). Figure~\ref{fig:score-comparison}B further groups test-set alignment scores by TM-score to assess performance under different levels of global similarity for protein pairs. Although all methods degrade as TM-score decreases, PLASMA and PLASMA-PF consistently maintain high ROC-AUC values above $0.9$, whereas baseline EBA, \textsc{CosineSim}, \new{Foldseek, and TM-align} deteriorate sharply on low-similarity samples \new{when TM-score is sufficiently small (\eg lower than $0.5$)}.

While both PLASMA variants demonstrate strong performance in score-based discrimination, their alignment quality differs. This is evident in Figure~\ref{fig:lms-comparison}, which compares their performance using the LMS score to evaluate correspondence between predicted alignments and annotated regions. PLASMA consistently outperforms PLASMA-PF across motifs, binding sites, and active sites, demonstrating that learning improves the prediction of \new{local structural motifs}. By contrast, while EBA also produces alignment matrices, it cannot be meaningfully assessed with LMS: its unconstrained formulation yields a maximal LMS of $1.0$ regardless of true alignment accuracy.

\subsection{Representative Alignment Examples}
\label{subsec:case-study}
The next experiment evaluates PLASMA’s utility in real biological applications. We examine three protein pairs of different substructure sizes (independent of the training set), including simple local motifs, complex cofactor-binding domains, and extended multi-element substructures. In each case, we provide UniProt identifiers, functional descriptions, alignment results, and visualizations from PLASMA and EBA, and corresponding analyses. Appendix~\ref{sec:alignment-matrix-vis-extra} provides additional visualizations that further illustrate the generality of these conclusions. Collectively, these cases show PLASMA detects biologically meaningful local similarities across diverse sequences, structures, and functions.

\begin{figure}[tb]
    \centering
    \includegraphics[width=0.95\linewidth]{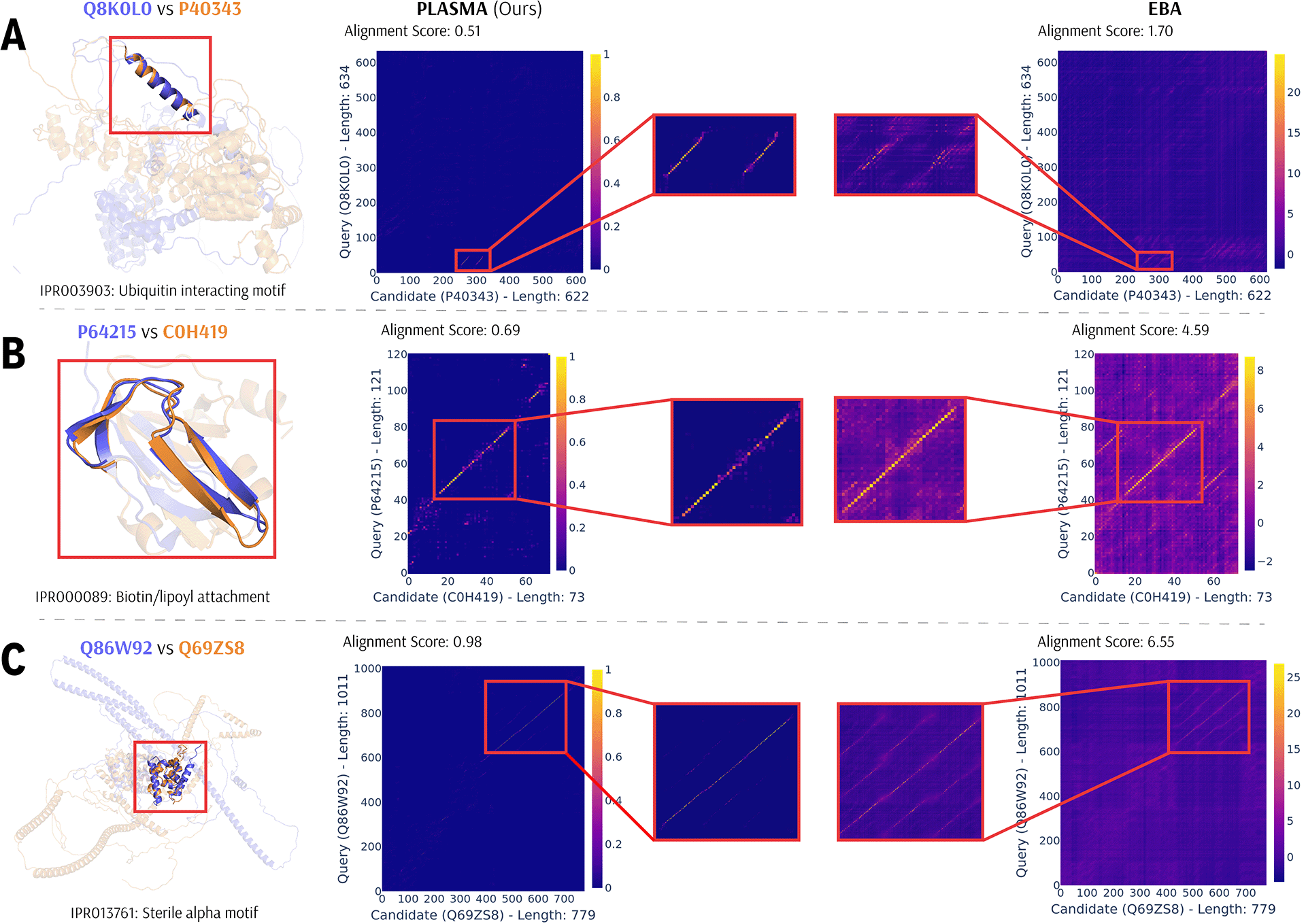}
    \vspace{-2mm}
    \caption{Representative alignment examples across three protein pairs. \textbf{A}, P40343 vs Q8K0L0. \textbf{B}, P64215 vs C0H419. \textbf{C}, Q69ZS8 vs Q86W92. Left: 3D structures with highlighted aligned regions. Center and right: alignment matrices from PLASMA and EBA with zoomed insets. A higher resolution version of this figure can be found at Appendix~\ref{sec:highres-case-study}.}
    \label{fig:case-study}
\end{figure}

\paragraph{Conserved Small Helical Motifs Across Functionally Diverse Protein Structures}
The first case matches local structures between P40343 (Vps27, a yeast ESCRT-0 complex component) and Q8K0L0 (ASB2, a mouse E3 ubiquitin ligase substrate-recognition component). The two proteins share no apparent sequence homology ($21.0\%$ identity) and participate in distinct cellular processes (endosomal sorting versus proteasomal degradation), yet both use analogous helical arrangements for protein-protein interactions: Vps27’s GAT domain forms coiled-coils for ESCRT-I recruitment \citep{curtiss2007efficient}, whereas ASB2 employs ankyrin repeat helices for substrate recognition in the E3 ligase complex. PLASMA assigns high-confidence scores to residues mediating these interactions (Figure~\ref{fig:case-study}A). The 3D structure visualization also confirms the alignment of the conserved Leu-X-X-Leu-Leu motif for both proteins \citep{ren2008doa1}, with an aligned RMSD of $0.18$ \AA. This finding suggests potential convergent evolution of helical protein-binding interfaces across distinct cellular machineries.
By contrast, EBA identifies multiple helices, but most correspond to nonfunctional scaffold regions rather than the relevant interaction motifs.

\paragraph{Structurally and Functionally Relevant motifs of Different Sizes and Metabolic Contexts}
The second case examines P64215 (GcvH, glycine cleavage system H protein from Mycobacterium tuberculosis) and C0H419 (YngHB, biotin/lipoyl attachment protein from Bacillus subtilis) \citep{cui2006identification}. These proteins have different overall sequences ($25.2\%$ sequence identity) and metabolic functions: GcvH shuttles methylamine groups in glycine catabolism, while YngHB accommodates both biotin and lipoic acid in a single-domain architecture. Despite these differences, both bind similar cofactors and exhibit conserved $\beta$-sheet arrangements necessary for post-translational modification. As shown in Figure~\ref{fig:case-study}B, PLASMA successfully aligns the four-stranded $\beta$-barrel architectures, highlighting the critical lysine-containing $\beta$-turns with an overall alignment score of $0.69$ and RMSD of $0.83$, whereas the baseline EBA misaligns nonfunctional regions. The alignment of complex conserved structural motifs across protein families demonstrates the potential of PLASMA in revealing modular evolution and conserved cofactor-binding architectures.

\paragraph{Extended Multi-Element Substructures in Cell Adhesion Regulators}
The third case investigates Q69ZS8 (Kazrin, a scaffold protein in Mus musculus) and Q86W92 (Liprin-$\beta$1/PPFIBP1, a human focal adhesion regulator). Despite their different cellular localizations and interaction partners, they regulate distinct but mechanistically related aspects of cell-cell adhesion: Kazrin organizes desmosomal components in keratinocytes, and Liprin-$\beta$1 modulates focal adhesion disassembly and cell migration. Yet both proteins rely on extended $\alpha$-helical regions for protein-protein interactions \citep{groot2004kazrin}. As in Figure~\ref{fig:case-study}C, PLASMA successfully aligns complex multi-coil substructures spanning multiple helical segments interspersed with flexible linkers, with an overall alignment score of $0.98$ and RMSD $0.82$ \AA. The alignment highlights conserved leucine-rich motifs and hinge regions that stabilize oligomerization interfaces, revealing analogous scaffolding strategies. In contrast, EBA identifies plausible structures but often misaligns helices or matches nonfunctional scaffold regions, failing to capture more than just biologically meaningful substructures.

\section{Related Works}
\paragraph{Protein Global Structure Alignment}
Global structure alignment methods evaluate overall protein similarity. Classic approaches like \textsc{TM-Align} \citep{zhangTMalignProteinStructure2005} are foundational, while modern methods increase efficiency by abstracting structures into 1D sequences (\textsc{Foldseek} \citep{vankempenFastAccurateProtein2024}), representing them as fixed vectors for rapid search (\textsc{TM-Vec} \citep{hamamsyProteinRemoteHomology2024}), or using advanced spatial indexing (\textsc{GTalign} \citep{margelevivcius2024gtalign}). The field has also expanded to align multiple structures (\textsc{mTM-align} \citep{dong2018mtmalign}), multi-chain complexes (\textsc{MM-align} \citep{mukherjee2009mmalign}), and diverse macromolecules universally (\textsc{US-align} \citep{zhang2022usalign}). However, their global nature limits the detection of conserved motifs in dissimilar proteins.

\paragraph{Substructure and Sequence-based Alignment}
To find local similarities, substructure-based methods use graph-based residue embeddings \citep{tanProteinRepresentationLearning2024}, \new{focus on active-site environments \citep{castillo2025actseek}, or apply linear-assignment formulations \citep{zhang2025eplsap}}. PLM-based residue representations are also widely used from raw embedding similarity scoring \citep{kaminskiPLMBLASTDistantHomology2023,liuPLMSearchProteinLanguage2024} to learned alignment models and embedding-aware dynamic programming \citep{llinares2023DEDAL,iovino2024peba}. OT-based differentiable graph matching has been used to learn structure/function-aware substitution matrices \citep{pellizzoni2024structure}, with a primary focus on learning matching costs. PLASMA instead targets residue-level local substructure alignment, producing explicit mappings with practical speed and interpretability. Meanwhile, embedding-score-based alignment methods remain hard to interpret quantitatively, as their scores are essentially unbounded \citep{pantoliniEmbeddingbasedAlignmentCombining2024}.

\section{Conclusion and Discussion}
\label{sec:conclusion}
This work presents PLASMA, a \new{local structural motif} alignment framework leveraging regularized optimal transport to detect biologically meaningful local similarities across proteins with diverse sequences, structures, and functions. PLASMA consistently outperforms baseline methods in accuracy, efficiency, and interpretability, capturing subtle structural correspondences often invisible to global alignments. Its trainable variant benefits from supervision to improve alignment precision, while the training-free variant achieves robust performance without task-specific labels.

Beyond quantitative performance, PLASMA provides clear, residue-level alignment matrices that support mechanistic insights into protein function, evolutionary relationships, and structure-guided protein engineering. Its ability to handle varying substructure sizes and complexities (\eg from short helices to extended multi-element domains) demonstrates versatility and practical relevance. Overall, PLASMA establishes a new standard for accurate, efficient, interpretable, and practically applicable \new{protein local structural motif alignment}.

\subsubsection*{Acknowledgments}
This work was supported by the grants from National Science Foundation of China (Grant Number 92451301; 62302291), the AI for Science Program by Shanghai Municipal Commission of Economy and Informatization (2025-GZL-RGZN-BTBX-02009), the National Key Research and Development Program of China (2024YFA0917603), and Computational Biology Key Program of Shanghai Science and Technology Commission (23JS1400600). Z.W.'s attendance at the conference is supported by his current affiliation, Sapient Intelligence.

\paragraph{Reproducibility Statement}
To promote reproducibility, we release all source code and trained models under an open-source license, which is available at \url{https://github.com/ZW471/PLASMA-Protein-Local-Alignment.git}. Details of data sources are provided in Appendix~\ref{sec:app:dataset}. Task definitions, evaluation protocols, and hyperparameter settings are described in Sections~\ref{sec:experimentSetup} and Appendices~\ref{sec:app:hyperparameters}. Implementation details and instructions for reproducing experiments are included in the project repository to facilitate independent verification.

\paragraph{Ethics Statement}
All experiments are conducted on publicly available protein sequence and structure databases. We follow established ethical guidelines in data usage and acknowledge that historical biases present in these resources may be reflected in our results, which is independent of model development.

\paragraph{The Use of Large Language Models (LLM)}
In the preparation of this manuscript, GPT-5 and GPT-4o were utilized as writing assistants. The usage was strictly limited to improving grammar, clarity, and overall readability. All scientific ideas, experimental results, and conclusions were conceived and formulated exclusively by the authors. All text polished or modified by the LLM was subsequently reviewed and edited by the authors to ensure that the original scientific meaning was accurately preserved.

\clearpage
\bibliography{iclr2026_conference}
\bibliographystyle{iclr2026_conference}

\clearpage
\appendix
\new{This appendix provides additional details, analyses, and results that complement the main paper.
\begin{itemize}[leftmargin=*]
    \item Appendix~\ref{sec:ot-formulation}  gives the full derivation of our OT objective.
    \item Appendix~\ref{sec:complexity-analysis} presents a more precise discussion of computational cost.
    \item Appendix~\ref{sec:detailed-experimental-setup} describes the benchmark datasets (\textbf{VenusX}) and the hyperparameter configuration.
    \item Appendix~\ref{sec:app:baselines} summarizes all comparison methods, including global structure alignment, global embedding-based alignment, local embedding-based alignment, and the backbone models.
    \item Appendix~\ref{sec:full-performance-comparision} and Appendix~\ref{sec:full-extrapolation-performance-comparison} report complete quantitative results for all backbones on \texttt{test\_inter} and \texttt{test\_extra} split.
    \item Appendix~\ref{sec:ablation} provides further insight into the contribution of individual components. 
    \item Appendices~\ref{sec:highres-case-study}-\ref{sec:alignment-matrix-visualizations} contains additional visualizations of alignment matrices and case studies.
	\item Appendices~\ref{sec:temperature-analysis}-\ref{sec:hyperparameter-analysis} offer more detailed quantitative analyses of model behaviour under different settings.
\end{itemize}}

\section{Optimal Transport Formulation for Protein Alignment}
\label{sec:ot-formulation}

To circumvent the computational bottleneck of explicit fragment enumeration, we reframe the alignment problem as finding optimal correspondences between individual residues rather than pre-defined fragments. This approach leverages optimal transport theory, which provides a principled framework for finding the most efficient assignment between two sets of points based on their similarity and a transportation cost function.

Specifically, we model protein substructure alignment as an entropy regularized optimal transport problem that determines how to optimally redistribute alignment weights from query residues to candidate residues. Instead of relying solely on explicit structural coordinates, this formulation operates on learned residue representations that encode local neighborhood properties, biochemical characteristics, and structural context. The optimal transport solver then identifies which residues should be matched by minimizing the total transportation cost---effectively the sum of dissimilarities between matched residue pairs---across the embedding space.

This approach naturally produces soft, many-to-many alignments where functionally and structurally similar residues are preferentially matched, while simultaneously identifying the corresponding aligned fragments without explicit enumeration. Mathematically, we formulate this as the following optimal transport problem with entropic constraints:
\begin{align}
\min_{\Omega} \quad & \sum_{i=1}^{N} \sum_{j=1}^{M} \Omega_{ij} \gC_{ij} - \lambda \sum_{i=1}^{N} \sum_{j=1}^{M} \Omega_{ij} \log(\Omega_{ij}) \label{eq:ot} \\
\text{subject to:} \quad & \sum_{j=1}^{M} \Omega_{ij} = 1, \quad \forall i \in \{1, \ldots, N\} \\
& \sum_{i=1}^{N} \Omega_{ij} = 1, \quad \forall j \in \{1, \ldots, M\} \\
& \Omega_{ij} \geq 0, \quad \forall i,j
\end{align}
Here, $\Omega \in \mathbb{R}^{N \times M}$ is the transport plan (alignment matrix), $\gC_{ij}$ represents the cost of aligning query residue $i$ to candidate residue $j$, and $\lambda$ is the entropic regularization parameter that controls the smoothness of the alignment. This optimization seeks to find the optimal transport plan that minimizes the total alignment cost while the entropic regularization term ($-\lambda$ term) encourages smooth, distributed assignments rather than hard one-to-one mappings. The equality constraints ensure each query residue distributes $1$ total weight and each candidate residue receives $1$ total weight.

\section{Complexity Analysis}
\label{sec:complexity-analysis}
PLASMA achieves optimal $O(N^2)$ complexity while maintaining full differentiability. The cost matrix computation dominates computational requirements, requiring $O(N \cdot M \cdot D) = O(N^2 \cdot D)$ operations for the hinge non-linearity between proteins of lengths $N$ and $M$, where $D$ represents the embedding dimension. The siamese network contributes $O(N \cdot D^2)$ operations per protein (if using a two-layer MLP), yielding total $O(N \cdot D^2)$ since $D \ll N$ in practice. The Sinkhorn algorithm requires $O(T \cdot N^2)$ operations where $T$ represents the number of iterations (typically $T \ll N$). The Plan Assessor contributes $O(N^2)$ for substructure similarity computation and $O(K^2 \cdot N^2)$ for confidence weight calculation via diagonal convolution with kernel size $K \ll N$. The overall complexity remains $O(N^2)$, matching the best achievable complexity of the methods based on dynamic programming.

\section{Detailed Experimental Setup}
\label{sec:detailed-experimental-setup}
\subsection{Benchmark Datasets: VenusX} 
\label{sec:app:dataset}
We construct our evaluation datasets from the \textbf{VenusX} \citep{tanVenusXUnlockingFinegrained2025} benchmark (https://github.com/ai4protein/VenusX), which provides protein pairs with annotated biologically important substructures curated from the InterPro \citep{blumInterProProteinSequence2025} database. We focus on three substructure types: activation sites, binding sites, and motifs, corresponding to the \texttt{VenusX\_Res\_\{Act/BindI/Motif\}\_MP50} datasets where protein pairs share less than 50\% sequence similarity. These datasets present increasing difficulty due to their substructure sizes: active sites (18.7 $\pm$ 7.0 residues), binding sites (26.6 $\pm$ 21.7 residues), and motifs (80.23 $\pm$ 73.8 residues). From each VenusX dataset, we generate 20,000 protein pairs with balanced labels: half sharing the same InterPro family ID (positive pairs, $y=1$) and half from different families (negative pairs, $y=0$). Each sample is represented as $(\gP_q, \gP_c, \mathbf{l}_q, \mathbf{l}_c, y)$, where $\gP_q$ and $\gP_c$ are the protein pair, $\mathbf{l}_q$ and $\mathbf{l}_c$ are their respective substructure annotations, and $y$ indicates family membership.

To evaluate all the embedding based methods' generalization capability across different evolutionary contexts, we create two complementary test scenarios using three different random seeds for robust evaluation. This dual evaluation is crucial for protein analysis since biological systems constantly encounter both familiar protein families with slight variations and entirely novel protein architectures through evolution, horizontal gene transfer, and structural convergence. First, we randomly exclude 10\% of InterPro family IDs and split the remaining data into training (75\%), validation (5\%), and \texttt{test\_inter} (20\%). \texttt{test\_inter} evaluates \textit{interpolation} performance---the model's ability to recognize substructure similarities within the distribution of known protein families, mimicking scenarios where researchers analyze variants of well-characterized proteins. Second, we create \texttt{test\_extra} by sampling an equivalent number of protein pairs exclusively from the excluded InterPro families (maintaining the same 50--50 balance between positive and negative pairs). \texttt{test\_extra} evaluates \textit{extrapolation} performance---the model's ability to identify functional similarities in completely novel protein families, which is critical for annotating newly discovered proteins, understanding convergent evolution, and predicting function in understudied organisms. For each test scenario, the data exclusion and splitting procedure is repeated across three different seeds ($1$, $42$, and $100$) to ensure statistical reliability.

\subsection{Hyperparameter Configuration}
\label{sec:app:hyperparameters}
For both PLASMA and PLASMA-PF variants, we employ the following hyperparameters: the siamese network uses a hidden dimension of $512$ to balance expressiveness with computational efficiency. To ensure computational feasibility while maintaining statistical significance, our training sets only use $1500$ protein pairs by sampling $10\%$ of the full training set. The Sinkhorn temperature parameter $\tau$ is set to $0.1$ to encourage sparse, focused alignments that highlight the most relevant correspondences. The diagonal convolution kernel size $K=10$ captures sequential patterns in alignment matrices, while the residue matching threshold $\rho=0.5$ defines when transport weights indicate meaningful correspondences between residue pairs. See Appendix~\ref{sec:hyperparameter-analysis} for detailed sensitivity analysis and justification of these choices.

\section{Baselines}
\label{sec:app:baselines}
\subsection{Global Structure Alignment Methods}
Traditional structural biology approaches rely on atomic coordinates to identify protein similarities:
\begin{itemize}[leftmargin=*]
    \item \textsc{TM-Align} \citep{zhangTMalignProteinStructure2005} represents the gold standard for protein structure alignment based on Template Modeling scores. This method performs geometric alignment of protein backbones to identify structurally similar regions.
    \item \textsc{Foldseek} \citep{vankempenFastAccurateProtein2024} performs structural alignment using 3Di tokenizations, converting 3D structural information into sequence-like representations for comparison.
\end{itemize}

\subsection{Global Embedding-based Alignment}

\textsc{CosineSim} methods employ direct cosine similarity between globally aggregated protein embeddings from the backbone models discussed in Appendix \ref{subsec:backbones}, similar to the approach used in TM-Vec \citep{hamamsyProteinRemoteHomology2024}. This approach provides a baseline for embedding-based similarity without explicit residue-level alignment, representing proteins as single vectors and measuring their similarity through cosine distance.

\subsection{Local Embedding-based Alignment}
\textsc{EBA} \citep{pantoliniEmbeddingbasedAlignmentCombining2024} represents the current state-of-the-art in local embedding-based alignment, combining statistical alignment with neural embeddings to identify similar substructures. This method performs local alignment at the residue level using learned representations.

\subsection{Backbones}\label{subsec:backbones}

We evaluate PLASMA with seven popular protein sequence and structure representation models, using the following specific versions and configurations:
\begin{itemize}[leftmargin=*]
    \item \textsc{Ankh} \citep{elnaggarAnkhOptimizedProtein2023}: We employ the \texttt{base} model variant, which is a compact encoder-decoder architecture optimized for protein sequences with 110 million parameters. This model was trained on protein sequences using a masked language modeling objective and represents one of the most parameter-efficient protein language models. \textit{Available at:} \url{https://huggingface.co/ElnaggarLab/ankh-base}
    \item \textsc{ESM2} \citep{linEvolutionaryscalePredictionAtomiclevel2023}: We utilize the \texttt{t33\_650M\_UR50D} variant, a 650-million parameter encoder-only transformer model with 33 layers. This model was trained on the UniRef50 database and represents one of the largest and most comprehensive protein language models available, providing rich contextual representations for protein analysis. \textit{Available at:} \url{https://huggingface.co/facebook/esm2_t33_650M_UR50D}
    \item \textsc{ProstT5} \citep{heinzingerBilingualLanguageModel2024}: We use the \texttt{AA2fold} checkpoint, which is specifically fine-tuned for protein folding applications. This bilingual language model can process both amino acid sequences and structural information, making it particularly well-suited for structure-aware protein analysis tasks. \textit{Available at:} \url{https://huggingface.co/Rostlab/ProstT5}
    \item \textsc{ProtT5} \citep{elnaggarProtTransCrackingLanguage2020}: We employ the \texttt{xl\_half\_uniref50-enc} model, which uses only the encoder component of the T5 architecture. This variant was trained on UniRef50 \citep{suzekUniRefComprehensiveNonredundant2007} sequences and provides balanced performance between computational efficiency and representation quality with approximately 3 billion parameters. \textit{Available at:} \url{https://huggingface.co/Rostlab/prot_t5_xl_half_uniref50-enc}
    \item \textsc{ProtSSN} \citep{tanSemanticalGeometricalProtein2025}: We utilize the \texttt{k20\_h512} configuration, which combines sequence and structural information through a hybrid architecture. The model uses $k=20$ nearest neighbors for structural context and hidden dimensions of $512$, enabling it to capture both sequential and geometric protein properties. \textit{Available at:} \url{https://github.com/tyang816/ProtSSN}
    \item \textsc{TM-Vec} \citep{hamamsyProteinRemoteHomology2024}: We employ the \texttt{cath\_model\_large} variant, which was specifically trained on the CATH structural classification database \citep{knudsenTheCATHDatabase2010}. This model specializes in learning structure-aware representations and is particularly effective for detecting remote homology relationships based on structural similarity. \textit{Available at:} \url{https://figshare.com/articles/dataset/TMvec_DeepBLAST_models/25810099}
    \item \textsc{ProtBERT} \citep{brandesProteinBERTUniversalDeeplearning2022}: We use the \texttt{bfd} checkpoint, which was trained on the Big Fantastic Database \citep{jumperHighlyAccurateProtein2021} containing over 2.1 billion protein sequences. This BERT-based model provides robust protein representations through bidirectional context modeling and large-scale pretraining. \textit{Available at:} \url{https://huggingface.co/Rostlab/prot_bert_bfd}
\end{itemize}

\clearpage
\section{Full Interpolation Performance Comparison}\label{sec:full-performance-comparision}
This section presents comprehensive experimental results using seven backbone protein representation learning models (\textsc{ProstT5}, \textsc{ProtT5}, \textsc{Ankh}, \textsc{ESM2}, \textsc{ProtSSN}, \textsc{TM-Vec}, and \textsc{ProtBERT}) across three substructure alignment tasks (motifs, binding sites, and active sites) on the \texttt{test\_inter} dataset. The key findings demonstrate that both PLASMA and PLASMA-PF consistently achieve superior performance across all backbone-task combinations, highlighting the robustness of our optimal transport framework regardless of the underlying protein representation model. Additionally, the Label Match Score (LMS) results show that the trainable PLASMA variant significantly outperforms the parameter-free PLASMA-PF in predicting precise locations of aligned substructures, validating the benefits of supervised learning for accurate residue-level alignment localization.
\begin{table}[htbp]
    \centering
    \caption{Comprehensive motif detection results on \texttt{test\_inter} dataset across seven protein representation models.}
    \renewcommand{\arraystretch}{1.2}
    \resizebox{\textwidth}{!}{
    \begin{tabular}{cc ccccccc}
        \toprule
        \midrule
        \multirow{2}[4]{*}{\textbf{Metrics}} & \multirow{2}[4]{*}{\textbf{Methods}} & \multicolumn{7}{c}{\textbf{Motif}} \\
        \cmidrule(lr){3-9}          &       & \textsc{ProstT5}  & \textsc{ProtT5} & \textsc{Ankh} & \textsc{ESM2} & \textsc{ProtSSN} & \textsc{TM-Vec}  & \textsc{ProtBERT}  \\
        \toprule
        \multirow{6}[4]{*}{\begin{sideways}ROC-AUC\end{sideways}} & \textsc{PLASMA} & \cellcolor[RGB]{255,201,92}$\mathbf{.97_{\pm.002}}$ & \cellcolor[RGB]{255,201,92}$\mathbf{.97_{\pm.002}}$ & \cellcolor[RGB]{255,208,113}$\mathbf{.95_{\pm.002}}$ & \cellcolor[RGB]{255,201,94}$\mathbf{.96_{\pm.002}}$ & \cellcolor[RGB]{255,206,108}$\mathbf{.96_{\pm.001}}$ & \cellcolor[RGB]{255,223,161}$\mathbf{.92_{\pm.004}}$ & \cellcolor[RGB]{255,242,215}$\mathbf{.87_{\pm.004}}$ \\
        & \textsc{PLASMA-PF} & \cellcolor[RGB]{255,213,130}$.94_{\pm.003}$ & \cellcolor[RGB]{255,206,107}$.96_{\pm.002}$ & \cellcolor[RGB]{255,209,117}$.95_{\pm.003}$ & \cellcolor[RGB]{255,220,151}$.93_{\pm.004}$ & \cellcolor[RGB]{255,230,180}$.91_{\pm.003}$ & \cellcolor[RGB]{255,240,212}$.87_{\pm.001}$ & \cellcolor[RGB]{255,246,227}$.85_{\pm.004}$ \\
        & \textsc{EBA}   & \cellcolor[RGB]{255,230,182}$.90_{\pm.004}$ & \cellcolor[RGB]{255,227,171}$.91_{\pm.004}$ & \cellcolor[RGB]{255,240,212}$.87_{\pm.005}$ & \cellcolor[RGB]{255,238,206}$.88_{\pm.003}$ & \cellcolor[RGB]{180,180,255}$.44_{\pm.002}$ & \cellcolor[RGB]{255,238,203}$.88_{\pm.004}$ & \cellcolor[RGB]{243,243,255}$.73_{\pm.006}$ \\
        & \textsc{CosineSim} & \cellcolor[RGB]{255,250,241}$.82_{\pm.008}$ & \cellcolor[RGB]{255,241,212}$.87_{\pm.003}$ & \cellcolor[RGB]{255,247,232}$.84_{\pm.006}$ & \cellcolor[RGB]{243,243,255}$.73_{\pm.009}$ & \cellcolor[RGB]{248,248,255}$.75_{\pm.006}$ & \cellcolor[RGB]{255,243,220}$.86_{\pm.005}$ & \cellcolor[RGB]{210,210,255}$.57_{\pm.014}$ \\
        \cmidrule{2-9}          & \textsc{Foldseek} & \multicolumn{7}{c}{\cellcolor[RGB]{255,249,237}$.83_{\pm.007}$} \\
        & \textsc{TM-Align} & \multicolumn{7}{c}{\cellcolor[RGB]{255,255,255}$.78_{\pm.003}$} \\
        \toprule
        \multirow{6}[4]{*}{\begin{sideways}PR-AUC\end{sideways}} & \textsc{PLASMA} & \cellcolor[RGB]{255,203,98}$\mathbf{.96_{\pm.002}}$ & \cellcolor[RGB]{255,203,99}$\mathbf{.96_{\pm.003}}$ & \cellcolor[RGB]{255,210,119}$\mathbf{.95_{\pm.002}}$ & \cellcolor[RGB]{255,201,92}$\mathbf{.97_{\pm.001}}$ & \cellcolor[RGB]{255,207,111}$\mathbf{.96_{\pm.001}}$ & \cellcolor[RGB]{255,225,164}$\mathbf{.93_{\pm.004}}$ & \cellcolor[RGB]{255,243,219}$\mathbf{.89_{\pm.003}}$ \\
        & \textsc{PLASMA-PF} & \cellcolor[RGB]{255,213,131}$.95_{\pm.003}$ & \cellcolor[RGB]{255,207,112}$.96_{\pm.002}$ & \cellcolor[RGB]{255,211,122}$.95_{\pm.003}$ & \cellcolor[RGB]{255,217,141}$.94_{\pm.002}$ & \cellcolor[RGB]{255,228,176}$.92_{\pm.001}$ & \cellcolor[RGB]{255,246,228}$.88_{\pm.002}$ & \cellcolor[RGB]{255,246,230}$.87_{\pm.003}$ \\
        & \textsc{EBA}   & \cellcolor[RGB]{255,229,179}$.92_{\pm.004}$ & \cellcolor[RGB]{255,226,168}$.93_{\pm.004}$ & \cellcolor[RGB]{255,239,207}$.90_{\pm.004}$ & \cellcolor[RGB]{255,237,202}$.90_{\pm.004}$ & \cellcolor[RGB]{180,180,255}$.45_{\pm.004}$ & \cellcolor[RGB]{255,235,197}$.91_{\pm.004}$ & \cellcolor[RGB]{244,244,255}$.78_{\pm.005}$ \\
        & \textsc{CosineSim} & \cellcolor[RGB]{255,252,246}$.85_{\pm.005}$ & \cellcolor[RGB]{255,244,223}$.88_{\pm.002}$ & \cellcolor[RGB]{255,249,238}$.86_{\pm.005}$ & \cellcolor[RGB]{241,241,255}$.76_{\pm.008}$ & \cellcolor[RGB]{246,246,255}$.78_{\pm.006}$ & \cellcolor[RGB]{254,244,222}$.88_{\pm.002}$ & \cellcolor[RGB]{216,216,255}$.63_{\pm.016}$ \\
        \cmidrule{2-9}          & \textsc{Foldseek} & \multicolumn{7}{c}{\cellcolor[RGB]{246,246,255}$.78_{\pm.008}$} \\
        & \textsc{TM-Align} & \multicolumn{7}{c}{\cellcolor[RGB]{255,255,255}$.83_{\pm.004}$} \\
        \toprule
        \multirow{6}[4]{*}{\begin{sideways}F1-MAX\end{sideways}} & \textsc{PLASMA} & \cellcolor[RGB]{255,206,109}$\mathbf{.92_{\pm.001}}$ & \cellcolor[RGB]{255,201,93}$\mathbf{.93_{\pm.001}}$ & \cellcolor[RGB]{255,201,92}$\mathbf{.93_{\pm.002}}$ & \cellcolor[RGB]{255,205,104}$\mathbf{.93_{\pm.004}}$ & \cellcolor[RGB]{255,211,123}$\mathbf{.91_{\pm.000}}$ & \cellcolor[RGB]{255,222,156}$\mathbf{.88_{\pm.004}}$ & \cellcolor[RGB]{255,243,221}$.80_{\pm.001}$ \\
        & \textsc{PLASMA-PF} & \cellcolor[RGB]{255,216,140}$.90_{\pm.005}$ & \cellcolor[RGB]{255,204,101}$.93_{\pm.002}$ & \cellcolor[RGB]{255,204,103}$.93_{\pm.004}$ & \cellcolor[RGB]{255,221,153}$.89_{\pm.004}$ & \cellcolor[RGB]{255,235,197}$.84_{\pm.004}$ & \cellcolor[RGB]{255,235,196}$.84_{\pm.002}$ & \cellcolor[RGB]{255,248,234}$.77_{\pm.002}$ \\
        & \textsc{EBA}   & \cellcolor[RGB]{255,236,198}$.84_{\pm.006}$ & \cellcolor[RGB]{255,230,180}$.86_{\pm.003}$ & \cellcolor[RGB]{255,242,218}$.80_{\pm.005}$ & \cellcolor[RGB]{255,242,216}$.81_{\pm.003}$ & \cellcolor[RGB]{180,180,255}$.00_{\pm.000}$ & \cellcolor[RGB]{254,240,211}$.82_{\pm.003}$ & \cellcolor[RGB]{253,253,255}$.69_{\pm.006}$ \\
        & \textsc{CosineSim} & \cellcolor[RGB]{255,251,244}$.74_{\pm.007}$ & \cellcolor[RGB]{255,244,224}$.79_{\pm.004}$ & \cellcolor[RGB]{255,249,238}$.76_{\pm.003}$ & \cellcolor[RGB]{253,253,255}$.69_{\pm.001}$ & \cellcolor[RGB]{254,254,255}$.70_{\pm.002}$ & \cellcolor[RGB]{255,246,230}$.78_{\pm.005}$ & \cellcolor[RGB]{251,251,255}$.67_{\pm.003}$ \\
        \cmidrule{2-9}          & \textsc{Foldseek} & \multicolumn{7}{c}{\cellcolor[RGB]{255,234,193}$.84_{\pm.007}$} \\
        & \textsc{TM-Align} & \multicolumn{7}{c}{\cellcolor[RGB]{255,255,255}$.70_{\pm.002}$} \\
        \toprule
        \multirow{2}[2]{*}{\begin{sideways}LMS\end{sideways} \vspace{3pt}} & \textsc{PLASMA} & \cellcolor[RGB]{255,201,92}$\mathbf{.91_{\pm.007}}$ & \cellcolor[RGB]{255,201,92}$\mathbf{.92_{\pm.001}}$ & \cellcolor[RGB]{255,201,92}$\mathbf{.92_{\pm.002}}$ & \cellcolor[RGB]{255,201,92}$\mathbf{.92_{\pm.005}}$ & \cellcolor[RGB]{255,201,92}$\mathbf{.73_{\pm.013}}$ & \cellcolor[RGB]{255,201,92}$\mathbf{.76_{\pm.005}}$ & \cellcolor[RGB]{255,201,92}$\mathbf{.71_{\pm.007}}$ \\
        & \textsc{PLASMA-PF} & \cellcolor[RGB]{255,255,255}$.57_{\pm.003}$ & \cellcolor[RGB]{255,255,255}$.37_{\pm.006}$ & \cellcolor[RGB]{255,255,255}$.75_{\pm.006}$ & \cellcolor[RGB]{255,255,255}$.46_{\pm.009}$ & \cellcolor[RGB]{255,255,255}$.21_{\pm.002}$ & \cellcolor[RGB]{255,255,255}$.45_{\pm.001}$ & \cellcolor[RGB]{255,255,255}$.39_{\pm.009}$ \\
        \midrule
        \bottomrule
    \end{tabular}%
}
    \label{tab:appendix_motif_test_frequent}%
\end{table}%
\begin{table}[htbp]
    \centering
    \caption{Comprehensive binding site detection results on \texttt{test\_inter} dataset across seven protein representation models.}
    \renewcommand{\arraystretch}{1.2}
    \resizebox{\textwidth}{!}{
    \begin{tabular}{cc ccccccc}
        \toprule
        \midrule
        \multirow{2}[4]{*}{\textbf{Metrics}} & \multirow{2}[4]{*}{\textbf{Methods}} & \multicolumn{7}{c}{\textbf{Binding Site}} \\
        \cmidrule(lr){3-9}          &       & \textsc{ProstT5}  & \textsc{ProtT5} & \textsc{Ankh} & \textsc{ESM2} & \textsc{ProtSSN} & \textsc{TM-Vec}  & \textsc{ProtBERT}  \\
        \toprule
        \multirow{6}[4]{*}{\begin{sideways}ROC-AUC\end{sideways}} & \textsc{PLASMA} & \cellcolor[RGB]{255,201,93}$\mathbf{.99_{\pm.001}}$ & \cellcolor[RGB]{255,202,95}$\mathbf{.99_{\pm.000}}$ & \cellcolor[RGB]{255,201,92}$\mathbf{.99_{\pm.000}}$ & \cellcolor[RGB]{255,202,96}$\mathbf{.99_{\pm.001}}$ & \cellcolor[RGB]{255,204,102}$\mathbf{.99_{\pm.001}}$ & \cellcolor[RGB]{255,223,159}$\mathbf{.96_{\pm.003}}$ & \cellcolor[RGB]{255,214,133}$\mathbf{.98_{\pm.001}}$ \\
        & \textsc{PLASMA-PF} & \cellcolor[RGB]{255,206,108}$.99_{\pm.001}$ & \cellcolor[RGB]{255,207,110}$.99_{\pm.001}$ & \cellcolor[RGB]{255,205,104}$.99_{\pm.000}$ & \cellcolor[RGB]{255,228,175}$.96_{\pm.003}$ & \cellcolor[RGB]{255,216,137}$.97_{\pm.001}$ & \cellcolor[RGB]{255,242,217}$.92_{\pm.004}$ & \cellcolor[RGB]{254,248,236}$.90_{\pm.003}$ \\
        & \textsc{EBA}   & \cellcolor[RGB]{255,218,144}$.97_{\pm.001}$ & \cellcolor[RGB]{255,217,140}$.97_{\pm.001}$ & \cellcolor[RGB]{255,217,141}$.97_{\pm.001}$ & \cellcolor[RGB]{255,220,150}$.97_{\pm.002}$ & \cellcolor[RGB]{180,180,255}$.40_{\pm.005}$ & \cellcolor[RGB]{254,233,189}$.95_{\pm.000}$ & \cellcolor[RGB]{250,250,255}$.84_{\pm.006}$ \\
        & \textsc{CosineSim} & \cellcolor[RGB]{254,254,255}$.87_{\pm.005}$ & \cellcolor[RGB]{255,252,248}$.88_{\pm.004}$ & \cellcolor[RGB]{255,225,164}$.96_{\pm.002}$ & \cellcolor[RGB]{241,241,255}$.79_{\pm.009}$ & \cellcolor[RGB]{235,235,255}$.75_{\pm.008}$ & \cellcolor[RGB]{255,244,222}$.92_{\pm.006}$ & \cellcolor[RGB]{220,220,255}$.66_{\pm.008}$ \\
        \cmidrule{2-9}          & \textsc{Foldseek} & \multicolumn{7}{c}{\cellcolor[RGB]{255,252,246}$.89_{\pm.001}$} \\
        & \textsc{TM-Align} & \multicolumn{7}{c}{\cellcolor[RGB]{255,255,255}$.87_{\pm.003}$} \\
        \toprule
        \multirow{6}[4]{*}{\begin{sideways}PR-AUC\end{sideways}} & \textsc{PLASMA} & \cellcolor[RGB]{255,202,96}$\mathbf{.99_{\pm.001}}$ & \cellcolor[RGB]{255,202,96}$\mathbf{.99_{\pm.001}}$ & \cellcolor[RGB]{255,201,92}$\mathbf{.99_{\pm.000}}$ & \cellcolor[RGB]{255,201,94}$\mathbf{.99_{\pm.001}}$ & \cellcolor[RGB]{255,204,101}$\mathbf{.99_{\pm.001}}$ & \cellcolor[RGB]{255,227,170}$\mathbf{.97_{\pm.002}}$ & \cellcolor[RGB]{255,216,139}$\mathbf{.98_{\pm.001}}$ \\
        & \textsc{PLASMA-PF} & \cellcolor[RGB]{255,208,115}$.99_{\pm.001}$ & \cellcolor[RGB]{255,210,120}$.99_{\pm.001}$ & \cellcolor[RGB]{255,206,108}$.99_{\pm.000}$ & \cellcolor[RGB]{255,229,179}$.97_{\pm.002}$ & \cellcolor[RGB]{255,217,142}$.98_{\pm.001}$ & \cellcolor[RGB]{255,251,244}$.93_{\pm.004}$ & \cellcolor[RGB]{255,251,244}$.93_{\pm.001}$ \\
        & \textsc{EBA}   & \cellcolor[RGB]{255,220,151}$.98_{\pm.000}$ & \cellcolor[RGB]{255,218,145}$.98_{\pm.001}$ & \cellcolor[RGB]{255,219,147}$.98_{\pm.001}$ & \cellcolor[RGB]{255,222,156}$.98_{\pm.001}$ & \cellcolor[RGB]{180,180,255}$.42_{\pm.004}$ & \cellcolor[RGB]{255,233,189}$.96_{\pm.001}$ & \cellcolor[RGB]{248,248,255}$.87_{\pm.003}$ \\
        & \textsc{CosineSim} & \cellcolor[RGB]{253,253,255}$.90_{\pm.005}$ & \cellcolor[RGB]{253,253,255}$.90_{\pm.003}$ & \cellcolor[RGB]{255,228,175}$.97_{\pm.002}$ & \cellcolor[RGB]{242,242,255}$.83_{\pm.007}$ & \cellcolor[RGB]{235,235,255}$.78_{\pm.005}$ & \cellcolor[RGB]{255,247,231}$.94_{\pm.004}$ & \cellcolor[RGB]{222,222,255}$.70_{\pm.006}$ \\
        \cmidrule{2-9}          & \textsc{Foldseek} & \multicolumn{7}{c}{\cellcolor[RGB]{242,242,255}$.83_{\pm.002}$} \\
        & \textsc{TM-Align} & \multicolumn{7}{c}{\cellcolor[RGB]{255,255,255}$.91_{\pm.002}$} \\
        \toprule
        \multirow{6}[4]{*}{\begin{sideways}F1-MAX\end{sideways}} & \textsc{PLASMA} & \cellcolor[RGB]{255,203,100}$\mathbf{.98_{\pm.002}}$ & \cellcolor[RGB]{255,201,92}$\mathbf{.98_{\pm.001}}$ & \cellcolor[RGB]{255,202,97}$\mathbf{.98_{\pm.001}}$ & \cellcolor[RGB]{255,202,95}$\mathbf{.98_{\pm.002}}$ & \cellcolor[RGB]{255,208,114}$\mathbf{.97_{\pm.002}}$ & \cellcolor[RGB]{255,222,157}$\mathbf{.95_{\pm.002}}$ & \cellcolor[RGB]{255,228,175}$\mathbf{.94_{\pm.002}}$ \\
        & \textsc{PLASMA-PF} & \cellcolor[RGB]{255,216,138}$.96_{\pm.001}$ & \cellcolor[RGB]{255,208,114}$.97_{\pm.001}$ & \cellcolor[RGB]{255,213,129}$.97_{\pm.001}$ & \cellcolor[RGB]{255,237,202}$.92_{\pm.003}$ & \cellcolor[RGB]{255,229,178}$.94_{\pm.001}$ & \cellcolor[RGB]{255,241,214}$.91_{\pm.005}$ & \cellcolor[RGB]{253,253,255}$.83_{\pm.002}$ \\
        & \textsc{EBA}   & \cellcolor[RGB]{255,229,178}$.94_{\pm.001}$ & \cellcolor[RGB]{255,228,174}$.94_{\pm.001}$ & \cellcolor[RGB]{255,228,175}$.94_{\pm.001}$ & \cellcolor[RGB]{255,232,188}$.93_{\pm.002}$ & \cellcolor[RGB]{180,180,255}$.00_{\pm.000}$ & \cellcolor[RGB]{255,235,195}$.93_{\pm.001}$ & \cellcolor[RGB]{249,249,255}$.78_{\pm.007}$ \\
        & \textsc{CosineSim} & \cellcolor[RGB]{251,251,255}$.80_{\pm.008}$ & \cellcolor[RGB]{251,251,255}$.80_{\pm.005}$ & \cellcolor[RGB]{255,242,216}$.91_{\pm.005}$ & \cellcolor[RGB]{244,244,255}$.73_{\pm.006}$ & \cellcolor[RGB]{242,242,254}$.69_{\pm.006}$ & \cellcolor[RGB]{255,253,249}$.86_{\pm.007}$ & \cellcolor[RGB]{239,239,255}$.67_{\pm.001}$ \\
        \cmidrule{2-9}          & \textsc{Foldseek} & \multicolumn{7}{c}{\cellcolor[RGB]{255,228,175}$.94_{\pm.001}$} \\
        & \textsc{TM-Align} & \multicolumn{7}{c}{\cellcolor[RGB]{255,255,255}$.84_{\pm.005}$} \\
        \toprule
        \multirow{2}[2]{*}{\begin{sideways}LMS\end{sideways} \vspace{3pt}} & \textsc{PLASMA} & \cellcolor[RGB]{255,201,92}$\mathbf{.93_{\pm.002}}$ & \cellcolor[RGB]{255,201,92}$\mathbf{.93_{\pm.003}}$ & \cellcolor[RGB]{255,201,92}$\mathbf{.93_{\pm.004}}$ & \cellcolor[RGB]{255,201,92}$\mathbf{.93_{\pm.003}}$ & \cellcolor[RGB]{255,201,92}$\mathbf{.85_{\pm.006}}$ & \cellcolor[RGB]{255,201,92}$\mathbf{.86_{\pm.002}}$ & \cellcolor[RGB]{255,201,92}$\mathbf{.84_{\pm.003}}$ \\
        & \textsc{PLASMA-PF} & \cellcolor[RGB]{255,255,255}$.80_{\pm.008}$ & \cellcolor[RGB]{255,255,255}$.59_{\pm.008}$ & \cellcolor[RGB]{255,255,255}$.85_{\pm.005}$ & \cellcolor[RGB]{255,255,255}$.57_{\pm.009}$ & \cellcolor[RGB]{255,255,255}$.36_{\pm.005}$ & \cellcolor[RGB]{255,255,255}$.60_{\pm.008}$ & \cellcolor[RGB]{255,255,255}$.44_{\pm.004}$ \\
        \midrule
        \bottomrule
    \end{tabular}%
}
    \label{tab:appendix_binding_site_test_frequent}%
\end{table}%
\begin{table}[htbp]
    \centering
    \caption{Comprehensive active site detection results on \texttt{test\_inter} dataset across seven protein representation models.}
    \renewcommand{\arraystretch}{1.2}
    \resizebox{\textwidth}{!}{
    \begin{tabular}{cc ccccccc}
        \toprule
        \midrule
        \multirow{2}[4]{*}{\textbf{Metrics}} & \multirow{2}[4]{*}{\textbf{Methods}} & \multicolumn{7}{c}{\textbf{Active Site}} \\
        \cmidrule(lr){3-9}          &       & \textsc{ProstT5}  & \textsc{ProtT5} & \textsc{Ankh} & \textsc{ESM2} & \textsc{ProtSSN} & \textsc{TM-Vec}  & \textsc{ProtBERT}  \\
        \toprule
        \multirow{6}[4]{*}{\begin{sideways}ROC-AUC\end{sideways}} & \textsc{PLASMA} & \cellcolor[RGB]{255,202,95}$\mathbf{.99_{\pm.001}}$ & \cellcolor[RGB]{255,202,96}$\mathbf{.99_{\pm.001}}$ & \cellcolor[RGB]{255,201,92}$\mathbf{.99_{\pm.001}}$ & \cellcolor[RGB]{255,205,104}$\mathbf{.99_{\pm.001}}$ & \cellcolor[RGB]{255,208,114}$\mathbf{.99_{\pm.002}}$ & \cellcolor[RGB]{255,209,117}$\mathbf{.99_{\pm.003}}$ & \cellcolor[RGB]{255,218,144}$\mathbf{.99_{\pm.004}}$ \\
        & \textsc{PLASMA-PF} & \cellcolor[RGB]{255,209,117}$.99_{\pm.002}$ & \cellcolor[RGB]{255,210,121}$.99_{\pm.003}$ & \cellcolor[RGB]{255,212,127}$.99_{\pm.003}$ & \cellcolor[RGB]{255,249,237}$.96_{\pm.002}$ & \cellcolor[RGB]{255,223,161}$.98_{\pm.002}$ & \cellcolor[RGB]{255,219,146}$.98_{\pm.003}$ & \cellcolor[RGB]{254,254,255}$.94_{\pm.006}$ \\
        & \textsc{EBA}   & \cellcolor[RGB]{255,214,132}$.99_{\pm.003}$ & \cellcolor[RGB]{255,215,137}$.99_{\pm.003}$ & \cellcolor[RGB]{255,216,138}$.99_{\pm.003}$ & \cellcolor[RGB]{255,217,142}$.99_{\pm.003}$ & \cellcolor[RGB]{180,180,255}$.43_{\pm.005}$ & \cellcolor[RGB]{255,215,134}$.99_{\pm.003}$ & \cellcolor[RGB]{248,248,254}$.90_{\pm.005}$ \\
        & \textsc{CosineSim} & \cellcolor[RGB]{249,249,255}$.91_{\pm.004}$ & \cellcolor[RGB]{249,249,255}$.91_{\pm.003}$ & \cellcolor[RGB]{255,235,194}$.97_{\pm.002}$ & \cellcolor[RGB]{231,231,255}$.78_{\pm.009}$ & \cellcolor[RGB]{225,225,255}$.74_{\pm.006}$ & \cellcolor[RGB]{255,230,182}$.98_{\pm.002}$ & \cellcolor[RGB]{212,212,255}$.66_{\pm.003}$ \\
        \cmidrule{2-9}          & \textsc{Foldseek} & \multicolumn{7}{c}{\cellcolor[RGB]{247,247,255}$.89_{\pm.001}$} \\
        & \textsc{TM-Align} & \multicolumn{7}{c}{\cellcolor[RGB]{255,255,255}$.94_{\pm.003}$} \\
        \toprule
        \multirow{6}[4]{*}{\begin{sideways}PR-AUC\end{sideways}} & \textsc{PLASMA} & \cellcolor[RGB]{255,203,100}$\mathbf{.99_{\pm.000}}$ & \cellcolor[RGB]{255,202,96}$\mathbf{.99_{\pm.001}}$ & \cellcolor[RGB]{255,201,92}$\mathbf{.99_{\pm.001}}$ & \cellcolor[RGB]{255,204,103}$\mathbf{.99_{\pm.000}}$ & \cellcolor[RGB]{255,206,107}$\mathbf{.99_{\pm.001}}$ & \cellcolor[RGB]{255,210,121}$\mathbf{.99_{\pm.002}}$ & \cellcolor[RGB]{255,220,150}$\mathbf{.99_{\pm.003}}$ \\
        & \textsc{PLASMA-PF} & \cellcolor[RGB]{255,210,119}$.99_{\pm.001}$ & \cellcolor[RGB]{255,214,131}$.99_{\pm.002}$ & \cellcolor[RGB]{255,210,121}$.99_{\pm.002}$ & \cellcolor[RGB]{255,247,230}$.97_{\pm.001}$ & \cellcolor[RGB]{255,221,155}$.99_{\pm.001}$ & \cellcolor[RGB]{255,226,168}$.98_{\pm.003}$ & \cellcolor[RGB]{254,254,254}$.95_{\pm.004}$ \\
        & \textsc{EBA}   & \cellcolor[RGB]{255,214,132}$.99_{\pm.003}$ & \cellcolor[RGB]{255,214,133}$.99_{\pm.002}$ & \cellcolor[RGB]{255,215,134}$.99_{\pm.002}$ & \cellcolor[RGB]{255,215,136}$.99_{\pm.002}$ & \cellcolor[RGB]{180,180,255}$.43_{\pm.006}$ & \cellcolor[RGB]{255,214,134}$.99_{\pm.003}$ & \cellcolor[RGB]{249,249,255}$.92_{\pm.003}$ \\
        & \textsc{CosineSim} & \cellcolor[RGB]{251,251,255}$.93_{\pm.002}$ & \cellcolor[RGB]{250,250,255}$.92_{\pm.001}$ & \cellcolor[RGB]{255,235,194}$.98_{\pm.001}$ & \cellcolor[RGB]{237,237,255}$.83_{\pm.004}$ & \cellcolor[RGB]{231,231,255}$.79_{\pm.002}$ & \cellcolor[RGB]{255,232,188}$.98_{\pm.001}$ & \cellcolor[RGB]{219,219,255}$.70_{\pm.007}$ \\
        \cmidrule{2-9}          & \textsc{Foldseek} & \multicolumn{7}{c}{\cellcolor[RGB]{237,237,255}$.83_{\pm.006}$} \\
        & \textsc{TM-Align} & \multicolumn{7}{c}{\cellcolor[RGB]{255,255,255}$.96_{\pm.001}$} \\
        \toprule
        \multirow{6}[4]{*}{\begin{sideways}F1-MAX\end{sideways}} & \textsc{PLASMA} & \cellcolor[RGB]{255,204,101}$\mathbf{.98_{\pm.003}}$ & \cellcolor[RGB]{255,204,101}$\mathbf{.98_{\pm.003}}$ & \cellcolor[RGB]{255,201,92}$\mathbf{.99_{\pm.003}}$ & \cellcolor[RGB]{255,207,111}$\mathbf{.98_{\pm.001}}$ & \cellcolor[RGB]{255,202,97}$\mathbf{.99_{\pm.002}}$ & \cellcolor[RGB]{255,209,116}$\mathbf{.98_{\pm.003}}$ & \cellcolor[RGB]{255,225,166}$.96_{\pm.004}$ \\
        & \textsc{PLASMA-PF} & \cellcolor[RGB]{255,213,131}$.98_{\pm.003}$ & \cellcolor[RGB]{255,212,127}$.98_{\pm.004}$ & \cellcolor[RGB]{255,214,132}$.98_{\pm.003}$ & \cellcolor[RGB]{255,247,231}$.93_{\pm.004}$ & \cellcolor[RGB]{255,227,171}$.96_{\pm.003}$ & \cellcolor[RGB]{255,217,140}$.97_{\pm.004}$ & \cellcolor[RGB]{254,254,254}$.89_{\pm.005}$ \\
        & \textsc{EBA}   & \cellcolor[RGB]{255,217,140}$.97_{\pm.005}$ & \cellcolor[RGB]{255,215,136}$.98_{\pm.004}$ & \cellcolor[RGB]{255,216,138}$.97_{\pm.003}$ & \cellcolor[RGB]{255,216,139}$.97_{\pm.003}$ & \cellcolor[RGB]{180,180,255}$.00_{\pm.000}$ & \cellcolor[RGB]{255,217,141}$.97_{\pm.005}$ & \cellcolor[RGB]{250,250,255}$.84_{\pm.004}$ \\
        & \textsc{CosineSim} & \cellcolor[RGB]{250,250,255}$.85_{\pm.004}$ & \cellcolor[RGB]{249,249,255}$.83_{\pm.002}$ & \cellcolor[RGB]{255,243,219}$.94_{\pm.003}$ & \cellcolor[RGB]{239,239,255}$.71_{\pm.006}$ & \cellcolor[RGB]{237,237,255}$.68_{\pm.001}$ & \cellcolor[RGB]{254,244,223}$.93_{\pm.002}$ & \cellcolor[RGB]{235,235,255}$.67_{\pm.006}$ \\
        \cmidrule{2-9}          & \textsc{Foldseek} & \multicolumn{7}{c}{\cellcolor[RGB]{255,219,147}$.97_{\pm.005}$} \\
        & \textsc{TM-Align} & \multicolumn{7}{c}{\cellcolor[RGB]{255,255,255}$.90_{\pm.003}$} \\
        \toprule
        \multirow{2}[2]{*}{\begin{sideways}LMS\end{sideways} \vspace{3pt}} & \textsc{PLASMA} & \cellcolor[RGB]{255,201,92}$\mathbf{.97_{\pm.004}}$ & \cellcolor[RGB]{255,201,92}$\mathbf{.97_{\pm.004}}$ & \cellcolor[RGB]{255,201,92}$\mathbf{.97_{\pm.003}}$ & \cellcolor[RGB]{255,201,92}$\mathbf{.97_{\pm.004}}$ & \cellcolor[RGB]{255,201,92}$\mathbf{.89_{\pm.016}}$ & \cellcolor[RGB]{255,201,92}$\mathbf{.93_{\pm.006}}$ & \cellcolor[RGB]{255,201,92}$\mathbf{.89_{\pm.008}}$ \\
        & \textsc{PLASMA-PF} & \cellcolor[RGB]{255,255,255}$.91_{\pm.010}$ & \cellcolor[RGB]{255,255,255}$.68_{\pm.003}$ & \cellcolor[RGB]{255,255,255}$.95_{\pm.006}$ & \cellcolor[RGB]{255,255,255}$.63_{\pm.013}$ & \cellcolor[RGB]{255,255,255}$.43_{\pm.007}$ & \cellcolor[RGB]{255,255,255}$.77_{\pm.011}$ & \cellcolor[RGB]{255,255,255}$.52_{\pm.004}$ \\
        \midrule
        \bottomrule
    \end{tabular}%
}
    \label{tab:appendix_active_site_test_frequent}%
\end{table}%
\renewcommand{\arraystretch}{2}
\begin{table}[htbp]
    \centering
    \caption{\new{Model performance on \texttt{test\_inter} (mean $\pm$ std over three independent seeds). Colors indicate relative performance versus TM-Align, percentage values report the associated specific relative performance difference.}}
    \resizebox{\textwidth}{!}{
    \begin{tabular}{cc ccc ccc ccc}
        \toprule
        \midrule
        \multirow{2}[4]{*}{\textbf{Metrics}} & \multirow{2}[4]{*}{\textbf{Methods}} & \multicolumn{3}{c}{\textbf{Motif}} & \multicolumn{3}{c}{\textbf{Binding Site}} & \multicolumn{3}{c}{\textbf{Active Site}} \\
        \cmidrule(lr){3-5}\cmidrule(lr){6-8}\cmidrule(lr){9-11}          &       & \textsc{Ankh}  & \textsc{ESM2}  & \textsc{ProtSSN} & \textsc{Ankh}  & \textsc{ESM2}  & \textsc{ProtSSN} & \textsc{Ankh}  & \textsc{ESM2}  & \textsc{ProtSSN} \\
        \toprule
        \multirow{6}[4]{*}{\begin{sideways}ROC-AUC\end{sideways}} & PLASMA & \cellcolor[RGB]{255,207,111}$\mathbf{.95_{\pm.002}^{\uparrow 21.8\%}}$ & \cellcolor[RGB]{255,201,92}$\mathbf{.96_{\pm.002}^{\uparrow 23.1\%}}$ & \cellcolor[RGB]{255,205,106}$\mathbf{.96_{\pm.001}^{\uparrow 23.1\%}}$ & \cellcolor[RGB]{255,201,92}$\mathbf{.99_{\pm.000}^{\uparrow 13.8\%}}$ & \cellcolor[RGB]{255,202,96}$\mathbf{.99_{\pm.001}^{\uparrow 13.8\%}}$ & \cellcolor[RGB]{255,204,102}$\mathbf{.99_{\pm.001}^{\uparrow 13.8\%}}$ & \cellcolor[RGB]{255,201,92}$\mathbf{.99_{\pm.001}^{\uparrow 5.3\%}}$ & \cellcolor[RGB]{255,205,104}$\mathbf{.99_{\pm.001}^{\uparrow 5.3\%}}$ & \cellcolor[RGB]{255,208,114}$\mathbf{.99_{\pm.002}^{\uparrow 5.3\%}}$ \\
        & PLASMA-PF & \cellcolor[RGB]{255,208,115}$.95_{\pm.003}^{\uparrow 21.8\%}$ & \cellcolor[RGB]{255,220,150}$.93_{\pm.004}^{\uparrow 19.2\%}$ & \cellcolor[RGB]{255,230,179}$.91_{\pm.003}^{\uparrow 16.7\%}$ & \cellcolor[RGB]{255,205,104}$.99_{\pm.000}^{\uparrow 13.8\%}$ & \cellcolor[RGB]{255,228,175}$.96_{\pm.003}^{\uparrow 10.3\%}$ & \cellcolor[RGB]{255,216,137}$.97_{\pm.001}^{\uparrow 11.5\%}$ & \cellcolor[RGB]{255,212,127}$.99_{\pm.003}^{\uparrow 5.3\%}$ & \cellcolor[RGB]{255,249,237}$.96_{\pm.002}^{\uparrow 2.1\%}$ & \cellcolor[RGB]{255,223,161}$.98_{\pm.002}^{\uparrow 4.3\%}$ \\
        & EBA   & \cellcolor[RGB]{255,240,211}$.87_{\pm.005}^{\uparrow 11.5\%}$ & \cellcolor[RGB]{255,238,206}$.88_{\pm.003}^{\uparrow 12.8\%}$ & \cellcolor[RGB]{180,180,255}$.44_{\pm.002}^{\downarrow 43.6\%}$ & \cellcolor[RGB]{255,217,141}$.97_{\pm.001}^{\uparrow 11.5\%}$ & \cellcolor[RGB]{255,220,150}$.97_{\pm.002}^{\uparrow 11.5\%}$ & \cellcolor[RGB]{180,180,255}$.40_{\pm.005}^{\downarrow 54.0\%}$ & \cellcolor[RGB]{255,216,138}$.99_{\pm.003}^{\uparrow 5.3\%}$ & \cellcolor[RGB]{255,217,142}$.99_{\pm.003}^{\uparrow 5.3\%}$ & \cellcolor[RGB]{180,180,255}$.43_{\pm.005}^{\downarrow 54.3\%}$ \\
        & Backbone & \cellcolor[RGB]{254,247,231}$.84_{\pm.006}^{\uparrow 7.7\%}$ & \cellcolor[RGB]{243,243,255}$.73_{\pm.009}^{\downarrow 6.4\%}$ & \cellcolor[RGB]{248,248,254}$.75_{\pm.006}^{\downarrow 3.8\%}$ & \cellcolor[RGB]{255,225,164}$.96_{\pm.002}^{\uparrow 10.3\%}$ & \cellcolor[RGB]{241,241,255}$.79_{\pm.009}^{\downarrow 9.2\%}$ & \cellcolor[RGB]{235,235,255}$.75_{\pm.008}^{\downarrow 13.8\%}$ & \cellcolor[RGB]{255,235,194}$.97_{\pm.002}^{\uparrow 3.2\%}$ & \cellcolor[RGB]{231,231,255}$.78_{\pm.009}^{\downarrow 17.0\%}$ & \cellcolor[RGB]{225,225,255}$.74_{\pm.006}^{\downarrow 21.3\%}$ \\
        \cmidrule{2-11}          & Foldseek & \multicolumn{3}{c}{\cellcolor[RGB]{254,249,237}$.83_{\pm.007}^{\uparrow 6.4\%}$} & \multicolumn{3}{c}{\cellcolor[RGB]{255,252,246}$.89_{\pm.001}^{\uparrow 2.3\%}$} & \multicolumn{3}{c}{\cellcolor[RGB]{247,247,255}$.89_{\pm.001}^{\downarrow 5.3\%}$} \\
        & TM-Align & \multicolumn{3}{c}{\cellcolor[RGB]{255,255,255}$.78_{\pm.003}$} & \multicolumn{3}{c}{\cellcolor[RGB]{255,255,255}$.87_{\pm.003}$} & \multicolumn{3}{c}{\cellcolor[RGB]{255,255,255}$.94_{\pm.003}$} \\
        \toprule
        \multirow{6}[4]{*}{\begin{sideways}PR-AUC\end{sideways}} & PLASMA & \cellcolor[RGB]{255,210,119}$\mathbf{.95_{\pm.002}^{\uparrow 14.5\%}}$ & \cellcolor[RGB]{255,201,92}$\mathbf{.97_{\pm.001}^{\uparrow 16.9\%}}$ & \cellcolor[RGB]{255,207,111}$\mathbf{.96_{\pm.001}^{\uparrow 15.7\%}}$ & \cellcolor[RGB]{255,201,92}$\mathbf{.99_{\pm.000}^{\uparrow 8.8\%}}$ & \cellcolor[RGB]{255,201,94}$\mathbf{.99_{\pm.001}^{\uparrow 8.8\%}}$ & \cellcolor[RGB]{255,204,101}$\mathbf{.99_{\pm.001}^{\uparrow 8.8\%}}$ & \cellcolor[RGB]{255,201,92}$\mathbf{.99_{\pm.001}^{\uparrow 3.1\%}}$ & \cellcolor[RGB]{255,204,103}$\mathbf{.99_{\pm.000}^{\uparrow 3.1\%}}$ & \cellcolor[RGB]{255,206,107}$\mathbf{.99_{\pm.001}^{\uparrow 3.1\%}}$ \\
        & PLASMA-PF & \cellcolor[RGB]{255,211,122}$.95_{\pm.003}^{\uparrow 14.5\%}$ & \cellcolor[RGB]{255,217,141}$.94_{\pm.002}^{\uparrow 13.3\%}$ & \cellcolor[RGB]{255,228,176}$.92_{\pm.001}^{\uparrow 10.8\%}$ & \cellcolor[RGB]{255,206,108}$.99_{\pm.000}^{\uparrow 8.8\%}$ & \cellcolor[RGB]{255,229,179}$.97_{\pm.002}^{\uparrow 6.6\%}$ & \cellcolor[RGB]{255,217,142}$.98_{\pm.001}^{\uparrow 7.7\%}$ & \cellcolor[RGB]{255,210,121}$.99_{\pm.002}^{\uparrow 3.1\%}$ & \cellcolor[RGB]{255,247,230}$.97_{\pm.001}^{\uparrow 1.0\%}$ & \cellcolor[RGB]{255,221,155}$.99_{\pm.001}^{\uparrow 3.1\%}$ \\
        & EBA   & \cellcolor[RGB]{255,239,207}$.90_{\pm.004}^{\uparrow 8.4\%}$ & \cellcolor[RGB]{255,237,202}$.90_{\pm.004}^{\uparrow 8.4\%}$ & \cellcolor[RGB]{180,180,255}$.45_{\pm.004}^{\downarrow 45.8\%}$ & \cellcolor[RGB]{255,219,147}$.98_{\pm.001}^{\uparrow 7.7\%}$ & \cellcolor[RGB]{255,222,156}$.98_{\pm.001}^{\uparrow 7.7\%}$ & \cellcolor[RGB]{180,180,255}$.42_{\pm.004}^{\downarrow 53.8\%}$ & \cellcolor[RGB]{255,215,134}$.99_{\pm.002}^{\uparrow 3.1\%}$ & \cellcolor[RGB]{255,215,136}$.99_{\pm.002}^{\uparrow 3.1\%}$ & \cellcolor[RGB]{180,180,255}$.43_{\pm.006}^{\downarrow 55.2\%}$ \\
        & Backbone & \cellcolor[RGB]{255,249,238}$.86_{\pm.005}^{\uparrow 3.6\%}$ & \cellcolor[RGB]{241,241,255}$.76_{\pm.008}^{\downarrow 8.4\%}$ & \cellcolor[RGB]{246,246,255}$.78_{\pm.006}^{\downarrow 6.0\%}$ & \cellcolor[RGB]{255,228,175}$.97_{\pm.002}^{\uparrow 6.6\%}$ & \cellcolor[RGB]{242,242,255}$.83_{\pm.007}^{\downarrow 8.8\%}$ & \cellcolor[RGB]{235,235,255}$.78_{\pm.005}^{\downarrow 14.3\%}$ & \cellcolor[RGB]{255,235,194}$.98_{\pm.001}^{\uparrow 2.1\%}$ & \cellcolor[RGB]{237,237,255}$.83_{\pm.004}^{\downarrow 13.5\%}$ & \cellcolor[RGB]{231,231,255}$.79_{\pm.002}^{\downarrow 17.7\%}$ \\
        \cmidrule{2-11}          & Foldseek & \multicolumn{3}{c}{\cellcolor[RGB]{246,246,255}$.78_{\pm.008}^{\downarrow 6.0\%}$} & \multicolumn{3}{c}{\cellcolor[RGB]{242,242,255}$.83_{\pm.002}^{\downarrow 8.8\%}$} & \multicolumn{3}{c}{\cellcolor[RGB]{237,237,255}$.83_{\pm.006}^{\downarrow 13.5\%}$} \\
        & TM-Align & \multicolumn{3}{c}{\cellcolor[RGB]{255,255,255}$.83_{\pm.004}$} & \multicolumn{3}{c}{\cellcolor[RGB]{255,255,255}$.91_{\pm.002}$} & \multicolumn{3}{c}{\cellcolor[RGB]{255,255,255}$.96_{\pm.001}$} \\
        \toprule
        \multirow{6}[4]{*}{\begin{sideways}F1-MAX\end{sideways}} & PLASMA & \cellcolor[RGB]{255,201,92}$\mathbf{.93_{\pm.002}^{\uparrow 32.9\%}}$ & \cellcolor[RGB]{255,205,104}$\mathbf{.93_{\pm.004}^{\uparrow 32.9\%}}$ & \cellcolor[RGB]{255,211,123}$\mathbf{.91_{\pm.000}^{\uparrow 30.0\%}}$ & \cellcolor[RGB]{255,201,93}$\mathbf{.98_{\pm.001}^{\uparrow 16.7\%}}$ & \cellcolor[RGB]{255,201,92}$\mathbf{.98_{\pm.002}^{\uparrow 16.7\%}}$ & \cellcolor[RGB]{255,207,111}$\mathbf{.97_{\pm.002}^{\uparrow 15.5\%}}$ & \cellcolor[RGB]{255,201,92}$\mathbf{.99_{\pm.003}^{\uparrow 10.0\%}}$ & \cellcolor[RGB]{255,207,111}$\mathbf{.98_{\pm.001}^{\uparrow 8.9\%}}$ & \cellcolor[RGB]{255,202,97}$\mathbf{.99_{\pm.002}^{\uparrow 10.0\%}}$ \\
        & PLASMA-PF & \cellcolor[RGB]{255,204,103}$.93_{\pm.004}^{\uparrow 32.9\%}$ & \cellcolor[RGB]{255,221,153}$.89_{\pm.004}^{\uparrow 27.1\%}$ & \cellcolor[RGB]{255,235,197}$.84_{\pm.004}^{\uparrow 20.0\%}$ & \cellcolor[RGB]{255,212,127}$.97_{\pm.001}^{\uparrow 15.5\%}$ & \cellcolor[RGB]{255,237,202}$.92_{\pm.003}^{\uparrow 9.5\%}$ & \cellcolor[RGB]{255,229,176}$.94_{\pm.001}^{\uparrow 11.9\%}$ & \cellcolor[RGB]{255,214,132}$.98_{\pm.003}^{\uparrow 8.9\%}$ & \cellcolor[RGB]{255,247,231}$.93_{\pm.004}^{\uparrow 3.3\%}$ & \cellcolor[RGB]{255,227,171}$.96_{\pm.003}^{\uparrow 6.7\%}$ \\
        & EBA   & \cellcolor[RGB]{255,242,218}$.80_{\pm.005}^{\uparrow 14.3\%}$ & \cellcolor[RGB]{255,242,216}$.81_{\pm.003}^{\uparrow 15.7\%}$ & \cellcolor[RGB]{180,180,255}$.00_{\pm.000}^{\downarrow 100.0\%}$ & \cellcolor[RGB]{255,228,174}$.94_{\pm.001}^{\uparrow 11.9\%}$ & \cellcolor[RGB]{255,232,187}$.93_{\pm.002}^{\uparrow 10.7\%}$ & \cellcolor[RGB]{180,180,255}$.00_{\pm.000}^{\downarrow 100.0\%}$ & \cellcolor[RGB]{255,216,138}$.97_{\pm.003}^{\uparrow 7.8\%}$ & \cellcolor[RGB]{255,216,139}$.97_{\pm.003}^{\uparrow 7.8\%}$ & \cellcolor[RGB]{180,180,255}$.00_{\pm.000}^{\downarrow 100.0\%}$ \\
        & Backbone & \cellcolor[RGB]{255,249,238}$.76_{\pm.003}^{\uparrow 8.6\%}$ & \cellcolor[RGB]{253,253,255}$.69_{\pm.001}^{\downarrow 1.4\%}$ & \cellcolor[RGB]{254,254,255}$.70_{\pm.002}^{\downarrow 0.0\%}$ & \cellcolor[RGB]{255,241,215}$.91_{\pm.005}^{\uparrow 8.3\%}$ & \cellcolor[RGB]{244,244,255}$.73_{\pm.006}^{\downarrow 13.1\%}$ & \cellcolor[RGB]{242,242,254}$.69_{\pm.006}^{\downarrow 17.9\%}$ & \cellcolor[RGB]{255,243,219}$.94_{\pm.003}^{\uparrow 4.4\%}$ & \cellcolor[RGB]{239,239,255}$.71_{\pm.006}^{\downarrow 21.1\%}$ & \cellcolor[RGB]{237,237,255}$.68_{\pm.001}^{\downarrow 24.4\%}$ \\
        \cmidrule{2-11}          & Foldseek & \multicolumn{3}{c}{\cellcolor[RGB]{255,234,193}$.84_{\pm.007}^{\uparrow 20.0\%}$} & \multicolumn{3}{c}{\cellcolor[RGB]{255,228,174}$.94_{\pm.001}^{\uparrow 11.9\%}$} & \multicolumn{3}{c}{\cellcolor[RGB]{255,219,147}$.97_{\pm.005}^{\uparrow 7.8\%}$} \\
        & TM-Align & \multicolumn{3}{c}{\cellcolor[RGB]{255,255,255}$.70_{\pm.002}$} & \multicolumn{3}{c}{\cellcolor[RGB]{255,255,255}$.84_{\pm.005}$} & \multicolumn{3}{c}{\cellcolor[RGB]{255,255,255}$.90_{\pm.003}$} \\
        \midrule
        \bottomrule
    \end{tabular}%
}
    \label{tab:main_results_test_inter_with_percentage}%
\end{table}%
\renewcommand{\arraystretch}{1.0}

\FloatBarrier
\section{Full Extrapolation Performance Comparison}\label{sec:full-extrapolation-performance-comparison}
This section evaluates PLASMA's generalization capability on the \texttt{test\_extra} dataset, which contains substructures never encountered during training. These experiments are crucial for assessing applicability in detecting unknown substructures. The results demonstrate that PLASMA maintains superior performance even when confronted with completely unseen substructures, achieving the highest scores for both detecting the existence of similar substructures and accurately localizing their positions for most of the cases. This robust extrapolation performance further validates that our optimal transport framework captures fundamental protein substructure similarity patterns that transcend specific training examples, making it highly valuable for analyzing newly discovered proteins and understudied organisms.

\begin{table}[htbp]
    \centering
    \caption{Comprehensive motif detection results on \texttt{test\_extra} dataset across seven protein representation models.}
    \renewcommand{\arraystretch}{1.2}
    \resizebox{\textwidth}{!}{
    \begin{tabular}{cc ccccccc}
        \toprule
        \midrule
        \multirow{2}[4]{*}{\textbf{Metrics}} & \multirow{2}[4]{*}{\textbf{Methods}} & \multicolumn{7}{c}{\textbf{Motif}} \\
        \cmidrule(lr){3-9}          &       & \textsc{ProstT5}  & \textsc{ProtT5} & \textsc{Ankh} & \textsc{ESM2} & \textsc{ProtSSN} & \textsc{TM-Vec}  & \textsc{ProtBERT}  \\
        \toprule
        \multirow{6}[4]{*}{\begin{sideways}ROC-AUC\end{sideways}} & \textsc{PLASMA} & \cellcolor[RGB]{255,206,109}$\mathbf{.97_{\pm.015}}$ & \cellcolor[RGB]{255,203,100}$\mathbf{.98_{\pm.012}}$ & \cellcolor[RGB]{255,201,92}$\mathbf{.98_{\pm.008}}$ & \cellcolor[RGB]{255,207,112}$\mathbf{.97_{\pm.013}}$ & \cellcolor[RGB]{255,215,135}$\mathbf{.96_{\pm.016}}$ & \cellcolor[RGB]{255,222,156}$\mathbf{.95_{\pm.023}}$ & \cellcolor[RGB]{251,251,255}$.79_{\pm.022}$ \\
        & \textsc{PLASMA-PF} & \cellcolor[RGB]{255,209,117}$.97_{\pm.014}$ & \cellcolor[RGB]{255,204,103}$.98_{\pm.010}$ & \cellcolor[RGB]{255,201,94}$.98_{\pm.009}$ & \cellcolor[RGB]{255,228,175}$.93_{\pm.004}$ & \cellcolor[RGB]{255,238,205}$.90_{\pm.005}$ & \cellcolor[RGB]{255,244,222}$.88_{\pm.039}$ & \cellcolor[RGB]{255,254,252}$.82_{\pm.016}$ \\
        & \textsc{EBA}   & \cellcolor[RGB]{255,225,167}$.94_{\pm.017}$ & \cellcolor[RGB]{255,220,150}$.95_{\pm.009}$ & \cellcolor[RGB]{255,240,211}$.90_{\pm.033}$ & \cellcolor[RGB]{255,231,184}$.92_{\pm.021}$ & \cellcolor[RGB]{180,180,255}$.32_{\pm.043}$ & \cellcolor[RGB]{255,224,163}$.94_{\pm.016}$ & \cellcolor[RGB]{247,247,255}$.76_{\pm.025}$ \\
        & \textsc{CosineSim} & \cellcolor[RGB]{255,252,246}$.84_{\pm.029}$ & \cellcolor[RGB]{255,242,216}$.89_{\pm.024}$ & \cellcolor[RGB]{254,250,240}$.85_{\pm.019}$ & \cellcolor[RGB]{243,243,255}$.74_{\pm.033}$ & \cellcolor[RGB]{252,252,255}$.79_{\pm.018}$ & \cellcolor[RGB]{255,253,249}$.83_{\pm.050}$ & \cellcolor[RGB]{225,225,255}$.62_{\pm.080}$ \\
        \cmidrule{2-9}          & \textsc{Foldseek} & \multicolumn{7}{c}{\cellcolor[RGB]{255,242,217}$.89_{\pm.033}$} \\
        & \textsc{TM-Align} & \multicolumn{7}{c}{\cellcolor[RGB]{255,255,255}$.81_{\pm.014}$} \\
        \toprule
        \multirow{6}[4]{*}{\begin{sideways}PR-AUC\end{sideways}} & \textsc{PLASMA} & \cellcolor[RGB]{255,209,116}$\mathbf{.97_{\pm.017}}$ & \cellcolor[RGB]{255,208,114}$\mathbf{.97_{\pm.018}}$ & \cellcolor[RGB]{255,201,92}$\mathbf{.98_{\pm.011}}$ & \cellcolor[RGB]{255,206,110}$\mathbf{.97_{\pm.014}}$ & \cellcolor[RGB]{255,215,135}$\mathbf{.96_{\pm.017}}$ & \cellcolor[RGB]{255,226,168}$\mathbf{.95_{\pm.025}}$ & \cellcolor[RGB]{252,252,255}$.84_{\pm.014}$ \\
        & \textsc{PLASMA-PF} & \cellcolor[RGB]{255,211,122}$.97_{\pm.015}$ & \cellcolor[RGB]{255,210,120}$.97_{\pm.016}$ & \cellcolor[RGB]{255,201,93}$.98_{\pm.010}$ & \cellcolor[RGB]{255,224,163}$.95_{\pm.005}$ & \cellcolor[RGB]{255,239,209}$.92_{\pm.007}$ & \cellcolor[RGB]{255,250,242}$.88_{\pm.040}$ & \cellcolor[RGB]{254,254,253}$\mathbf{.86_{\pm.012}}$ \\
        & \textsc{EBA}   & \cellcolor[RGB]{255,227,172}$.94_{\pm.018}$ & \cellcolor[RGB]{255,220,151}$.96_{\pm.010}$ & \cellcolor[RGB]{255,242,217}$.91_{\pm.035}$ & \cellcolor[RGB]{255,233,189}$.93_{\pm.019}$ & \cellcolor[RGB]{180,180,255}$.38_{\pm.014}$ & \cellcolor[RGB]{255,225,167}$.95_{\pm.014}$ & \cellcolor[RGB]{246,246,255}$.80_{\pm.029}$ \\
        & \textsc{CosineSim} & \cellcolor[RGB]{254,254,255}$.85_{\pm.028}$ & \cellcolor[RGB]{255,246,228}$.90_{\pm.017}$ & \cellcolor[RGB]{255,254,254}$.86_{\pm.023}$ & \cellcolor[RGB]{241,241,255}$.77_{\pm.041}$ & \cellcolor[RGB]{249,249,255}$.82_{\pm.027}$ & \cellcolor[RGB]{255,254,254}$.86_{\pm.036}$ & \cellcolor[RGB]{223,223,255}$.66_{\pm.090}$ \\
        \cmidrule{2-9}          & \textsc{Foldseek} & \multicolumn{7}{c}{\cellcolor[RGB]{252,252,255}$.84_{\pm.031}$} \\
        & \textsc{TM-Align} & \multicolumn{7}{c}{\cellcolor[RGB]{255,255,255}$.86_{\pm.020}$} \\
        \toprule
        \multirow{6}[4]{*}{\begin{sideways}F1-MAX\end{sideways}} & \textsc{PLASMA} & \cellcolor[RGB]{255,207,111}$\mathbf{.95_{\pm.011}}$ & \cellcolor[RGB]{255,204,103}$\mathbf{.96_{\pm.010}}$ & \cellcolor[RGB]{255,201,92}$\mathbf{.97_{\pm.009}}$ & \cellcolor[RGB]{255,212,125}$\mathbf{.95_{\pm.018}}$ & \cellcolor[RGB]{255,223,158}$\mathbf{.92_{\pm.022}}$ & \cellcolor[RGB]{255,222,156}$\mathbf{.92_{\pm.022}}$ & \cellcolor[RGB]{251,251,255}$.72_{\pm.017}$ \\
        & \textsc{PLASMA-PF} & \cellcolor[RGB]{255,217,143}$.93_{\pm.019}$ & \cellcolor[RGB]{255,207,110}$.96_{\pm.006}$ & \cellcolor[RGB]{255,207,110}$.96_{\pm.013}$ & \cellcolor[RGB]{255,231,183}$.90_{\pm.006}$ & \cellcolor[RGB]{255,245,226}$.84_{\pm.008}$ & \cellcolor[RGB]{255,242,217}$.85_{\pm.041}$ & \cellcolor[RGB]{254,254,255}$.75_{\pm.017}$ \\
        & \textsc{EBA}   & \cellcolor[RGB]{255,235,197}$.88_{\pm.027}$ & \cellcolor[RGB]{254,229,179}$.90_{\pm.014}$ & \cellcolor[RGB]{255,241,213}$.86_{\pm.035}$ & \cellcolor[RGB]{255,238,206}$.87_{\pm.024}$ & \cellcolor[RGB]{180,180,255}$.00_{\pm.000}$ & \cellcolor[RGB]{255,237,202}$.87_{\pm.019}$ & \cellcolor[RGB]{251,251,255}$.73_{\pm.008}$ \\
        & \textsc{CosineSim} & \cellcolor[RGB]{255,253,250}$.77_{\pm.020}$ & \cellcolor[RGB]{255,247,231}$.82_{\pm.025}$ & \cellcolor[RGB]{255,252,246}$.79_{\pm.008}$ & \cellcolor[RGB]{249,249,255}$.70_{\pm.014}$ & \cellcolor[RGB]{252,252,255}$.73_{\pm.013}$ & \cellcolor[RGB]{255,254,252}$.77_{\pm.040}$ & \cellcolor[RGB]{247,247,255}$.68_{\pm.015}$ \\
        \cmidrule{2-9}          & \textsc{Foldseek} & \multicolumn{7}{c}{\cellcolor[RGB]{255,228,175}$.91_{\pm.046}$} \\
        & \textsc{TM-Align} & \multicolumn{7}{c}{\cellcolor[RGB]{255,255,255}$.76_{\pm.015}$} \\
        \toprule
        \multirow{2}[2]{*}{\begin{sideways}LMS\end{sideways} \vspace{3pt}} & \textsc{PLASMA} & \cellcolor[RGB]{255,201,92}$\mathbf{.72_{\pm.022}}$ & \cellcolor[RGB]{255,201,92}$\mathbf{.70_{\pm.022}}$ & \cellcolor[RGB]{255,255,255}$.75_{\pm.045}$ & \cellcolor[RGB]{255,201,92}$\mathbf{.69_{\pm.019}}$ & \cellcolor[RGB]{255,201,92}$\mathbf{.52_{\pm.046}}$ & \cellcolor[RGB]{255,201,92}$\mathbf{.60_{\pm.021}}$ & \cellcolor[RGB]{255,201,92}$\mathbf{.48_{\pm.052}}$ \\
        & \textsc{PLASMA-PF} & \cellcolor[RGB]{255,255,255}$.62_{\pm.042}$ & \cellcolor[RGB]{255,255,255}$.38_{\pm.057}$ & \cellcolor[RGB]{255,201,92}$\mathbf{.78_{\pm.055}}$ & \cellcolor[RGB]{255,255,255}$.48_{\pm.074}$ & \cellcolor[RGB]{255,255,255}$.23_{\pm.021}$ & \cellcolor[RGB]{255,255,255}$.44_{\pm.026}$ & \cellcolor[RGB]{255,255,255}$.41_{\pm.066}$ \\
        \midrule
        \bottomrule
    \end{tabular}%
}
    \label{tab:appendix_motif_test_hard}%
\end{table}%
\begin{table}[htbp]
    \centering
    \caption{Comprehensive binding site detection results on \texttt{test\_extra} dataset across seven protein representation models.}
    \renewcommand{\arraystretch}{1.2}
    \resizebox{\textwidth}{!}{
    \begin{tabular}{cc ccccccc}
        \toprule
        \midrule
        \multirow{2}[4]{*}{\textbf{Metrics}} & \multirow{2}[4]{*}{\textbf{Methods}} & \multicolumn{7}{c}{\textbf{Binding Site}} \\
        \cmidrule(lr){3-9}          &       & \textsc{ProstT5}  & \textsc{ProtT5} & \textsc{Ankh} & \textsc{ESM2} & \textsc{ProtSSN} & \textsc{TM-Vec}  & \textsc{ProtBERT}  \\
        \toprule
        \multirow{6}[4]{*}{\begin{sideways}ROC-AUC\end{sideways}} & \textsc{PLASMA} & \cellcolor[RGB]{255,210,121}$\mathbf{.98_{\pm.009}}$ & \cellcolor[RGB]{255,208,113}$.98_{\pm.009}$ & \cellcolor[RGB]{255,206,107}$\mathbf{.99_{\pm.008}}$ & \cellcolor[RGB]{255,213,131}$\mathbf{.98_{\pm.013}}$ & \cellcolor[RGB]{255,218,143}$\mathbf{.98_{\pm.014}}$ & \cellcolor[RGB]{255,213,131}$\mathbf{.98_{\pm.008}}$ & \cellcolor[RGB]{255,250,242}$\mathbf{.92_{\pm.019}}$ \\
        & \textsc{PLASMA-PF} & \cellcolor[RGB]{255,207,110}$.98_{\pm.008}$ & \cellcolor[RGB]{255,208,114}$.98_{\pm.010}$ & \cellcolor[RGB]{255,201,93}$.99_{\pm.006}$ & \cellcolor[RGB]{254,252,247}$.92_{\pm.052}$ & \cellcolor[RGB]{255,230,180}$.96_{\pm.012}$ & \cellcolor[RGB]{255,242,216}$.95_{\pm.019}$ & \cellcolor[RGB]{250,250,255}$.87_{\pm.032}$ \\
        & \textsc{EBA}   & \cellcolor[RGB]{255,213,128}$.98_{\pm.013}$ & \cellcolor[RGB]{255,205,105}$\mathbf{.99_{\pm.009}}$ & \cellcolor[RGB]{255,201,92}$.99_{\pm.007}$ & \cellcolor[RGB]{255,221,153}$.97_{\pm.021}$ & \cellcolor[RGB]{180,180,255}$.30_{\pm.060}$ & \cellcolor[RGB]{255,214,133}$.98_{\pm.014}$ & \cellcolor[RGB]{245,245,255}$.83_{\pm.072}$ \\
        & \textsc{CosineSim} & \cellcolor[RGB]{253,253,254}$.89_{\pm.038}$ & \cellcolor[RGB]{249,249,254}$.86_{\pm.059}$ & \cellcolor[RGB]{255,217,141}$.98_{\pm.010}$ & \cellcolor[RGB]{231,231,255}$.72_{\pm.060}$ & \cellcolor[RGB]{229,229,255}$.70_{\pm.070}$ & \cellcolor[RGB]{255,244,224}$.94_{\pm.021}$ & \cellcolor[RGB]{212,212,255}$.56_{\pm.029}$ \\
        \cmidrule{2-9}          & \textsc{Foldseek} & \multicolumn{7}{c}{\cellcolor[RGB]{254,254,255}$.90_{\pm.013}$} \\
        & \textsc{TM-Align} & \multicolumn{7}{c}{\cellcolor[RGB]{255,255,255}$.91_{\pm.040}$} \\
        \toprule
        \multirow{6}[4]{*}{\begin{sideways}PR-AUC\end{sideways}} & \textsc{PLASMA} & \cellcolor[RGB]{255,207,110}$\mathbf{.98_{\pm.011}}$ & \cellcolor[RGB]{255,202,97}$\mathbf{.98_{\pm.010}}$ & \cellcolor[RGB]{255,204,101}$\mathbf{.98_{\pm.011}}$ & \cellcolor[RGB]{255,213,130}$\mathbf{.97_{\pm.019}}$ & \cellcolor[RGB]{255,215,136}$\mathbf{.97_{\pm.019}}$ & \cellcolor[RGB]{255,213,131}$\mathbf{.97_{\pm.012}}$ & \cellcolor[RGB]{255,253,249}$\mathbf{.90_{\pm.043}}$ \\
        & \textsc{PLASMA-PF} & \cellcolor[RGB]{255,208,114}$.98_{\pm.013}$ & \cellcolor[RGB]{255,208,113}$.98_{\pm.014}$ & \cellcolor[RGB]{255,206,108}$.98_{\pm.012}$ & \cellcolor[RGB]{255,254,253}$.90_{\pm.079}$ & \cellcolor[RGB]{254,233,188}$.95_{\pm.026}$ & \cellcolor[RGB]{255,243,219}$.93_{\pm.022}$ & \cellcolor[RGB]{248,248,255}$.84_{\pm.078}$ \\
        & \textsc{EBA}   & \cellcolor[RGB]{255,206,109}$.98_{\pm.014}$ & \cellcolor[RGB]{255,203,99}$.98_{\pm.014}$ & \cellcolor[RGB]{255,201,92}$.98_{\pm.012}$ & \cellcolor[RGB]{255,227,171}$.96_{\pm.035}$ & \cellcolor[RGB]{180,180,255}$.28_{\pm.063}$ & \cellcolor[RGB]{255,213,131}$.97_{\pm.020}$ & \cellcolor[RGB]{242,242,255}$.79_{\pm.115}$ \\
        & \textsc{CosineSim} & \cellcolor[RGB]{250,250,255}$.86_{\pm.076}$ & \cellcolor[RGB]{245,245,255}$.82_{\pm.099}$ & \cellcolor[RGB]{255,225,164}$.96_{\pm.023}$ & \cellcolor[RGB]{227,227,255}$.67_{\pm.093}$ & \cellcolor[RGB]{225,225,255}$.65_{\pm.118}$ & \cellcolor[RGB]{255,246,228}$.93_{\pm.029}$ & \cellcolor[RGB]{206,206,255}$.49_{\pm.076}$ \\
        \cmidrule{2-9}          & \textsc{Foldseek} & \multicolumn{7}{c}{\cellcolor[RGB]{239,239,255}$.76_{\pm.065}$} \\
        & \textsc{TM-Align} & \multicolumn{7}{c}{\cellcolor[RGB]{255,255,255}$.89_{\pm.064}$} \\
        \toprule
        \multirow{6}[4]{*}{\begin{sideways}F1-MAX\end{sideways}} & \textsc{PLASMA} & \cellcolor[RGB]{255,203,100}$\mathbf{.97_{\pm.016}}$ & \cellcolor[RGB]{255,201,92}$\mathbf{.97_{\pm.011}}$ & \cellcolor[RGB]{255,209,118}$.96_{\pm.022}$ & \cellcolor[RGB]{255,222,157}$.95_{\pm.030}$ & \cellcolor[RGB]{254,232,187}$.93_{\pm.026}$ & \cellcolor[RGB]{255,213,128}$.96_{\pm.014}$ & \cellcolor[RGB]{251,251,255}$.83_{\pm.046}$ \\
        & \textsc{PLASMA-PF} & \cellcolor[RGB]{255,214,132}$.96_{\pm.023}$ & \cellcolor[RGB]{255,207,110}$.97_{\pm.017}$ & \cellcolor[RGB]{255,215,137}$.96_{\pm.027}$ & \cellcolor[RGB]{253,253,255}$.85_{\pm.082}$ & \cellcolor[RGB]{255,247,233}$.90_{\pm.031}$ & \cellcolor[RGB]{255,236,197}$.93_{\pm.018}$ & \cellcolor[RGB]{246,246,255}$.76_{\pm.073}$ \\
        & \textsc{EBA}   & \cellcolor[RGB]{255,210,119}$.96_{\pm.021}$ & \cellcolor[RGB]{255,216,138}$.96_{\pm.026}$ & \cellcolor[RGB]{255,206,108}$\mathbf{.97_{\pm.021}}$ & \cellcolor[RGB]{255,234,192}$.93_{\pm.049}$ & \cellcolor[RGB]{180,180,255}$.00_{\pm.000}$ & \cellcolor[RGB]{255,228,173}$.94_{\pm.034}$ & \cellcolor[RGB]{242,242,255}$.73_{\pm.108}$ \\
        & \textsc{CosineSim} & \cellcolor[RGB]{247,247,255}$.78_{\pm.081}$ & \cellcolor[RGB]{246,246,255}$.76_{\pm.089}$ & \cellcolor[RGB]{255,242,218}$.91_{\pm.034}$ & \cellcolor[RGB]{233,233,255}$.62_{\pm.087}$ & \cellcolor[RGB]{231,231,255}$.60_{\pm.107}$ & \cellcolor[RGB]{254,254,255}$.86_{\pm.046}$ & \cellcolor[RGB]{227,227,255}$.55_{\pm.092}$ \\
        \cmidrule{2-9}          & \textsc{Foldseek} & \multicolumn{7}{c}{\cellcolor[RGB]{255,206,108}$.97_{\pm.014}$} \\
        & \textsc{TM-Align} & \multicolumn{7}{c}{\cellcolor[RGB]{255,255,255}$.87_{\pm.063}$} \\
        \toprule
        \multirow{2}[2]{*}{\begin{sideways}LMS\end{sideways} \vspace{3pt}} & \textsc{PLASMA} & \cellcolor[RGB]{255,201,92}$\mathbf{.84_{\pm.050}}$ & \cellcolor[RGB]{255,201,92}$\mathbf{.83_{\pm.051}}$ & \cellcolor[RGB]{255,255,255}$.82_{\pm.062}$ & \cellcolor[RGB]{255,201,92}$\mathbf{.77_{\pm.105}}$ & \cellcolor[RGB]{255,201,92}$\mathbf{.65_{\pm.088}}$ & \cellcolor[RGB]{255,201,92}$\mathbf{.75_{\pm.071}}$ & \cellcolor[RGB]{255,201,92}$\mathbf{.56_{\pm.075}}$ \\
        & \textsc{PLASMA-PF} & \cellcolor[RGB]{255,255,255}$.79_{\pm.098}$ & \cellcolor[RGB]{255,255,255}$.55_{\pm.079}$ & \cellcolor[RGB]{255,201,92}$\mathbf{.85_{\pm.058}}$ & \cellcolor[RGB]{255,255,255}$.49_{\pm.082}$ & \cellcolor[RGB]{255,255,255}$.36_{\pm.055}$ & \cellcolor[RGB]{255,255,255}$.65_{\pm.070}$ & \cellcolor[RGB]{255,255,255}$.43_{\pm.038}$ \\
        \midrule
        \bottomrule
    \end{tabular}%
}
    \label{tab:appendix_binding_site_test_hard}%
\end{table}%
\begin{table}[htbp]
    \centering
    \caption{Comprehensive active site detection results on \texttt{test\_extra} dataset across seven protein representation models.}
    \renewcommand{\arraystretch}{1.2}
    \resizebox{\textwidth}{!}{
    \begin{tabular}{cc ccccccc}
        \toprule
        \midrule
        \multirow{2}[4]{*}{\textbf{Metrics}} & \multirow{2}[4]{*}{\textbf{Methods}} & \multicolumn{7}{c}{\textbf{Active Site}} \\
        \cmidrule(lr){3-9}          &       & \textsc{ProstT5}  & \textsc{ProtT5} & \textsc{Ankh} & \textsc{ESM2} & \textsc{ProtSSN} & \textsc{TM-Vec}  & \textsc{ProtBERT}  \\
        \toprule
        \multirow{6}[4]{*}{\begin{sideways}ROC-AUC\end{sideways}} & \textsc{PLASMA} & \cellcolor[RGB]{255,205,106}$\mathbf{.98_{\pm.011}}$ & \cellcolor[RGB]{255,201,92}$\mathbf{.98_{\pm.010}}$ & \cellcolor[RGB]{255,210,119}$\mathbf{.98_{\pm.012}}$ & \cellcolor[RGB]{255,205,105}$\mathbf{.98_{\pm.010}}$ & \cellcolor[RGB]{255,214,131}$\mathbf{.97_{\pm.011}}$ & \cellcolor[RGB]{255,217,142}$\mathbf{.97_{\pm.013}}$ & \cellcolor[RGB]{255,246,228}$\mathbf{.95_{\pm.026}}$ \\
        & \textsc{PLASMA-PF} & \cellcolor[RGB]{255,202,95}$.98_{\pm.010}$ & \cellcolor[RGB]{255,203,100}$.98_{\pm.011}$ & \cellcolor[RGB]{255,215,134}$.97_{\pm.015}$ & \cellcolor[RGB]{254,238,203}$.96_{\pm.006}$ & \cellcolor[RGB]{255,225,164}$.97_{\pm.008}$ & \cellcolor[RGB]{255,223,159}$.97_{\pm.014}$ & \cellcolor[RGB]{254,254,252}$.93_{\pm.024}$ \\
        & \textsc{EBA}   & \cellcolor[RGB]{255,213,129}$.98_{\pm.012}$ & \cellcolor[RGB]{255,212,126}$.98_{\pm.012}$ & \cellcolor[RGB]{255,216,137}$.97_{\pm.013}$ & \cellcolor[RGB]{255,216,139}$.97_{\pm.012}$ & \cellcolor[RGB]{180,180,255}$.43_{\pm.066}$ & \cellcolor[RGB]{255,216,137}$.97_{\pm.013}$ & \cellcolor[RGB]{250,250,255}$.91_{\pm.027}$ \\
        & \textsc{CosineSim} & \cellcolor[RGB]{245,245,255}$.87_{\pm.032}$ & \cellcolor[RGB]{251,251,255}$.91_{\pm.011}$ & \cellcolor[RGB]{255,230,181}$.96_{\pm.012}$ & \cellcolor[RGB]{233,233,255}$.79_{\pm.068}$ & \cellcolor[RGB]{229,229,255}$.76_{\pm.033}$ & \cellcolor[RGB]{255,234,191}$.96_{\pm.013}$ & \cellcolor[RGB]{221,221,255}$.71_{\pm.012}$ \\
        \cmidrule{2-9}          & \textsc{Foldseek} & \multicolumn{7}{c}{\cellcolor[RGB]{245,245,255}$.87_{\pm.022}$} \\
        & \textsc{TM-Align} & \multicolumn{7}{c}{\cellcolor[RGB]{255,255,255}$.93_{\pm.009}$} \\
        \toprule
        \multirow{6}[4]{*}{\begin{sideways}PR-AUC\end{sideways}} & \textsc{PLASMA} & \cellcolor[RGB]{255,216,139}$.97_{\pm.014}$ & \cellcolor[RGB]{255,201,92}$\mathbf{.98_{\pm.010}}$ & \cellcolor[RGB]{255,220,151}$\mathbf{.97_{\pm.014}}$ & \cellcolor[RGB]{255,209,117}$\mathbf{.98_{\pm.011}}$ & \cellcolor[RGB]{255,217,142}$\mathbf{.97_{\pm.012}}$ & \cellcolor[RGB]{255,226,170}$\mathbf{.97_{\pm.016}}$ & \cellcolor[RGB]{255,238,204}$\mathbf{.96_{\pm.019}}$ \\
        & \textsc{PLASMA-PF} & \cellcolor[RGB]{255,215,135}$\mathbf{.98_{\pm.013}}$ & \cellcolor[RGB]{255,209,118}$.98_{\pm.011}$ & \cellcolor[RGB]{255,224,162}$.97_{\pm.015}$ & \cellcolor[RGB]{255,237,201}$.96_{\pm.006}$ & \cellcolor[RGB]{255,226,168}$.97_{\pm.009}$ & \cellcolor[RGB]{255,237,203}$.96_{\pm.017}$ & \cellcolor[RGB]{255,250,240}$.95_{\pm.017}$ \\
        & \textsc{EBA}   & \cellcolor[RGB]{255,218,146}$.97_{\pm.013}$ & \cellcolor[RGB]{255,225,164}$.97_{\pm.014}$ & \cellcolor[RGB]{255,215,136}$.97_{\pm.012}$ & \cellcolor[RGB]{255,223,159}$.97_{\pm.012}$ & \cellcolor[RGB]{180,180,255}$.43_{\pm.032}$ & \cellcolor[RGB]{255,220,151}$.97_{\pm.014}$ & \cellcolor[RGB]{252,252,255}$.93_{\pm.019}$ \\
        & \textsc{CosineSim} & \cellcolor[RGB]{249,249,255}$.90_{\pm.031}$ & \cellcolor[RGB]{252,252,255}$.92_{\pm.017}$ & \cellcolor[RGB]{255,239,208}$.96_{\pm.016}$ & \cellcolor[RGB]{239,239,255}$.84_{\pm.059}$ & \cellcolor[RGB]{234,234,255}$.80_{\pm.038}$ & \cellcolor[RGB]{255,236,200}$.96_{\pm.015}$ & \cellcolor[RGB]{227,227,255}$.75_{\pm.010}$ \\
        \cmidrule{2-9}          & \textsc{Foldseek} & \multicolumn{7}{c}{\cellcolor[RGB]{235,235,255}$.81_{\pm.026}$} \\
        & \textsc{TM-Align} & \multicolumn{7}{c}{\cellcolor[RGB]{255,255,255}$.94_{\pm.012}$} \\
        \toprule
        \multirow{6}[4]{*}{\begin{sideways}F1-MAX\end{sideways}} & \textsc{PLASMA} & \cellcolor[RGB]{255,204,102}$\mathbf{.97_{\pm.012}}$ & \cellcolor[RGB]{255,202,96}$\mathbf{.98_{\pm.013}}$ & \cellcolor[RGB]{255,201,92}$\mathbf{.98_{\pm.013}}$ & \cellcolor[RGB]{255,207,111}$\mathbf{.97_{\pm.011}}$ & \cellcolor[RGB]{255,211,124}$\mathbf{.97_{\pm.011}}$ & \cellcolor[RGB]{255,213,128}$\mathbf{.97_{\pm.015}}$ & \cellcolor[RGB]{255,248,235}$.92_{\pm.036}$ \\
        & \textsc{PLASMA-PF} & \cellcolor[RGB]{255,213,130}$.97_{\pm.015}$ & \cellcolor[RGB]{255,213,129}$.97_{\pm.020}$ & \cellcolor[RGB]{255,215,136}$.97_{\pm.018}$ & \cellcolor[RGB]{255,242,217}$.94_{\pm.016}$ & \cellcolor[RGB]{255,228,175}$.95_{\pm.012}$ & \cellcolor[RGB]{255,219,147}$.96_{\pm.011}$ & \cellcolor[RGB]{254,254,255}$.89_{\pm.032}$ \\
        & \textsc{EBA}   & \cellcolor[RGB]{255,208,115}$.97_{\pm.014}$ & \cellcolor[RGB]{255,213,129}$.97_{\pm.013}$ & \cellcolor[RGB]{255,212,126}$.97_{\pm.013}$ & \cellcolor[RGB]{255,211,124}$.97_{\pm.008}$ & \cellcolor[RGB]{180,180,255}$.00_{\pm.000}$ & \cellcolor[RGB]{255,217,140}$.97_{\pm.020}$ & \cellcolor[RGB]{252,252,255}$.87_{\pm.026}$ \\
        & \textsc{CosineSim} & \cellcolor[RGB]{248,248,255}$.83_{\pm.033}$ & \cellcolor[RGB]{250,250,255}$.84_{\pm.013}$ & \cellcolor[RGB]{255,248,234}$.92_{\pm.020}$ & \cellcolor[RGB]{242,242,255}$.75_{\pm.044}$ & \cellcolor[RGB]{239,239,255}$.71_{\pm.018}$ & \cellcolor[RGB]{255,248,236}$.92_{\pm.010}$ & \cellcolor[RGB]{236,236,255}$.68_{\pm.008}$ \\
        \cmidrule{2-9}          & \textsc{Foldseek} & \multicolumn{7}{c}{\cellcolor[RGB]{255,221,153}$.96_{\pm.015}$} \\
        & \textsc{TM-Align} & \multicolumn{7}{c}{\cellcolor[RGB]{255,255,255}$.90_{\pm.014}$} \\
        \toprule
        \multirow{2}[2]{*}{\begin{sideways}LMS\end{sideways} \vspace{3pt}} & \textsc{PLASMA} & \cellcolor[RGB]{255,255,255}$.89_{\pm.044}$ & \cellcolor[RGB]{255,201,92}$\mathbf{.83_{\pm.030}}$ & \cellcolor[RGB]{255,255,255}$.90_{\pm.034}$ & \cellcolor[RGB]{255,201,92}$\mathbf{.87_{\pm.038}}$ & \cellcolor[RGB]{255,201,92}$\mathbf{.67_{\pm.044}}$ & \cellcolor[RGB]{255,201,92}$\mathbf{.84_{\pm.053}}$ & \cellcolor[RGB]{255,201,92}$\mathbf{.60_{\pm.024}}$ \\
        & \textsc{PLASMA-PF} & \cellcolor[RGB]{255,201,92}$\mathbf{.90_{\pm.043}}$ & \cellcolor[RGB]{255,255,255}$.70_{\pm.014}$ & \cellcolor[RGB]{255,201,92}$\mathbf{.94_{\pm.029}}$ & \cellcolor[RGB]{255,255,255}$.68_{\pm.067}$ & \cellcolor[RGB]{255,255,255}$.43_{\pm.032}$ & \cellcolor[RGB]{255,255,255}$.78_{\pm.048}$ & \cellcolor[RGB]{255,255,255}$.50_{\pm.021}$ \\
        \midrule
        \bottomrule
    \end{tabular}%
}
    \label{tab:appendix_active_site_test_hard}%
\end{table}%

\renewcommand{\arraystretch}{2}
\begin{table}[htbp]
    \centering
    \caption{\new{Model performance on \texttt{test\_extra} (mean $\pm$ std over three independent seeds). Colors indicate relative performance versus TM-Align, percentage values report the associated specific relative performance difference.}}
    \resizebox{\textwidth}{!}{
    \begin{tabular}{cc ccc ccc ccc}
        \toprule
        \midrule
        \multirow{2}[4]{*}{\textbf{Metrics}} & \multirow{2}[4]{*}{\textbf{Methods}} & \multicolumn{3}{c}{\textbf{Motif}} & \multicolumn{3}{c}{\textbf{Binding Site}} & \multicolumn{3}{c}{\textbf{Active Site}} \\
        \cmidrule(lr){3-5}\cmidrule(lr){6-8}\cmidrule(lr){9-11}          &       & \textsc{Ankh}  & \textsc{ESM2}  & \textsc{ProtSSN} & \textsc{Ankh}  & \textsc{ESM2}  & \textsc{ProtSSN} & \textsc{Ankh}  & \textsc{ESM2}  & \textsc{ProtSSN} \\
        \toprule
        \multirow{6}[4]{*}{\begin{sideways}ROC-AUC\end{sideways}} & PLASMA & \cellcolor[RGB]{255,201,92}$\mathbf{.98_{\pm.008}^{\uparrow 21.0\%}}$ & \cellcolor[RGB]{255,207,112}$\mathbf{.97_{\pm.013}^{\uparrow 19.8\%}}$ & \cellcolor[RGB]{255,215,135}$\mathbf{.96_{\pm.016}^{\uparrow 18.5\%}}$ & \cellcolor[RGB]{255,206,107}$\mathbf{.99_{\pm.008}^{\uparrow 8.8\%}}$ & \cellcolor[RGB]{255,213,131}$\mathbf{.98_{\pm.013}^{\uparrow 7.7\%}}$ & \cellcolor[RGB]{255,218,143}$\mathbf{.98_{\pm.014}^{\uparrow 7.7\%}}$ & \cellcolor[RGB]{255,206,108}$\mathbf{.98_{\pm.012}^{\uparrow 5.4\%}}$ & \cellcolor[RGB]{255,201,92}$\mathbf{.98_{\pm.010}^{\uparrow 5.4\%}}$ & \cellcolor[RGB]{255,210,121}$\mathbf{.97_{\pm.011}^{\uparrow 4.3\%}}$ \\
        & PLASMA-PF & \cellcolor[RGB]{255,201,94}$.98_{\pm.009}^{\uparrow 21.0\%}$ & \cellcolor[RGB]{255,228,175}$.93_{\pm.004}^{\uparrow 14.8\%}$ & \cellcolor[RGB]{255,238,205}$.90_{\pm.005}^{\uparrow 11.1\%}$ & \cellcolor[RGB]{255,201,93}$.99_{\pm.006}^{\uparrow 8.8\%}$ & \cellcolor[RGB]{254,252,247}$.92_{\pm.052}^{\uparrow 1.1\%}$ & \cellcolor[RGB]{255,230,180}$.96_{\pm.012}^{\uparrow 5.5\%}$ & \cellcolor[RGB]{255,211,124}$.97_{\pm.015}^{\uparrow 4.3\%}$ & \cellcolor[RGB]{255,237,200}$.96_{\pm.006}^{\uparrow 3.2\%}$ & \cellcolor[RGB]{255,222,158}$.97_{\pm.008}^{\uparrow 4.3\%}$ \\
        & EBA   & \cellcolor[RGB]{255,240,211}$.90_{\pm.033}^{\uparrow 11.1\%}$ & \cellcolor[RGB]{255,231,184}$.92_{\pm.021}^{\uparrow 13.6\%}$ & \cellcolor[RGB]{180,180,255}$.32_{\pm.043}^{\downarrow 60.5\%}$ & \cellcolor[RGB]{255,201,92}$.99_{\pm.007}^{\uparrow 8.8\%}$ & \cellcolor[RGB]{255,221,153}$.97_{\pm.021}^{\uparrow 6.6\%}$ & \cellcolor[RGB]{180,180,255}$.30_{\pm.060}^{\downarrow 67.0\%}$ & \cellcolor[RGB]{255,212,127}$.97_{\pm.013}^{\uparrow 4.3\%}$ & \cellcolor[RGB]{255,213,129}$.97_{\pm.012}^{\uparrow 4.3\%}$ & \cellcolor[RGB]{180,180,255}$.43_{\pm.066}^{\downarrow 53.8\%}$ \\
        & Backbone & \cellcolor[RGB]{254,250,240}$.85_{\pm.019}^{\uparrow 4.9\%}$ & \cellcolor[RGB]{243,243,255}$.74_{\pm.033}^{\downarrow 8.6\%}$ & \cellcolor[RGB]{252,252,255}$.79_{\pm.018}^{\downarrow 2.5\%}$ & \cellcolor[RGB]{255,217,141}$.98_{\pm.010}^{\uparrow 7.7\%}$ & \cellcolor[RGB]{231,231,255}$.72_{\pm.060}^{\downarrow 20.9\%}$ & \cellcolor[RGB]{229,229,255}$.70_{\pm.070}^{\downarrow 23.1\%}$ & \cellcolor[RGB]{255,229,176}$.96_{\pm.012}^{\uparrow 3.2\%}$ & \cellcolor[RGB]{233,233,255}$.79_{\pm.068}^{\downarrow 15.1\%}$ & \cellcolor[RGB]{229,229,255}$.76_{\pm.033}^{\downarrow 18.3\%}$ \\
        \cmidrule{2-11}          & Foldseek & \multicolumn{3}{c}{\cellcolor[RGB]{255,242,217}$.89_{\pm.033}^{\uparrow 9.9\%}$} & \multicolumn{3}{c}{\cellcolor[RGB]{254,254,255}$.90_{\pm.013}^{\downarrow 1.1\%}$} & \multicolumn{3}{c}{\cellcolor[RGB]{245,245,255}$.87_{\pm.022}^{\downarrow 6.5\%}$} \\
        & TM-Align & \multicolumn{3}{c}{\cellcolor[RGB]{255,255,255}$.81_{\pm.014}$} & \multicolumn{3}{c}{\cellcolor[RGB]{255,255,255}$.91_{\pm.040}$} & \multicolumn{3}{c}{\cellcolor[RGB]{255,255,255}$.93_{\pm.009}$} \\
        \toprule
        \multirow{6}[4]{*}{\begin{sideways}PR-AUC\end{sideways}} & PLASMA & \cellcolor[RGB]{255,201,92}$\mathbf{.98_{\pm.011}^{\uparrow 14.0\%}}$ & \cellcolor[RGB]{255,206,110}$\mathbf{.97_{\pm.014}^{\uparrow 12.8\%}}$ & \cellcolor[RGB]{255,215,135}$\mathbf{.96_{\pm.017}^{\uparrow 11.6\%}}$ & \cellcolor[RGB]{255,204,101}$\mathbf{.98_{\pm.011}^{\uparrow 10.1\%}}$ & \cellcolor[RGB]{255,213,130}$\mathbf{.97_{\pm.019}^{\uparrow 9.0\%}}$ & \cellcolor[RGB]{255,215,136}$\mathbf{.97_{\pm.019}^{\uparrow 9.0\%}}$ & \cellcolor[RGB]{255,215,134}$\mathbf{.97_{\pm.014}^{\uparrow 3.2\%}}$ & \cellcolor[RGB]{255,201,92}$\mathbf{.98_{\pm.011}^{\uparrow 4.3\%}}$ & \cellcolor[RGB]{255,211,123}$\mathbf{.97_{\pm.012}^{\uparrow 3.2\%}}$ \\
        & PLASMA-PF & \cellcolor[RGB]{255,201,93}$.98_{\pm.010}^{\uparrow 14.0\%}$ & \cellcolor[RGB]{255,224,163}$.95_{\pm.005}^{\uparrow 10.5\%}$ & \cellcolor[RGB]{255,239,209}$.92_{\pm.007}^{\uparrow 7.0\%}$ & \cellcolor[RGB]{255,206,108}$.98_{\pm.012}^{\uparrow 10.1\%}$ & \cellcolor[RGB]{255,254,253}$.90_{\pm.079}^{\uparrow 1.1\%}$ & \cellcolor[RGB]{254,233,188}$.95_{\pm.026}^{\uparrow 6.7\%}$ & \cellcolor[RGB]{255,219,148}$.97_{\pm.015}^{\uparrow 3.2\%}$ & \cellcolor[RGB]{255,234,193}$.96_{\pm.006}^{\uparrow 2.1\%}$ & \cellcolor[RGB]{255,221,155}$.97_{\pm.009}^{\uparrow 3.2\%}$ \\
        & EBA   & \cellcolor[RGB]{255,242,217}$.91_{\pm.035}^{\uparrow 5.8\%}$ & \cellcolor[RGB]{255,233,189}$.93_{\pm.019}^{\uparrow 8.1\%}$ & \cellcolor[RGB]{180,180,255}$.38_{\pm.014}^{\downarrow 55.8\%}$ & \cellcolor[RGB]{255,201,92}$.98_{\pm.012}^{\uparrow 10.1\%}$ & \cellcolor[RGB]{255,227,171}$.96_{\pm.035}^{\uparrow 7.9\%}$ & \cellcolor[RGB]{180,180,255}$.28_{\pm.063}^{\downarrow 68.5\%}$ & \cellcolor[RGB]{255,208,115}$.97_{\pm.012}^{\uparrow 3.2\%}$ & \cellcolor[RGB]{255,218,144}$.97_{\pm.012}^{\uparrow 3.2\%}$ & \cellcolor[RGB]{180,180,255}$.43_{\pm.032}^{\downarrow 54.3\%}$ \\
        & Backbone & \cellcolor[RGB]{255,254,254}$.86_{\pm.023}^{\downarrow 0.0\%}$ & \cellcolor[RGB]{241,241,255}$.77_{\pm.041}^{\downarrow 10.5\%}$ & \cellcolor[RGB]{249,249,255}$.82_{\pm.027}^{\downarrow 4.7\%}$ & \cellcolor[RGB]{255,225,164}$.96_{\pm.023}^{\uparrow 7.9\%}$ & \cellcolor[RGB]{227,227,255}$.67_{\pm.093}^{\downarrow 24.7\%}$ & \cellcolor[RGB]{225,225,255}$.65_{\pm.118}^{\downarrow 27.0\%}$ & \cellcolor[RGB]{255,237,202}$.96_{\pm.016}^{\uparrow 2.1\%}$ & \cellcolor[RGB]{239,239,255}$.84_{\pm.059}^{\downarrow 10.6\%}$ & \cellcolor[RGB]{234,234,255}$.80_{\pm.038}^{\downarrow 14.9\%}$ \\
        \cmidrule{2-11}          & Foldseek & \multicolumn{3}{c}{\cellcolor[RGB]{252,252,255}$.84_{\pm.031}^{\downarrow 2.3\%}$} & \multicolumn{3}{c}{\cellcolor[RGB]{239,239,255}$.76_{\pm.065}^{\downarrow 14.6\%}$} & \multicolumn{3}{c}{\cellcolor[RGB]{235,235,255}$.81_{\pm.026}^{\downarrow 13.8\%}$} \\
        & TM-Align & \multicolumn{3}{c}{\cellcolor[RGB]{255,255,255}$.86_{\pm.020}$} & \multicolumn{3}{c}{\cellcolor[RGB]{255,255,255}$.89_{\pm.064}$} & \multicolumn{3}{c}{\cellcolor[RGB]{255,255,255}$.94_{\pm.012}$} \\
        \toprule
        \multirow{6}[4]{*}{\begin{sideways}F1-MAX\end{sideways}} & PLASMA & \cellcolor[RGB]{255,201,92}$\mathbf{.97_{\pm.009}^{\uparrow 27.6\%}}$ & \cellcolor[RGB]{255,212,125}$\mathbf{.95_{\pm.018}^{\uparrow 25.0\%}}$ & \cellcolor[RGB]{255,223,158}$\mathbf{.92_{\pm.022}^{\uparrow 21.1\%}}$ & \cellcolor[RGB]{255,204,103}$.96_{\pm.022}^{\uparrow 10.3\%}$ & \cellcolor[RGB]{255,219,147}$.95_{\pm.030}^{\uparrow 9.2\%}$ & \cellcolor[RGB]{255,230,181}$.93_{\pm.026}^{\uparrow 6.9\%}$ & \cellcolor[RGB]{255,201,92}$\mathbf{.98_{\pm.013}^{\uparrow 8.9\%}}$ & \cellcolor[RGB]{255,207,111}$\mathbf{.97_{\pm.011}^{\uparrow 7.8\%}}$ & \cellcolor[RGB]{255,211,124}$\mathbf{.97_{\pm.011}^{\uparrow 7.8\%}}$ \\
        & PLASMA-PF & \cellcolor[RGB]{255,207,110}$.96_{\pm.013}^{\uparrow 26.3\%}$ & \cellcolor[RGB]{255,231,183}$.90_{\pm.006}^{\uparrow 18.4\%}$ & \cellcolor[RGB]{255,245,226}$.84_{\pm.008}^{\uparrow 10.5\%}$ & \cellcolor[RGB]{255,211,124}$.96_{\pm.027}^{\uparrow 10.3\%}$ & \cellcolor[RGB]{253,253,255}$.85_{\pm.082}^{\downarrow 2.3\%}$ & \cellcolor[RGB]{255,247,232}$.90_{\pm.031}^{\uparrow 3.4\%}$ & \cellcolor[RGB]{255,215,136}$.97_{\pm.018}^{\uparrow 7.8\%}$ & \cellcolor[RGB]{255,242,217}$.94_{\pm.016}^{\uparrow 4.4\%}$ & \cellcolor[RGB]{255,228,175}$.95_{\pm.012}^{\uparrow 5.6\%}$ \\
        & EBA   & \cellcolor[RGB]{255,241,213}$.86_{\pm.035}^{\uparrow 13.2\%}$ & \cellcolor[RGB]{255,238,206}$.87_{\pm.024}^{\uparrow 14.5\%}$ & \cellcolor[RGB]{180,180,255}$.00_{\pm.000}^{\downarrow 100.0\%}$ & \cellcolor[RGB]{255,201,92}$\mathbf{.97_{\pm.021}^{\uparrow 11.5\%}}$ & \cellcolor[RGB]{255,232,187}$.93_{\pm.049}^{\uparrow 6.9\%}$ & \cellcolor[RGB]{180,180,255}$.00_{\pm.000}^{\downarrow 100.0\%}$ & \cellcolor[RGB]{255,212,126}$.97_{\pm.013}^{\uparrow 7.8\%}$ & \cellcolor[RGB]{255,211,124}$.97_{\pm.008}^{\uparrow 7.8\%}$ & \cellcolor[RGB]{180,180,255}$.00_{\pm.000}^{\downarrow 100.0\%}$ \\
        & Backbone & \cellcolor[RGB]{255,252,246}$.79_{\pm.008}^{\uparrow 3.9\%}$ & \cellcolor[RGB]{249,249,255}$.70_{\pm.014}^{\downarrow 7.9\%}$ & \cellcolor[RGB]{252,252,255}$.73_{\pm.013}^{\downarrow 3.9\%}$ & \cellcolor[RGB]{255,241,215}$.91_{\pm.034}^{\uparrow 4.6\%}$ & \cellcolor[RGB]{233,233,255}$.62_{\pm.087}^{\downarrow 28.7\%}$ & \cellcolor[RGB]{231,231,255}$.60_{\pm.107}^{\downarrow 31.0\%}$ & \cellcolor[RGB]{255,248,234}$.92_{\pm.020}^{\uparrow 2.2\%}$ & \cellcolor[RGB]{242,242,255}$.75_{\pm.044}^{\downarrow 16.7\%}$ & \cellcolor[RGB]{239,239,255}$.71_{\pm.018}^{\downarrow 21.1\%}$ \\
        \cmidrule{2-11}          & Foldseek & \multicolumn{3}{c}{\cellcolor[RGB]{255,228,175}$.91_{\pm.046}^{\uparrow 19.7\%}$} & \multicolumn{3}{c}{\cellcolor[RGB]{255,201,92}$.97_{\pm.014}^{\uparrow 11.5\%}$} & \multicolumn{3}{c}{\cellcolor[RGB]{255,221,153}$.96_{\pm.015}^{\uparrow 6.7\%}$} \\
        & TM-Align & \multicolumn{3}{c}{\cellcolor[RGB]{255,255,255}$.76_{\pm.015}$} & \multicolumn{3}{c}{\cellcolor[RGB]{255,255,255}$.87_{\pm.063}$} & \multicolumn{3}{c}{\cellcolor[RGB]{255,255,255}$.90_{\pm.014}$} \\
        \midrule
        \bottomrule
    \end{tabular}%
}
    \label{tab:main_results_test_extra_with_percentage}%
\end{table}%
\renewcommand{\arraystretch}{1}

\clearpage
\section{Ablation Study}
\label{sec:ablation}
\new{This section analyzes the contribution of the two plan-assessor components: the local-motif loss (LML) and the weight-correction term (WC) derived from the diagonal kernel. The combined ROC-AUC and LMS results across seven protein backbones and three tasks show two clear trends.

First, both LML and WC improve PLASMA’s alignment quality. Adding LML yields consistently higher ROC-AUC, confirming that it helps the model concentrate alignment mass on the task-relevant functional substructures it is trained to detect. We also observe that LML can slightly reduce performance on \texttt{test\_extra}, indicating a mild trade-off between specialization and generalization.

Second, WC is essential for ensuring stable alignment behavior, especially for the parameter-free PLASMA-PF variant. Removing WC causes a substantial performance drop on several backbones (notably ESM2 and ProtBERT), demonstrating that continuity weighting is crucial for suppressing fragmented correspondences and producing coherent alignment plans.

Overall, these results show that LML shapes the model toward identifying the desired functional motifs, while WC is indispensable for robust and stable alignment across architectures, particularly in the parameter-free setting.}



\begin{table}[htbp]
    \centering
    \caption{Ablation study results. Here we ablate two cases: not using the Label Matching Loss (w/o LML) and not using weight correction (w/o WC). }
    \renewcommand{\arraystretch}{1.2}
    \resizebox{\textwidth}{!}{
    \begin{tabular}{llccccccc}
        \toprule
        \textbf{Task} & \textbf{Method} & \textsc{ProstT5} & \textsc{ProtT5} & \textsc{Ankh} & \textsc{ESM2} & \textsc{ProtSSN} & \textsc{TM-Vec} & \textsc{ProtBERT} \\
        \midrule
        \multicolumn{9}{c}{\textbf{ROC-AUC}} \\
        \midrule
        \multirow{5}{*}{Motif} & PLASMA & $\mathbf{.97_{\pm.002}}$ & $\mathbf{.97_{\pm.002}}$ & $\mathbf{.95_{\pm.002}}$ & $\mathbf{.96_{\pm.002}}$ & $\mathbf{.96_{\pm.001}}$ & $\mathbf{.92_{\pm.004}}$ & $\mathbf{.87_{\pm.004}}$ \\
         & PLASMA-PF & $.94_{\pm.003}$ & $.96_{\pm.002}$ & $\mathbf{.95_{\pm.003}}$ & $.93_{\pm.004}$ & $.91_{\pm.003}$ & $.87_{\pm.001}$ & $.85_{\pm.004}$ \\
         & PLASMA (w/o LML) & $.95_{\pm.008}$ & $.95_{\pm.006}$ & $.93_{\pm.004}$ & $.91_{\pm.022}$ & $.89_{\pm.018}$ & $.89_{\pm.033}$ & $.85_{\pm.004}$ \\
         & PLASMA (w/o WC) & $.91_{\pm.005}$ & $.95_{\pm.004}$ & $.91_{\pm.004}$ & $.87_{\pm.019}$ & $.84_{\pm.003}$ & $.86_{\pm.006}$ & $.73_{\pm.009}$ \\
         & PLASMA-PF (w/o WC) & $.74_{\pm.004}$ & $.87_{\pm.002}$ & $.85_{\pm.006}$ & $.44_{\pm.009}$ & $.75_{\pm.009}$ & $.84_{\pm.007}$ & $.60_{\pm.012}$ \\
        \midrule
        \multirow{5}{*}{Binding Site} & PLASMA & $\mathbf{.99_{\pm.001}}$ & $\mathbf{.99_{\pm.000}}$ & $\mathbf{.99_{\pm.000}}$ & $\mathbf{.99_{\pm.001}}$ & $\mathbf{.99_{\pm.001}}$ & $.96_{\pm.003}$ & $\mathbf{.98_{\pm.001}}$ \\
         & PLASMA-PF & $\mathbf{.99_{\pm.001}}$ & $\mathbf{.99_{\pm.001}}$ & $\mathbf{.99_{\pm.000}}$ & $.96_{\pm.003}$ & $.97_{\pm.001}$ & $.92_{\pm.004}$ & $.90_{\pm.003}$ \\
         & PLASMA (w/o LML) & $\mathbf{.99_{\pm.002}}$ & $\mathbf{.99_{\pm.001}}$ & $\mathbf{.99_{\pm.002}}$ & $.97_{\pm.001}$ & $\mathbf{.99_{\pm.000}}$ & $\mathbf{.98_{\pm.001}}$ & $\mathbf{.98_{\pm.001}}$ \\
         & PLASMA (w/o WC) & $\mathbf{.99_{\pm.002}}$ & $\mathbf{.99_{\pm.001}}$ & $\mathbf{.99_{\pm.001}}$ & $.98_{\pm.001}$ & $.92_{\pm.008}$ & $.97_{\pm.002}$ & $.77_{\pm.004}$ \\
         & PLASMA-PF (w/o WC) & $.91_{\pm.002}$ & $.97_{\pm.001}$ & $.97_{\pm.002}$ & $.49_{\pm.003}$ & $.85_{\pm.006}$ & $.96_{\pm.003}$ & $.67_{\pm.006}$ \\
        \midrule
        \multirow{5}{*}{Active Site} & PLASMA & $\mathbf{.99_{\pm.001}}$ & $\mathbf{.99_{\pm.001}}$ & $\mathbf{.99_{\pm.001}}$ & $\mathbf{.99_{\pm.001}}$ & $\mathbf{.99_{\pm.002}}$ & $\mathbf{.99_{\pm.003}}$ & $\mathbf{.99_{\pm.004}}$ \\
         & PLASMA-PF & $\mathbf{.99_{\pm.002}}$ & $\mathbf{.99_{\pm.003}}$ & $\mathbf{.99_{\pm.003}}$ & $.96_{\pm.002}$ & $.98_{\pm.002}$ & $.98_{\pm.003}$ & $.94_{\pm.006}$ \\
         & PLASMA (w/o LML) & $\mathbf{.99_{\pm.001}}$ & $\mathbf{.99_{\pm.000}}$ & $\mathbf{.99_{\pm.001}}$ & $\mathbf{.99_{\pm.005}}$ & $\mathbf{.99_{\pm.000}}$ & $.98_{\pm.009}$ & $\mathbf{.99_{\pm.000}}$ \\
         & PLASMA (w/o WC) & $\mathbf{.99_{\pm.001}}$ & $\mathbf{.99_{\pm.001}}$ & $\mathbf{.99_{\pm.001}}$ & $.98_{\pm.000}$ & $.93_{\pm.008}$ & $\mathbf{.99_{\pm.001}}$ & $.81_{\pm.033}$ \\
         & PLASMA-PF (w/o WC) & $.95_{\pm.002}$ & $.97_{\pm.001}$ & $.98_{\pm.001}$ & $.55_{\pm.008}$ & $.87_{\pm.005}$ & $\mathbf{.99_{\pm.001}}$ & $.67_{\pm.009}$ \\
        \midrule
        \multicolumn{9}{c}{\textbf{LMS}} \\
        \midrule
        \multirow{2}{*}{Motif} & PLASMA & $\mathbf{.91_{\pm.007}}$ & $\mathbf{.92_{\pm.001}}$ & $\mathbf{.92_{\pm.002}}$ & $\mathbf{.92_{\pm.005}}$ & $\mathbf{.73_{\pm.013}}$ & $\mathbf{.76_{\pm.005}}$ & $.71_{\pm.007}$ \\
         & PLASMA (w/o LML) & $.66_{\pm.135}$ & $.65_{\pm.142}$ & $\mathbf{.92_{\pm.012}}$ & $.77_{\pm.170}$ & $.48_{\pm.136}$ & $.68_{\pm.167}$ & $\mathbf{.74_{\pm.012}}$ \\
        \midrule
        \multirow{2}{*}{Binding Site} & PLASMA & $\mathbf{.93_{\pm.002}}$ & $\mathbf{.93_{\pm.003}}$ & $\mathbf{.93_{\pm.004}}$ & $\mathbf{.93_{\pm.003}}$ & $.85_{\pm.006}$ & $.86_{\pm.002}$ & $.84_{\pm.003}$ \\
         & PLASMA (w/o LML) & $.87_{\pm.080}$ & $.84_{\pm.110}$ & $.79_{\pm.081}$ & $.49_{\pm.004}$ & $\mathbf{.88_{\pm.000}}$ & $\mathbf{.90_{\pm.011}}$ & $\mathbf{.89_{\pm.012}}$ \\
        \midrule
        \multirow{2}{*}{Active Site} & PLASMA & $\mathbf{.97_{\pm.004}}$ & $\mathbf{.97_{\pm.004}}$ & $\mathbf{.97_{\pm.003}}$ & $\mathbf{.97_{\pm.004}}$ & $.89_{\pm.016}$ & $\mathbf{.93_{\pm.006}}$ & $\mathbf{.89_{\pm.008}}$ \\
         & PLASMA (w/o LML) & $.89_{\pm.080}$ & $.84_{\pm.131}$ & $.91_{\pm.065}$ & $.79_{\pm.187}$ & $\mathbf{.90_{\pm.000}}$ & $.79_{\pm.143}$ & $\mathbf{.89_{\pm.007}}$ \\
        \bottomrule
    \end{tabular}%
}
    \label{tab:combined_ablation_frequent}%
\end{table}%

\clearpage
\section{Case Study}\label{sec:highres-case-study}
\new{To provide a clearer view of the residue-level alignment patterns, we include enlarged versions of the alignment matrices corresponding to Figure~\ref{fig:case-study} in the main text. These zoomed-in visualizations highlight how PLASMA identifies coherent local structural motifs across proteins with different folds, lengths, and sequence identities.}

\begin{figure}[h]
    \centering
    \includegraphics[width=\textwidth]{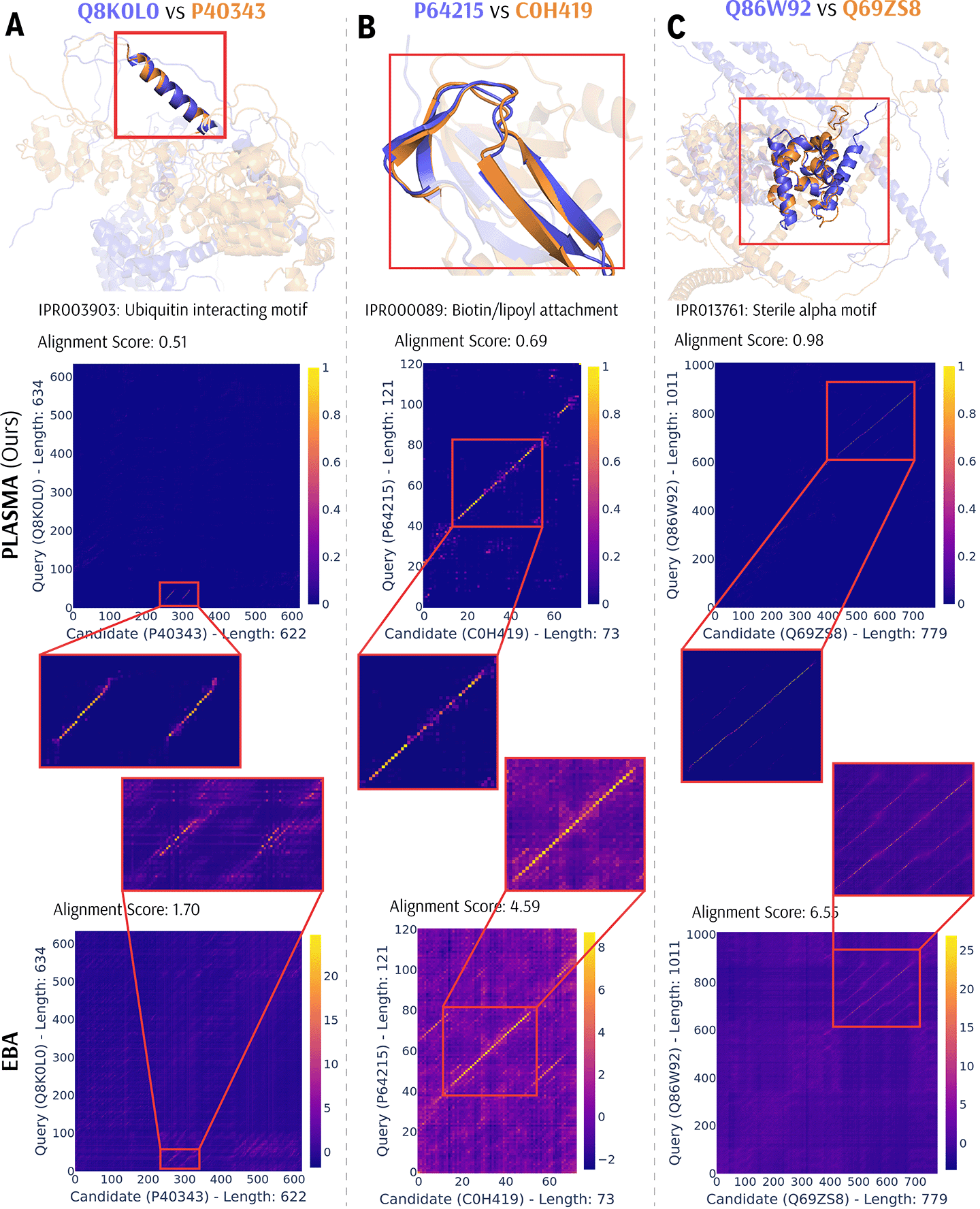}
    \vspace{-2mm}
    \caption{Representative alignment examples across three protein pairs. \textbf{A}, P40343 vs Q8K0L0. \textbf{B}, P64215 vs C0H419. \textbf{C}, Q69ZS8 vs Q86W92. Left: 3D structures with highlighted aligned regions. Center and right: alignment matrices from PLASMA and EBA with zoomed insets. This figure is the higher resolution version of Figure~\ref{fig:case-study}.}
    \label{fig:case-study-highres}
\end{figure}
\FloatBarrier

\clearpage
\section{Alignment Matrix Visualizations}\label{sec:alignment-matrix-visualizations}
\begin{figure}[htbp]
    \centering
    \includegraphics[width=0.75\textwidth]{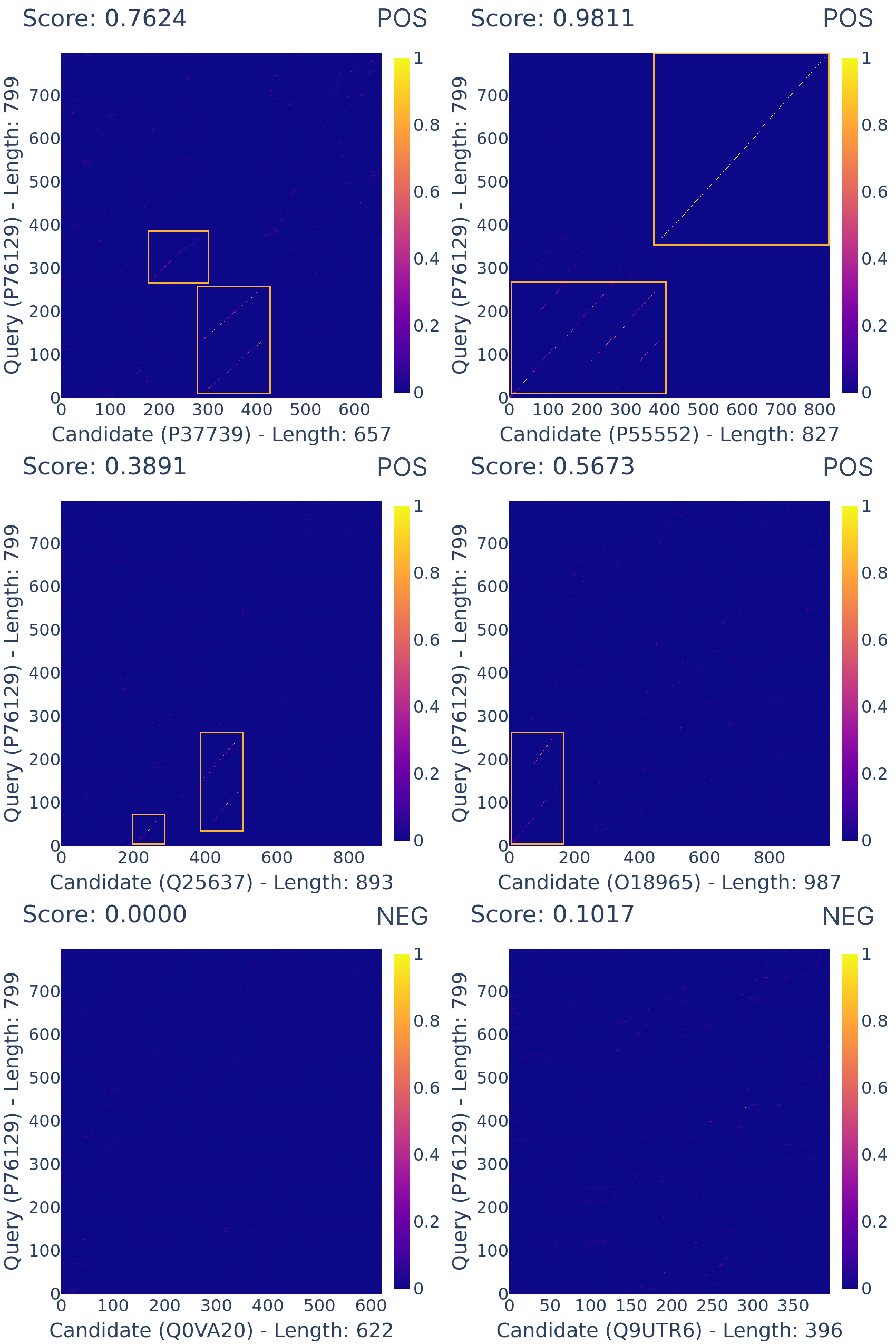}
    \caption{Representative alignment matrices comparing query protein P76129 against six candidate proteins. The visualization shows four positive pairs (POS) with shared substructures and two negative pairs (NEG) without substructure similarity. Orange regions highlight aligned substructures.}
    \label{fig:matrix-vis-p76129}
\end{figure}
Figure \ref{fig:matrix-vis-p76129} demonstrate PLASMA's interpretability by showing clear patterns that correspond to different levels of substructure similarity. The matrices were generated by comparing a single query protein (InterPro ID: P76129) against six different candidate proteins, including four positive pairs sharing functional substructures and two negative pairs without similar functional substructures. The orange-highlighted regions indicate aligned substructures, where larger and more intensely colored blocks correspond to stronger and more extensive alignments. Notably, positive pairs exhibit prominent diagonal patterns reflecting substructure correspondences, while negative pairs show minimal coherent structures and low alignment scores. This visualization validates that PLASMA's alignment scores accurately reflect the underlying biological relationships between protein substructures.

\clearpage
\section{Temperature Parameter Analysis}\label{sec:temperature-analysis}
\begin{figure}[htbp]
    \centering
    \includegraphics[width=\textwidth]{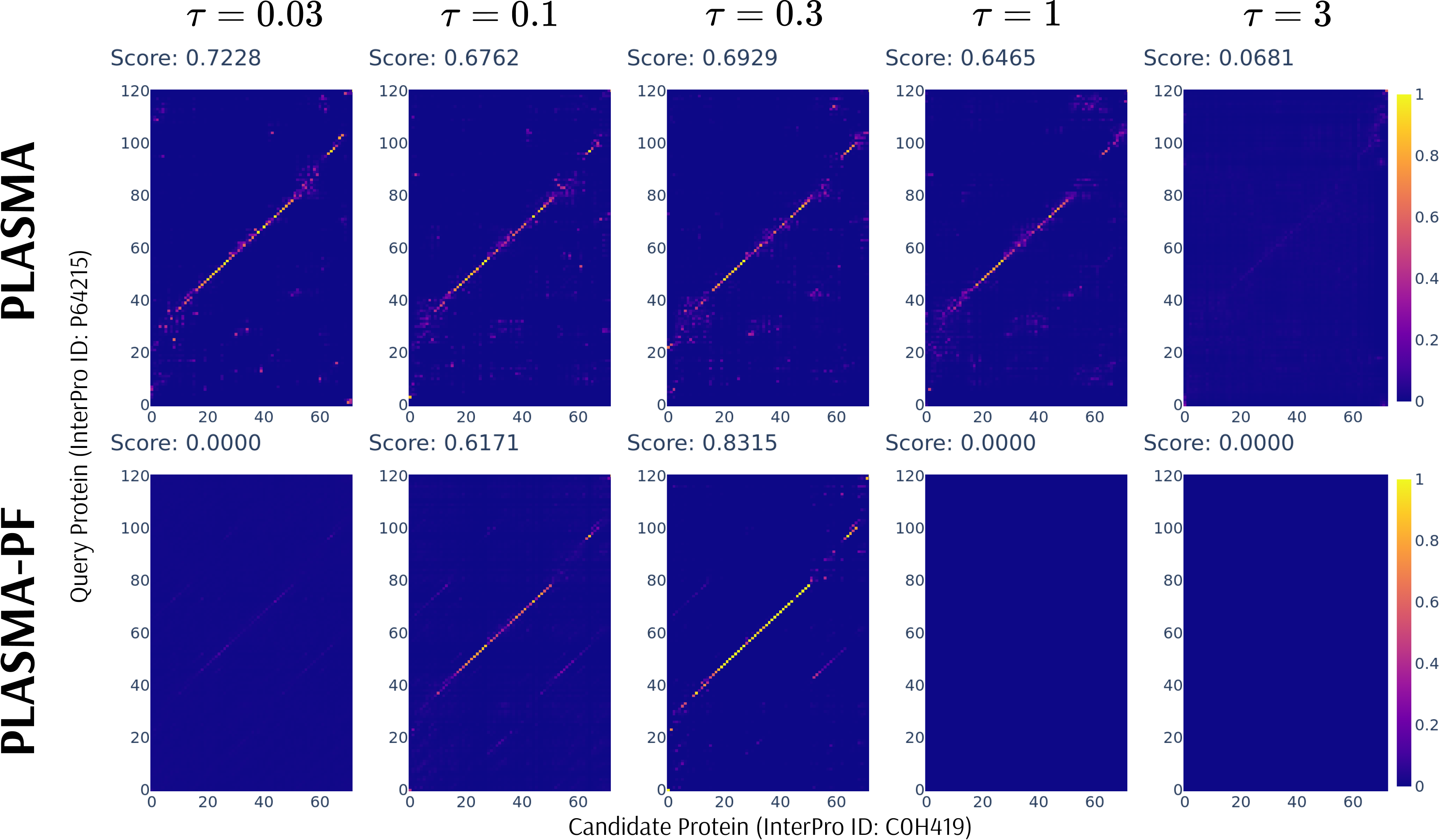}
    \caption{Effect of Sinkhorn temperature parameter $\tau$ on alignment matrix and score for both PLASMA and PLASMA-PF variants.}
    \label{fig:temp-vis}
\end{figure}
Figure \ref{fig:temp-vis} illustrates how the Sinkhorn temperature parameter $\tau$ impacts the alignment matrix in both PLASMA variants. The supervised PLASMA variant demonstrates greater stability and maintains meaningful alignment patterns across a wider range of temperature settings compared to PLASMA-PF, highlighting the robustness benefits of end-to-end training.

\section{Performance Evaluation at Different Structural Similarity Threshold}\label{sec:tm-score}
\new{We report the detailed values of the performance at different TM-score thresholds visualized in Figure~\ref{fig:score-comparison}. PLASMA consistently outperforms other baseline methods, especially in low similarity settings (\eg TM-score $\leq0.5$ and TM-score $\leq0.3$).}
\begin{table}[h]
\centering
\caption{Numerical results of the ROC-AUC Performance at different TM-Align thresholds.}
\resizebox{\textwidth}{!}{\begin{tabular}{llcccccc}
\toprule
Task & TM Score & \textsc{PLASMA} & \textsc{PLASMA-PF} & \textsc{EBA} & \textsc{CosineSim} & \textsc{TM-Align} & \textsc{Foldseek} \\
\midrule
\multirow{4}{*}{Motif} & $\leq$ 1.0 & \textbf{.96$_{\pm.002}$} & .95$_{\pm.002}$ & .87$_{\pm.003}$ & .84$_{\pm.004}$ & .78$_{\pm.004}$ & .83$_{\pm.004}$ \\
 & $\leq$ 0.7 & \textbf{.95$_{\pm.002}$} & .94$_{\pm.002}$ & .84$_{\pm.004}$ & .81$_{\pm.004}$ & .73$_{\pm.005}$ & .81$_{\pm.004}$ \\
 & $\leq$ 0.5 & \textbf{.93$_{\pm.003}$} & .93$_{\pm.003}$ & .81$_{\pm.005}$ & .78$_{\pm.005}$ & .66$_{\pm.006}$ & .79$_{\pm.005}$ \\
 & $\leq$ 0.3 & \textbf{.92$_{\pm.004}$} & .91$_{\pm.004}$ & .74$_{\pm.006}$ & .73$_{\pm.006}$ & .58$_{\pm.007}$ & .74$_{\pm.006}$ \\
\midrule
\multirow{4}{*}{Binding Site} & $\leq$ 1.0 & \textbf{.99$_{\pm.001}$} & .99$_{\pm.001}$ & .97$_{\pm.002}$ & .96$_{\pm.002}$ & .87$_{\pm.003}$ & .90$_{\pm.003}$ \\
 & $\leq$ 0.7 & \textbf{.99$_{\pm.001}$} & .98$_{\pm.002}$ & .95$_{\pm.003}$ & .93$_{\pm.003}$ & .76$_{\pm.006}$ & .88$_{\pm.004}$ \\
 & $\leq$ 0.5 & \textbf{.98$_{\pm.002}$} & .97$_{\pm.003}$ & .93$_{\pm.004}$ & .91$_{\pm.004}$ & .62$_{\pm.007}$ & .85$_{\pm.006}$ \\
 & $\leq$ 0.3 & \textbf{.97$_{\pm.004}$} & .96$_{\pm.004}$ & .89$_{\pm.007}$ & .88$_{\pm.007}$ & .45$_{\pm.010}$ & .80$_{\pm.009}$ \\
\midrule
\multirow{4}{*}{Active Site} & $\leq$ 1.0 & \textbf{.99$_{\pm.001}$} & .99$_{\pm.001}$ & .99$_{\pm.001}$ & .97$_{\pm.001}$ & .94$_{\pm.002}$ & .92$_{\pm.003}$ \\
 & $\leq$ 0.7 & \textbf{.99$_{\pm.001}$} & .98$_{\pm.002}$ & .97$_{\pm.002}$ & .95$_{\pm.003}$ & .88$_{\pm.005}$ & .91$_{\pm.004}$ \\
 & $\leq$ 0.5 & \textbf{.98$_{\pm.003}$} & .96$_{\pm.004}$ & .95$_{\pm.004}$ & .92$_{\pm.005}$ & .76$_{\pm.008}$ & .89$_{\pm.006}$ \\
 & $\leq$ 0.3 & \textbf{.96$_{\pm.007}$} & .90$_{\pm.010}$ & .89$_{\pm.011}$ & .83$_{\pm.013}$ & .59$_{\pm.016}$ & .84$_{\pm.013}$ \\
\bottomrule
\end{tabular}}
\label{tab:tmscore}
\end{table}

\begin{figure}[htbp]
    \centering
    \includegraphics[width=\textwidth]{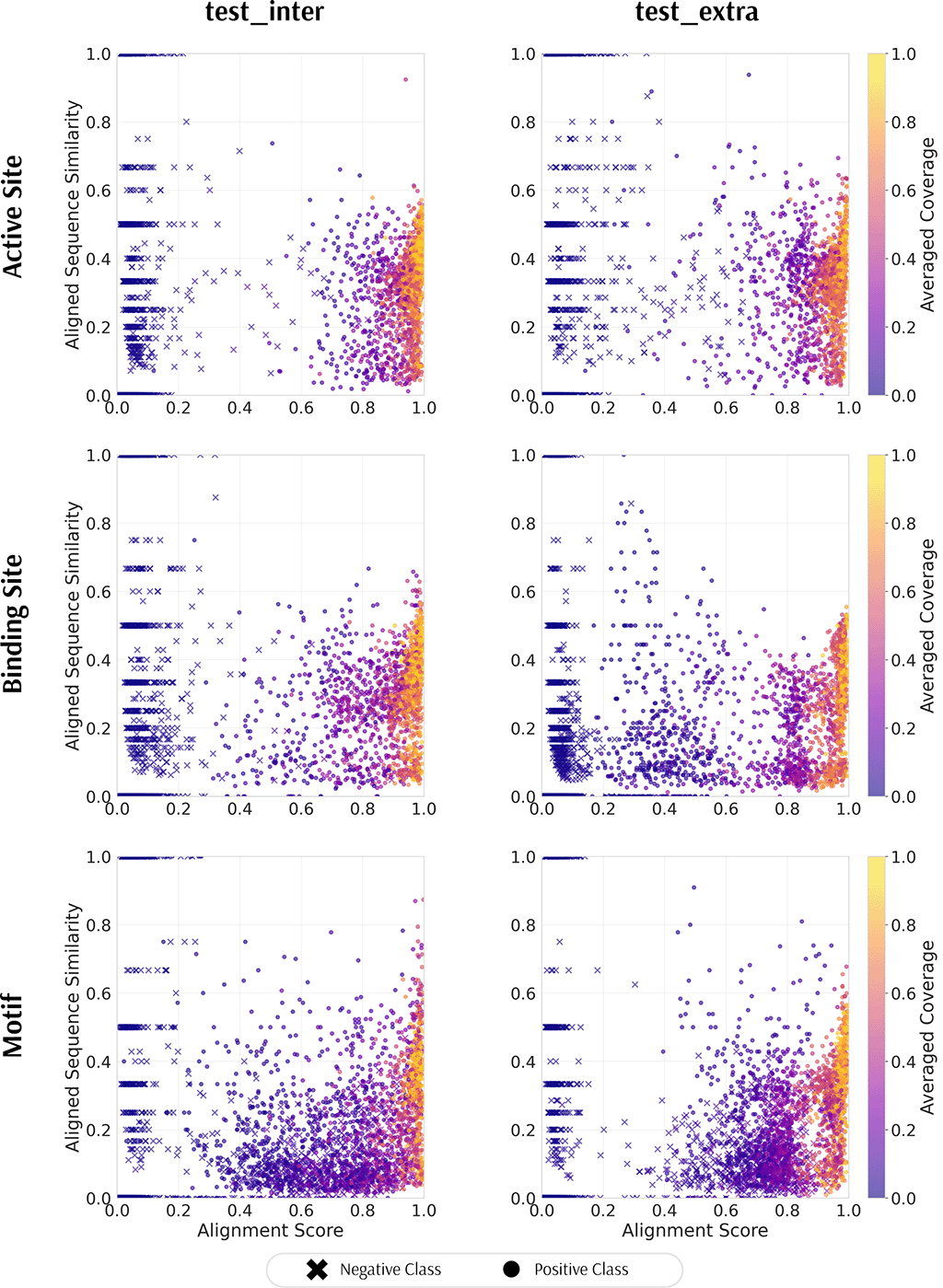}
    \caption{Sequence-similarity patterns of aligned substructures. Each panel shows how the aligned-sequence similarity varies with PLASMA’s alignment score, colored by the averaged coverage between the query and candidate proteins. These plots illustrate that high alignment scores do not simply arise from high sequence similarity; the alignment quality is driven by structural correspondence rather than sequence identity. All results use embeddings from Ankh.}
    \label{fig:seq-sim}
\end{figure}

\clearpage
\new{\section{Sequence Similarity Analysis}\label{sec:seq-sim-analysis}
To further examine whether PLASMA’s alignment performance is influenced by unintended global similarity, we analyze how PLASMA’s alignment score relates to the sequence similarity of the aligned residues. Same as before, we define sequence similarity as the percentage of aligned residue pairs that share the same amino acid type.

Figure~\ref{fig:seq-sim} presents the distribution of alignment scores and sequence-similarity values across all test pairs. The results show that high alignment scores typically coincide with high alignment coverage rather than high sequence similarity. Many correctly aligned substructures exhibit low sequence similarity despite high PLASMA scores, indicating that the method is driven by shared local 3D geometry rather than residue identity. For negative test pairs, the sequence-similarity values appear highly dispersed, which arises from their extremely low alignment coverage; with very few aligned residue pairs, the resulting sequence-similarity statistic becomes unstable and effectively uninformative. The upper-right region of the plot remains sparse, reflecting our dataset construction protocol that limits the global sequence identity of all test proteins to below 50\%.

Overall, this analysis demonstrates that strong PLASMA alignment scores do not depend on high sequence similarity. The method therefore does not rely on global homology signals and is not affected by unintended data leakage.}

\clearpage
\section{Hyperparameter Analysis}\label{sec:hyperparameter-analysis}

\begin{figure}[hbtp]
    \centering
    \begin{subfigure}[b]{\textwidth}
        \centering
        \includegraphics[width=\textwidth]{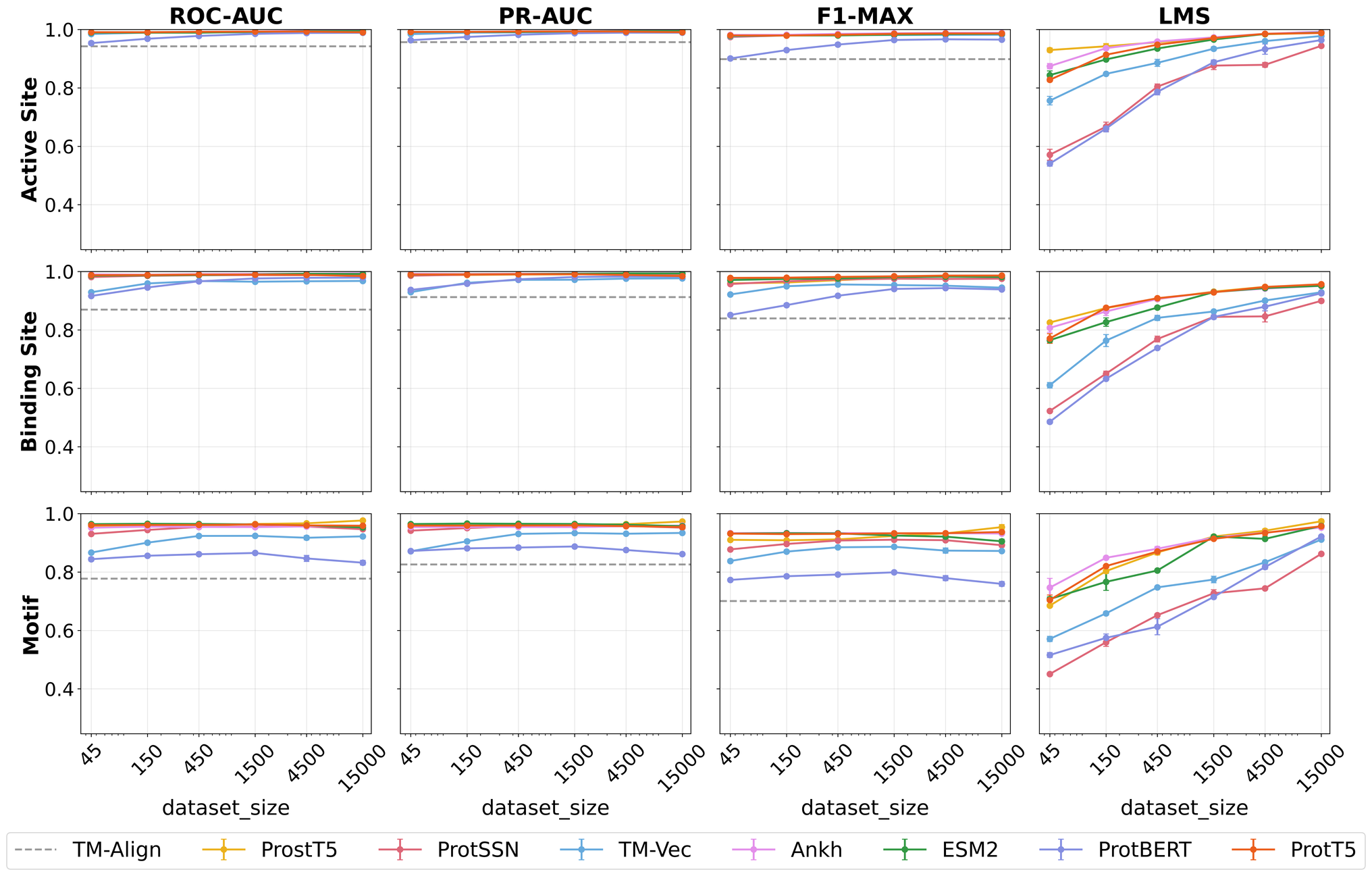}
        \caption{\texttt{test\_inter}}
        \label{fig:dataset-fraction-frequent}
    \end{subfigure}
    \hfill
    \begin{subfigure}[b]{\textwidth}
        \centering
        \includegraphics[width=\textwidth]{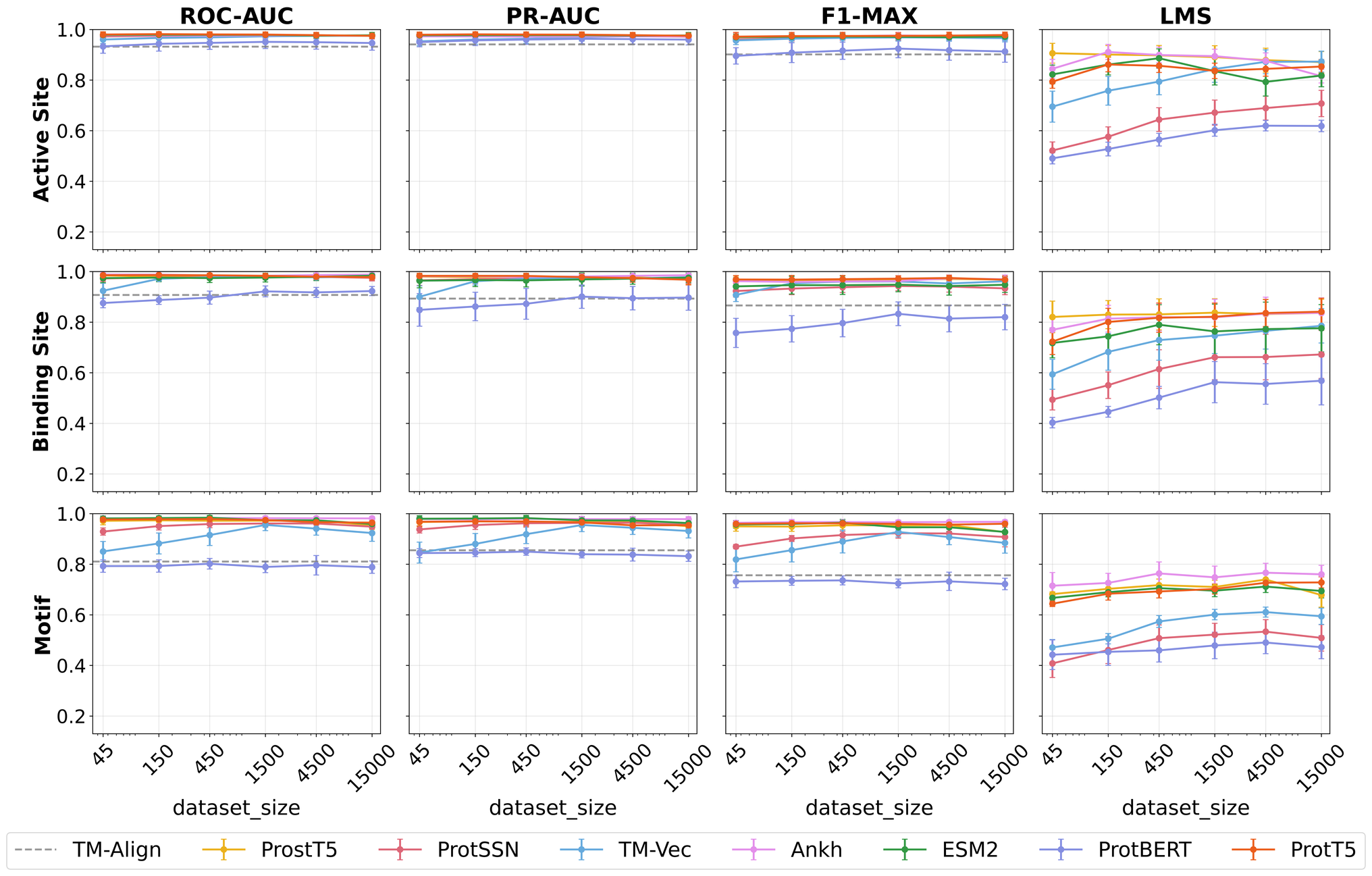}
        \caption{\texttt{test\_extra}}
        \label{fig:dataset-fraction-hard}
    \end{subfigure}
    \caption{Performance vs dataset fraction. PLASMA demonstrates high performance in predicting the existence of substructure similarities even with minimal training data (45 samples), and, in most cases, this ability remains stable when the dataset size increases. However, the LMS of PLASMA noticeably improves as dataset size increases, indicating that training is important for predicting the precise local of similar substructures.}
    \label{fig:dataset-fraction}
\end{figure}

\begin{figure}[p]
    \centering
    \begin{subfigure}[b]{\textwidth}
        \centering
        \includegraphics[width=\textwidth]{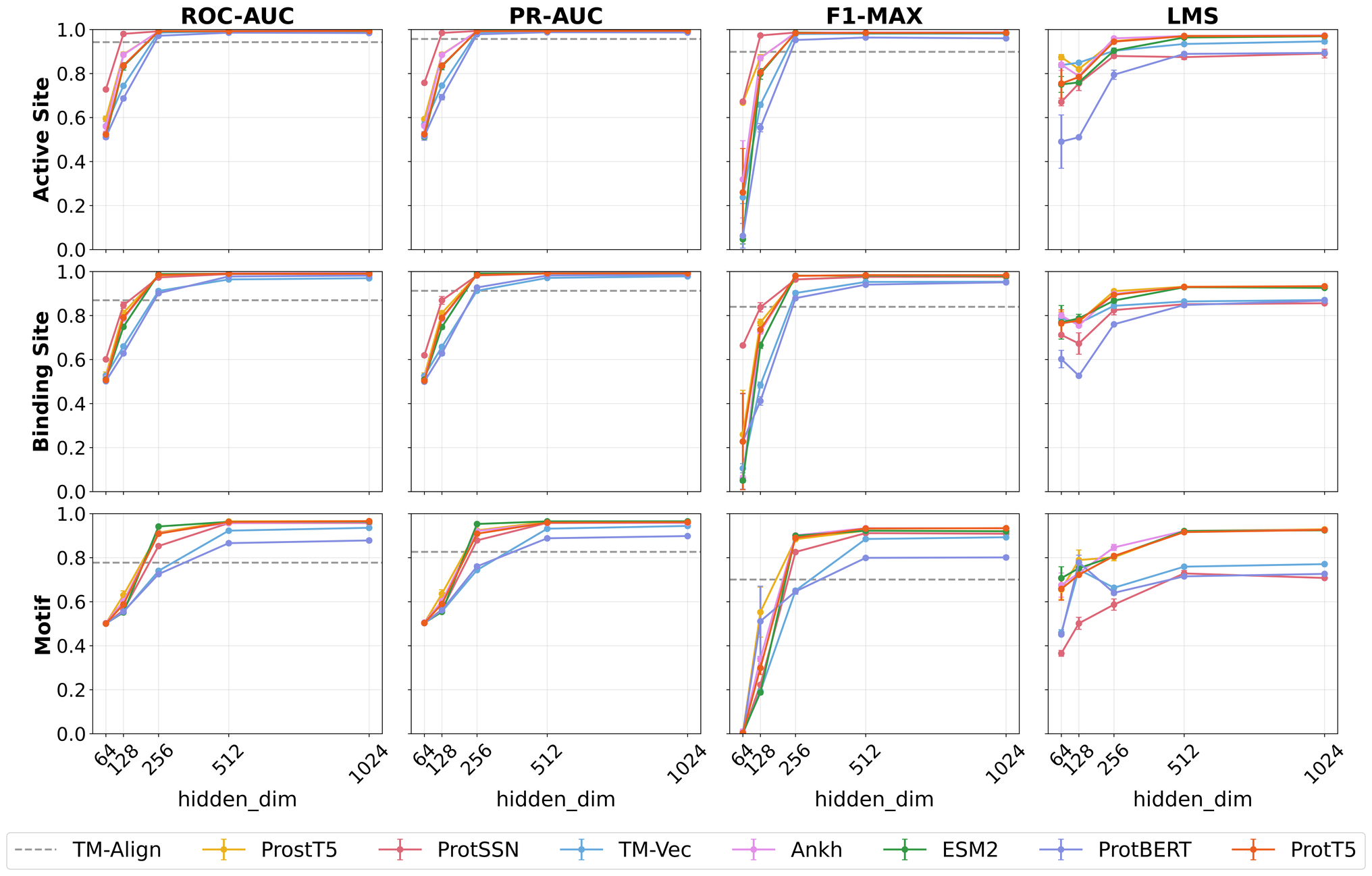}
        \caption{\texttt{test\_inter}}
        \label{fig:eta-hidden-dim-frequent}
    \end{subfigure}
    \hfill
    \begin{subfigure}[b]{\textwidth}
        \centering
        \includegraphics[width=\textwidth]{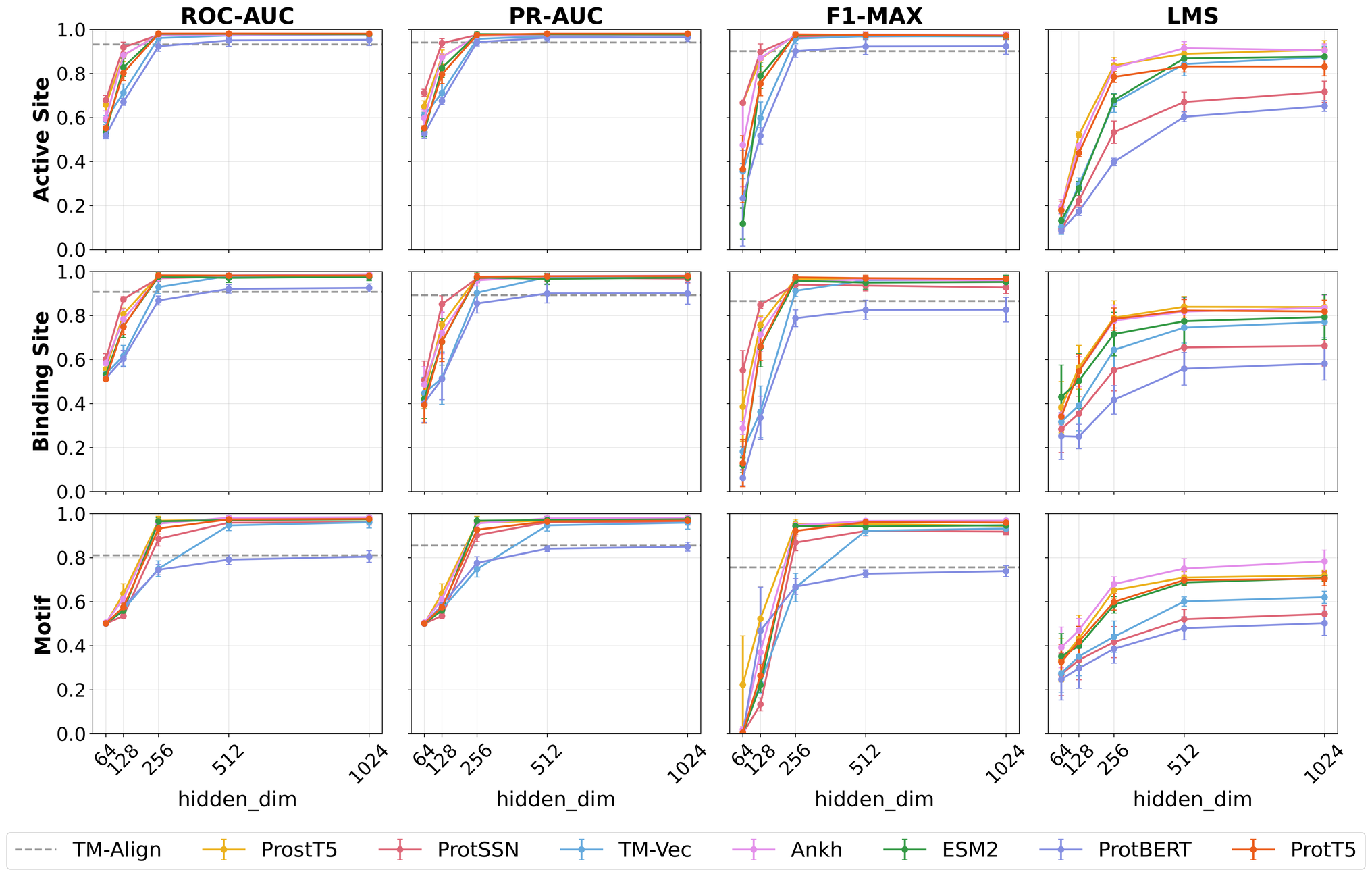}
        \caption{\texttt{test\_extra}}
        \label{fig:eta-hidden-dim-hard}
    \end{subfigure}
    \caption{Performance vs hidden dimension size of the siamese network. While PLASMA's performance remains stable when the hidden dimension size is greater than $256$, it would significantly drop when the hidden dimension size is less than this number.}
    \label{fig:eta-hidden-dim}
\end{figure}

\begin{figure}[p]
    \centering
    \begin{subfigure}[b]{\textwidth}
        \centering
        \includegraphics[width=\textwidth]{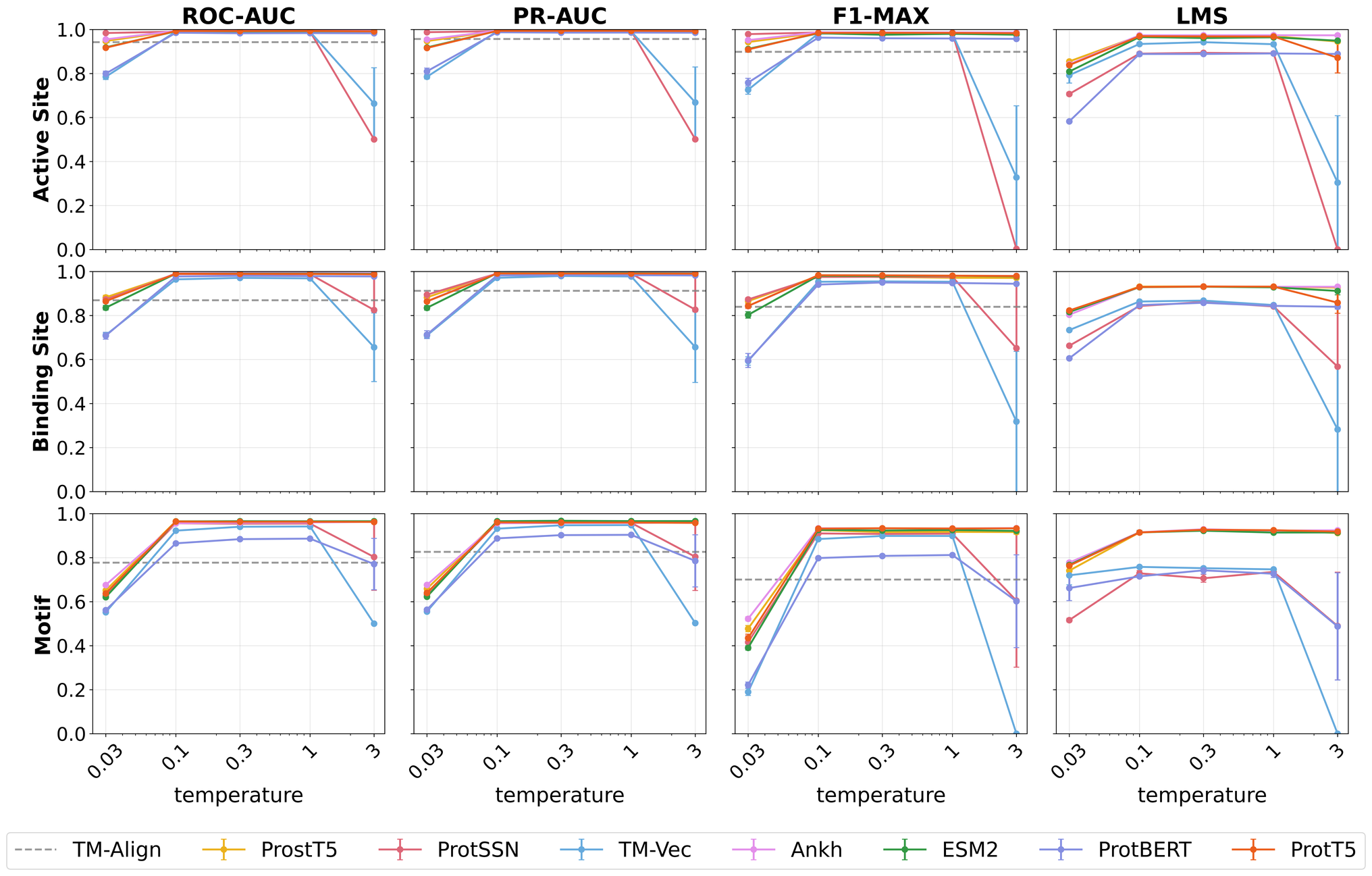}
        \caption{\texttt{test\_inter}}
        \label{fig:omega-temperature-frequent}
    \end{subfigure}
    \hfill
    \begin{subfigure}[b]{\textwidth}
        \centering
        \includegraphics[width=\textwidth]{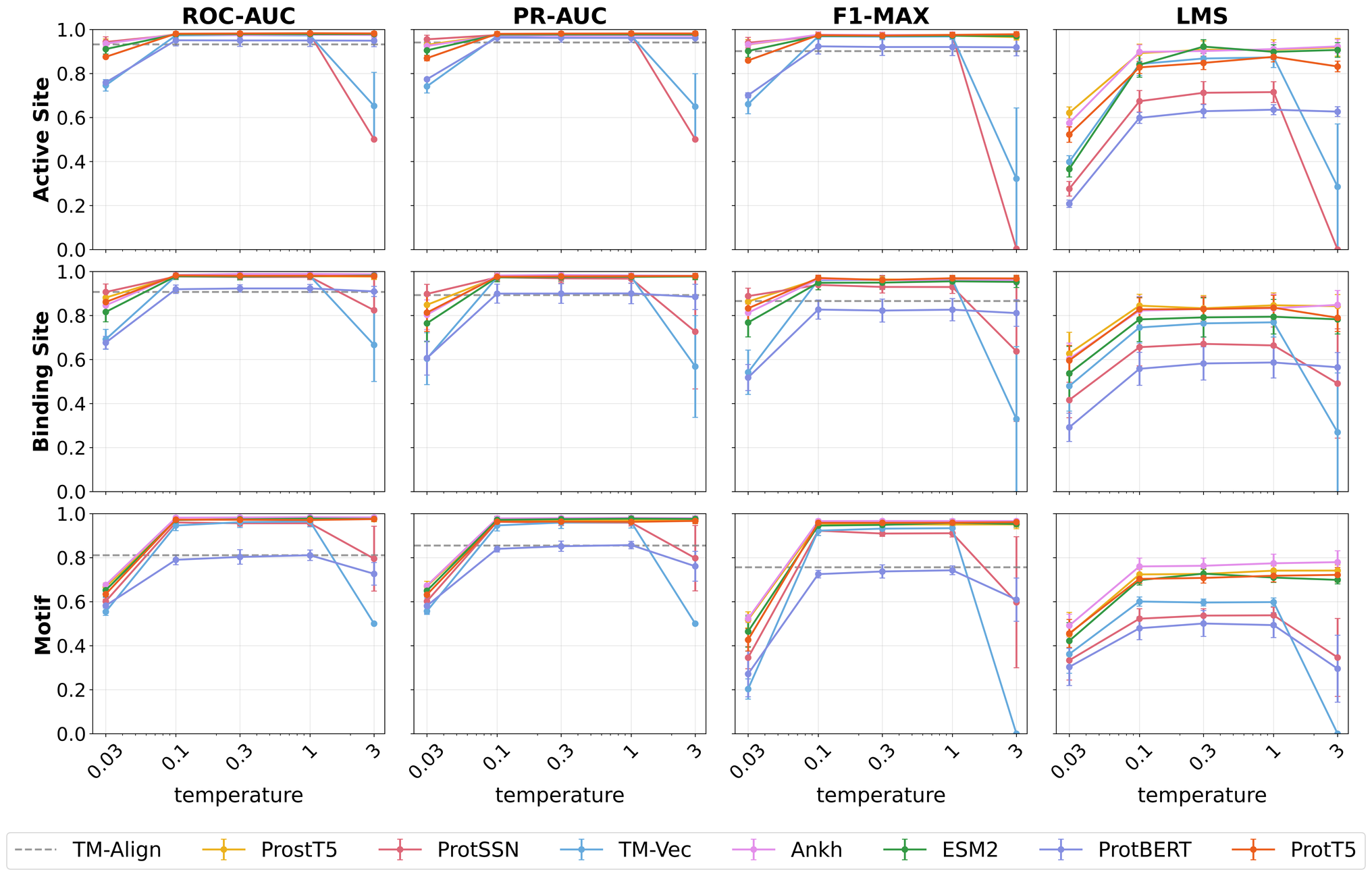}
        \caption{\texttt{test\_extra}}
        \label{fig:omega-temperature-hard}
    \end{subfigure}
    \caption{Performance vs Sinkhorn temperature ($\tau$). PLASMA's performance remains stably high within the $0.1$--$1$ range, but when out of this range, PLASMA's performance noticeably drops.}
    \label{fig:omega-temperature}
\end{figure}

\begin{figure}[p]
    \centering
    \begin{subfigure}[b]{\textwidth}
        \centering
        \includegraphics[width=\textwidth]{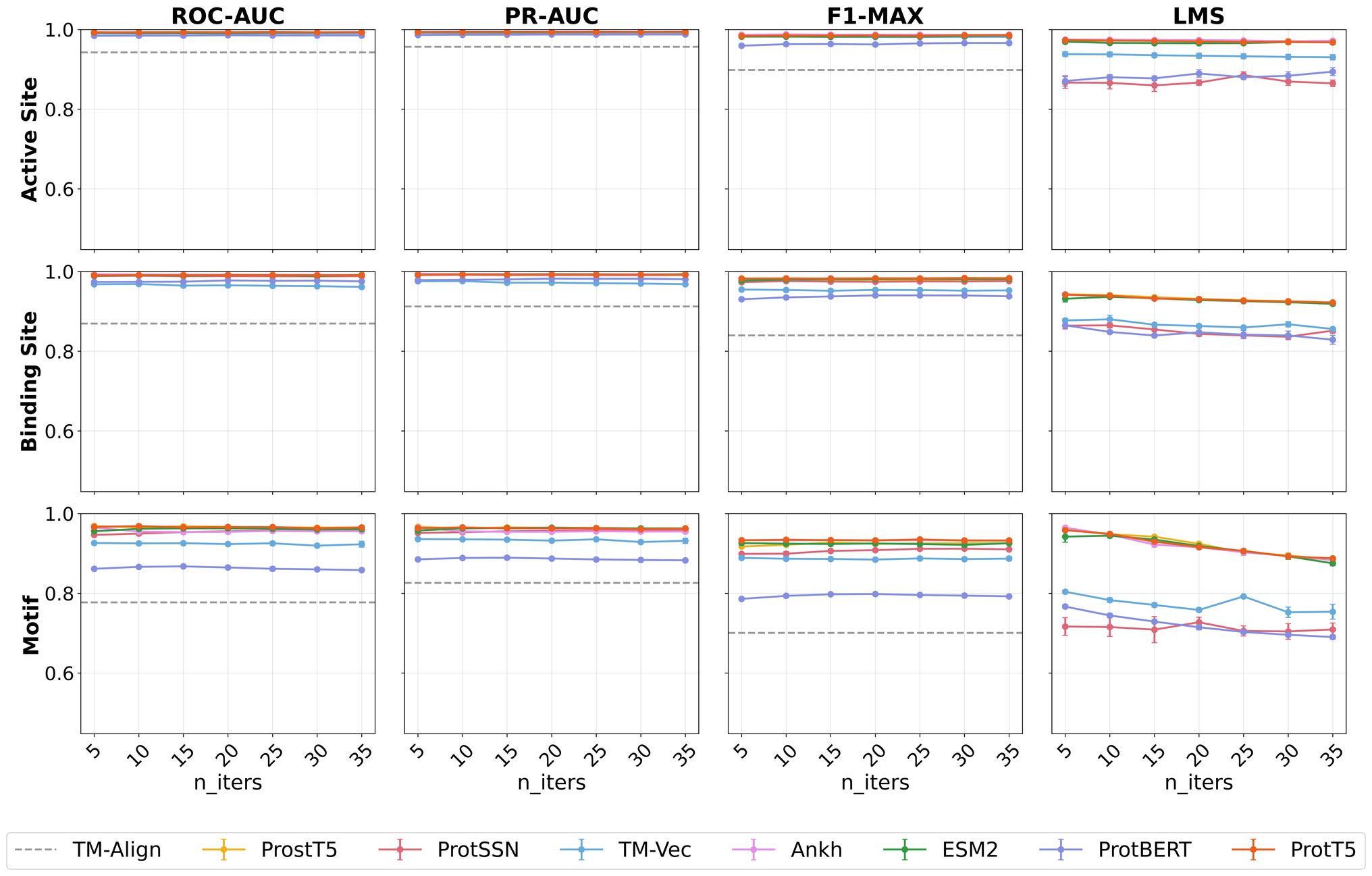}
        \caption{\texttt{test\_inter}}
        \label{fig:omega-n_iters-frequent}
    \end{subfigure}
    \hfill
    \begin{subfigure}[b]{\textwidth}
        \centering
        \includegraphics[width=\textwidth]{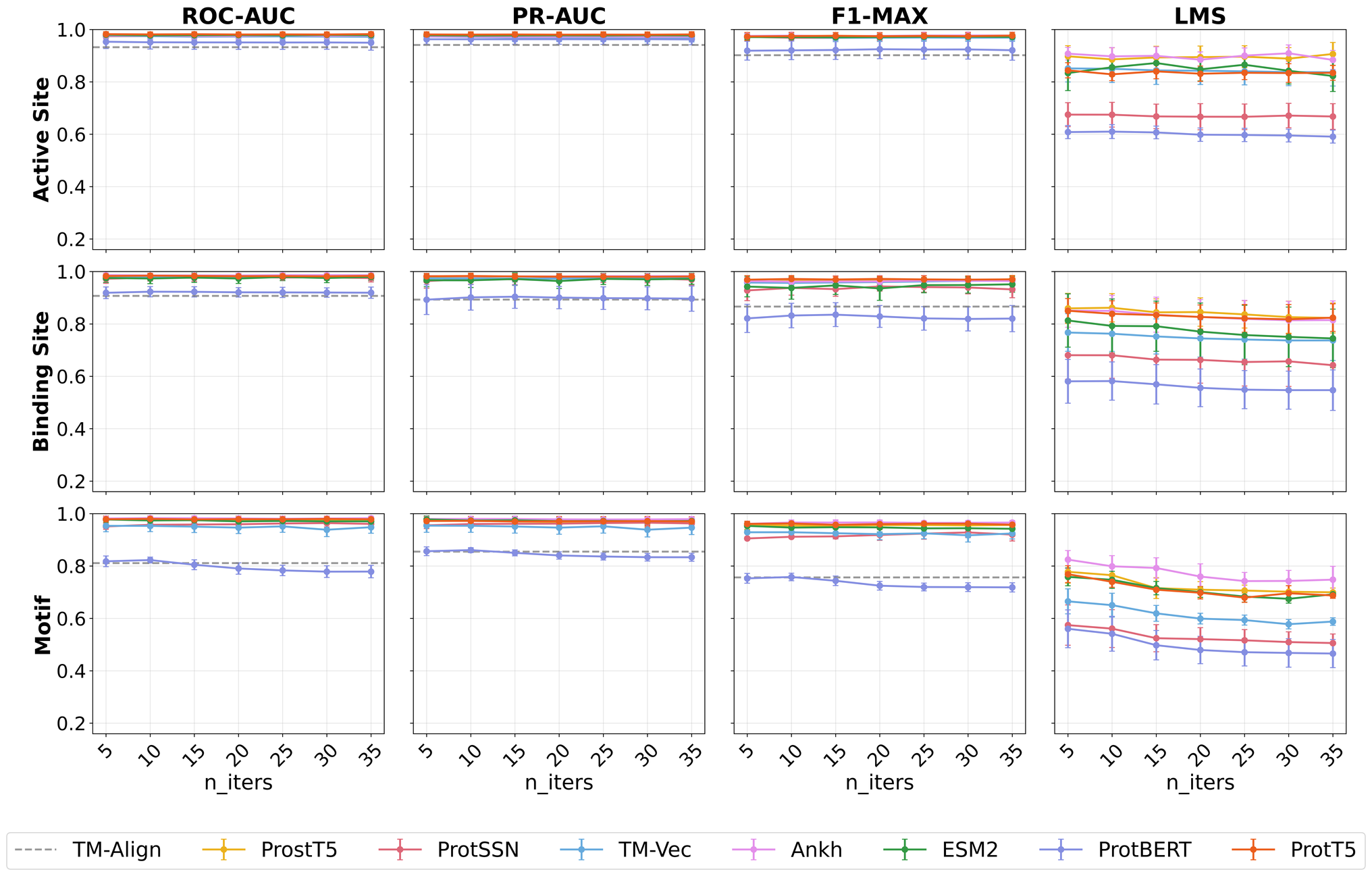}
        \caption{\texttt{test\_extra}}
        \label{fig:omega-n_iters-hard}
    \end{subfigure}
    \caption{Performance vs number of Sinkhorn iterations $T$. In most cases, PLASMA's performance is insensitive of the setting of $T$, but for analyzing motifs, we can see a subtle decreasing trend as the number of iteration increases.}
    \label{fig:omega-T}
\end{figure}

\begin{figure}[p]
    \centering
    \begin{subfigure}[b]{\textwidth}
        \centering
        \includegraphics[width=\textwidth]{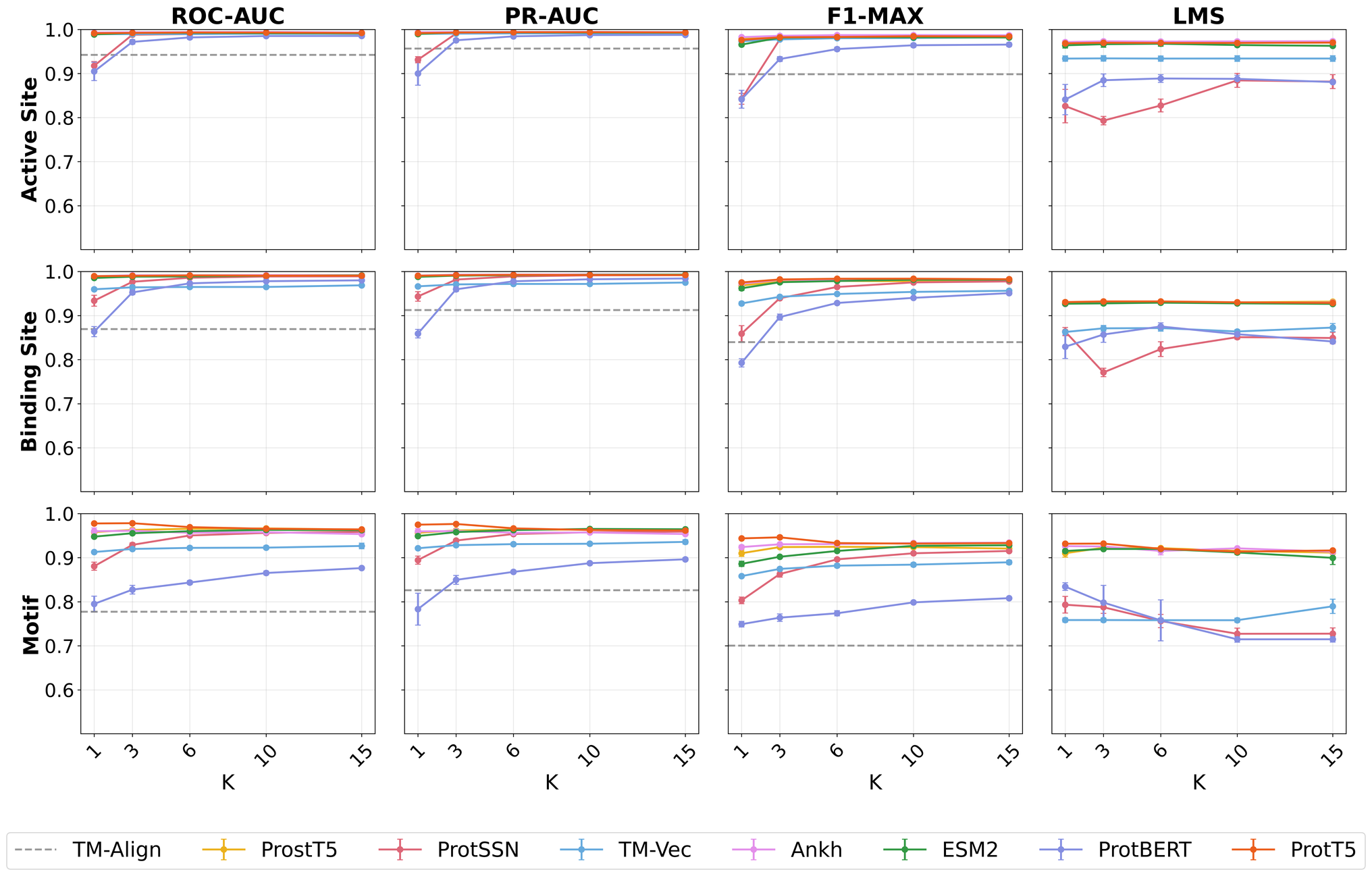}
        \caption{\texttt{test\_inter}}
        \label{fig:score-k-frequent}
    \end{subfigure}
    \hfill
    \begin{subfigure}[b]{\textwidth}
        \centering
        \includegraphics[width=\textwidth]{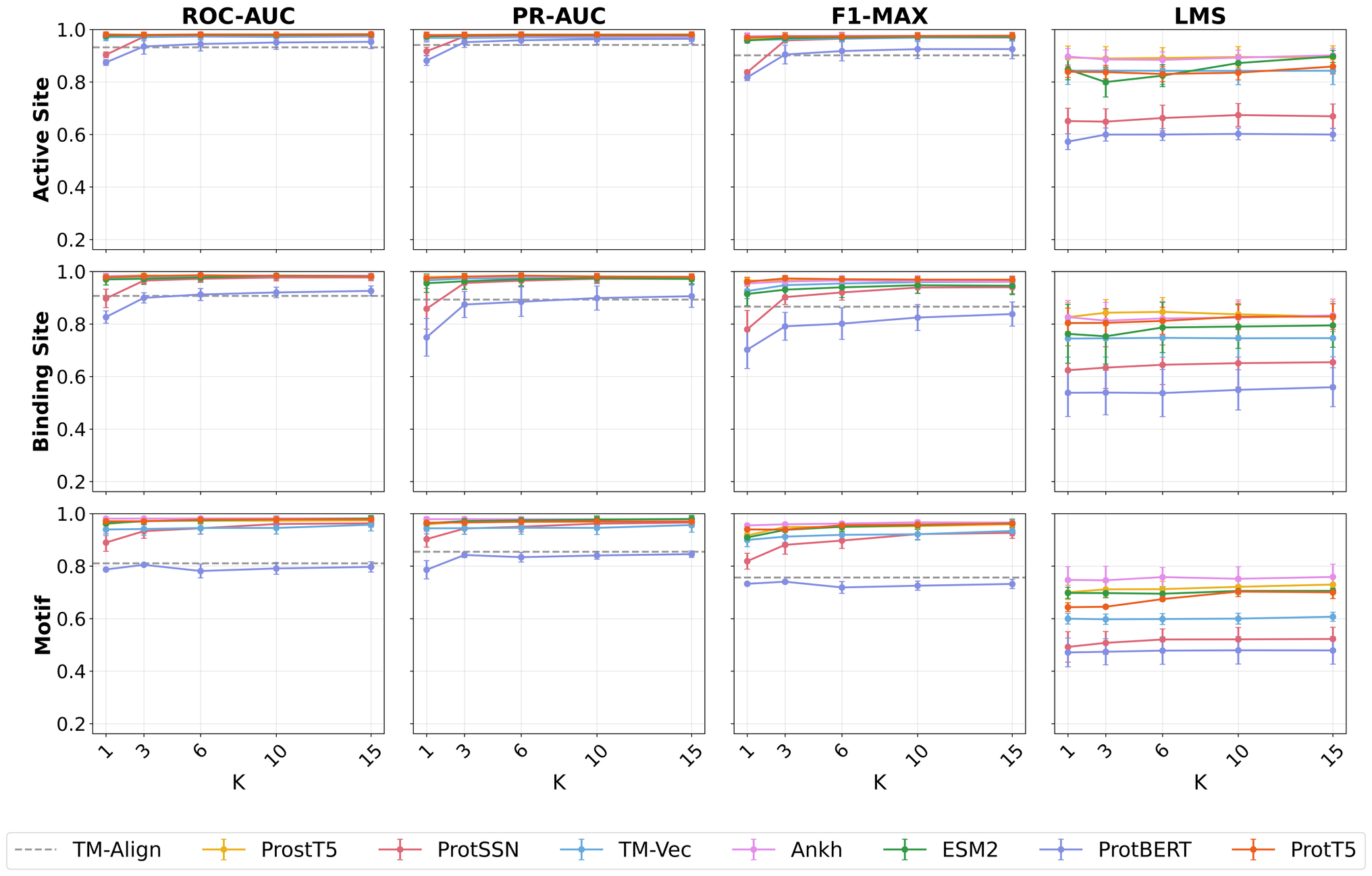}
        \caption{\texttt{test\_extra}}
        \label{fig:score-k-hard}
    \end{subfigure}
    \caption{Performance vs the kernel size of the diagonal convolution ($k$). For interpolation tasks and in particular when using \textsc{ProtSSN}, \textsc{ProtBERT}, or \textsc{TM-Vec} as the backbone, there is a trade-off between detecting the existence of substructure similarities and predicting the precise location of similar regions---the former prefers higher $k$ while the latter prefers lower $k$. However, for other cases, PLASMA demonstrates stable performance regardless the choice of $k$.}
    \label{fig:score-k}
\end{figure}

\begin{figure}[p]
    \centering
    \begin{subfigure}[b]{\textwidth}
        \centering
        \includegraphics[width=\textwidth]{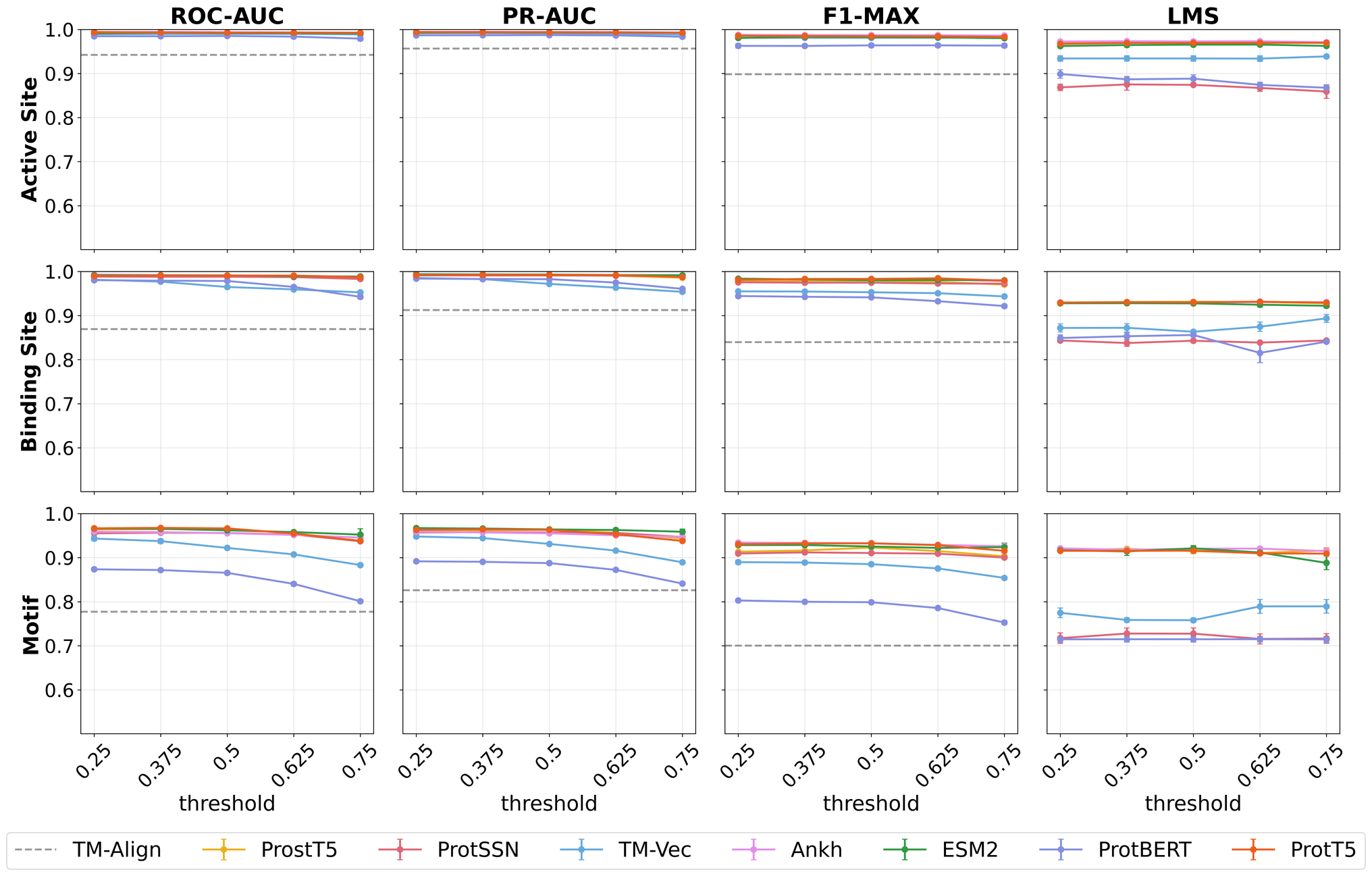}
        \caption{\texttt{test\_inter}}
        \label{fig:score-threshold-frequent}
    \end{subfigure}
    \hfill
    \begin{subfigure}[b]{\textwidth}
        \centering
        \includegraphics[width=\textwidth]{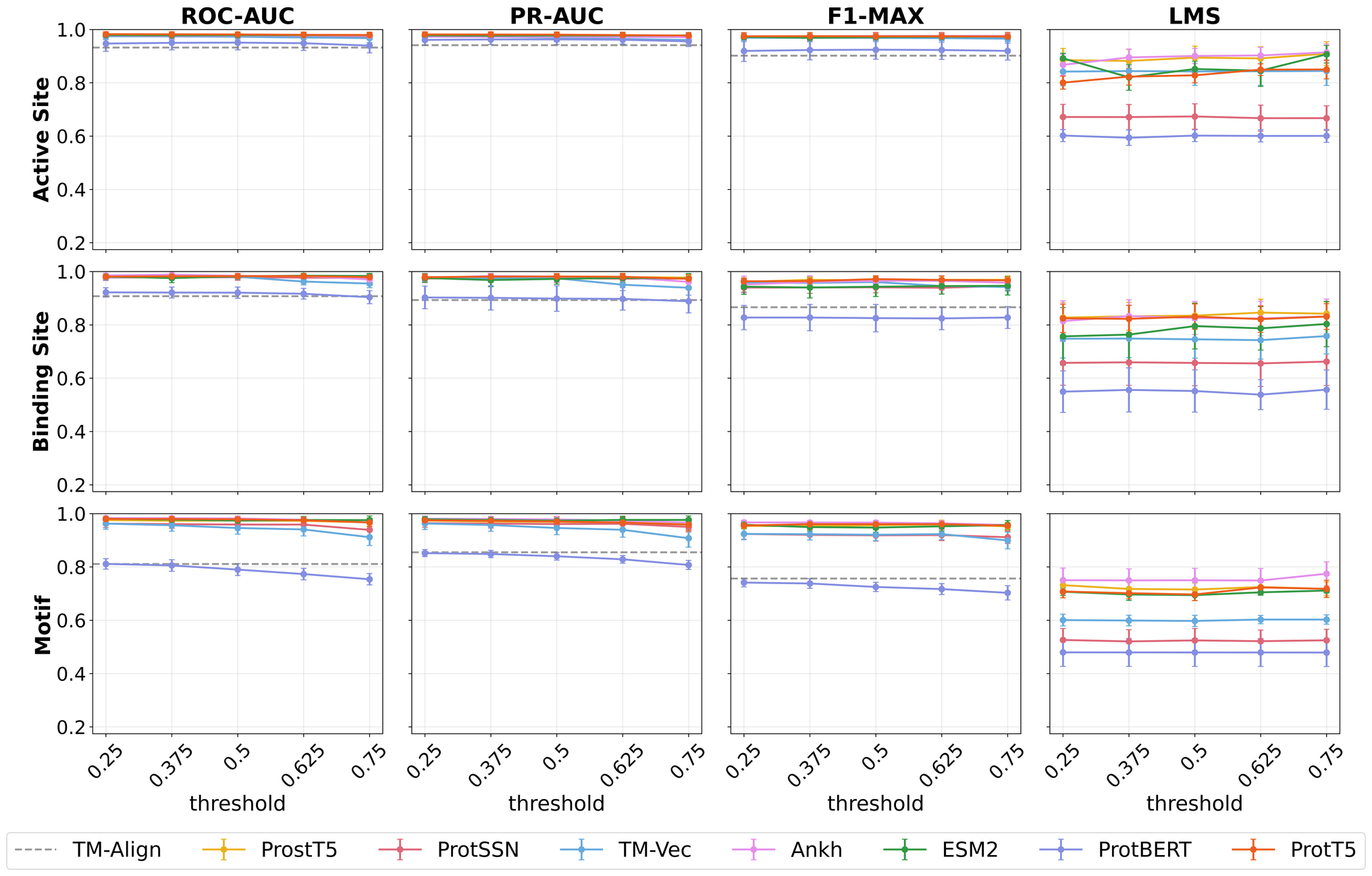}
        \caption{\texttt{test\_extra}}
        \label{fig:score-threshold-hard}
    \end{subfigure}
    \caption{Performance vs residue matching threshold ($\rho$). PLASMA's performance remains stable overall when choosing different $\rho$ values, but for some backbone choices, such as \textsc{TM-Vec} and \textsc{ProtBERT}, PLASMA shows a slight preference over lower $\rho$ values.}
    \label{fig:score-threshold}
\end{figure}

\clearpage
\section{Further Alignment Matrix Visualizations}
\label{sec:alignment-matrix-vis-extra}
\begin{figure}[htb]
    \centering
    \includegraphics[width=0.7\textwidth]{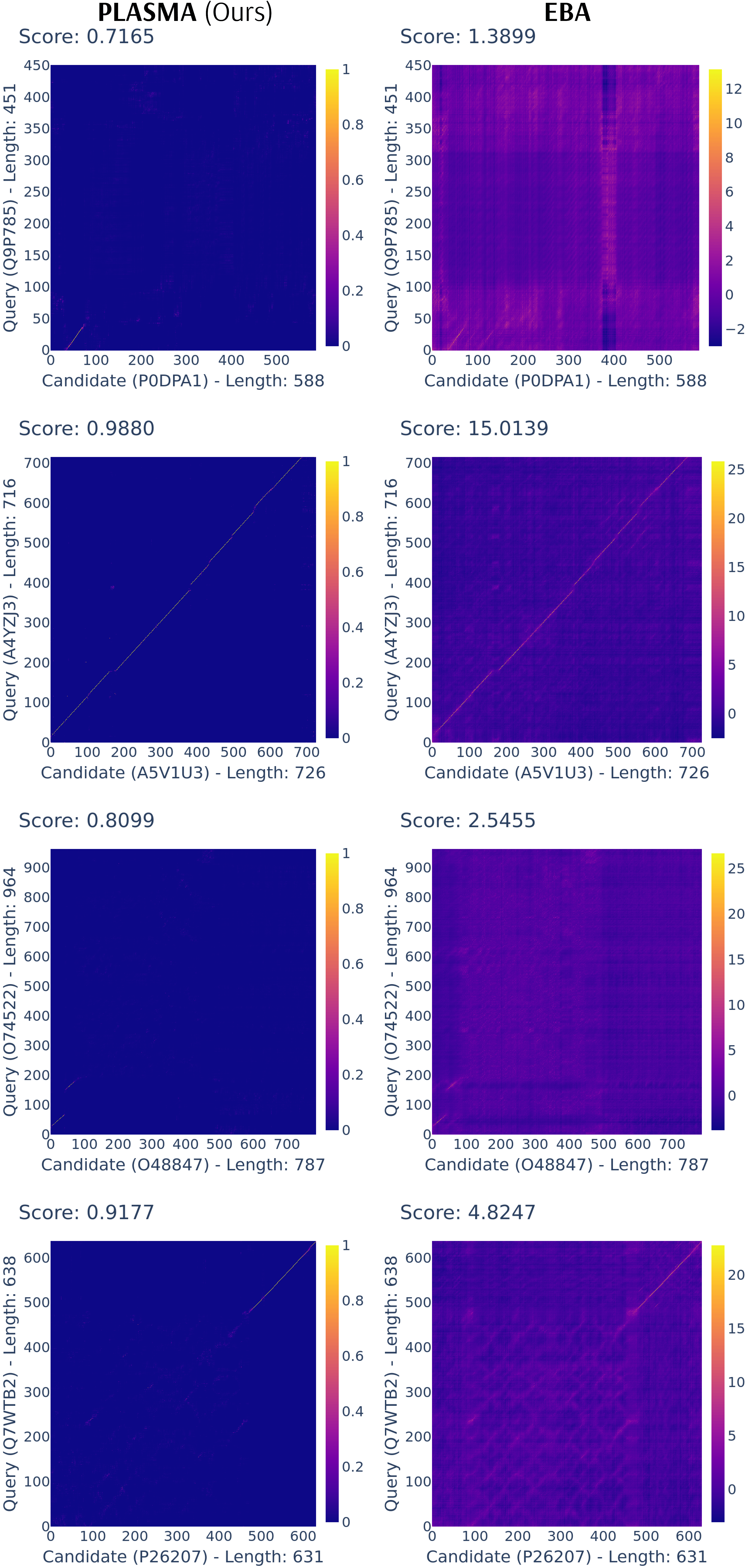}
    \caption{Alignment matrix visualizations of random positive pairs from \texttt{test\_inter}. (Part 1)}
    \label{fig:matrix-vis-1}
\end{figure}

\begin{figure}[p]
    \centering
    \includegraphics[width=0.7\textwidth]{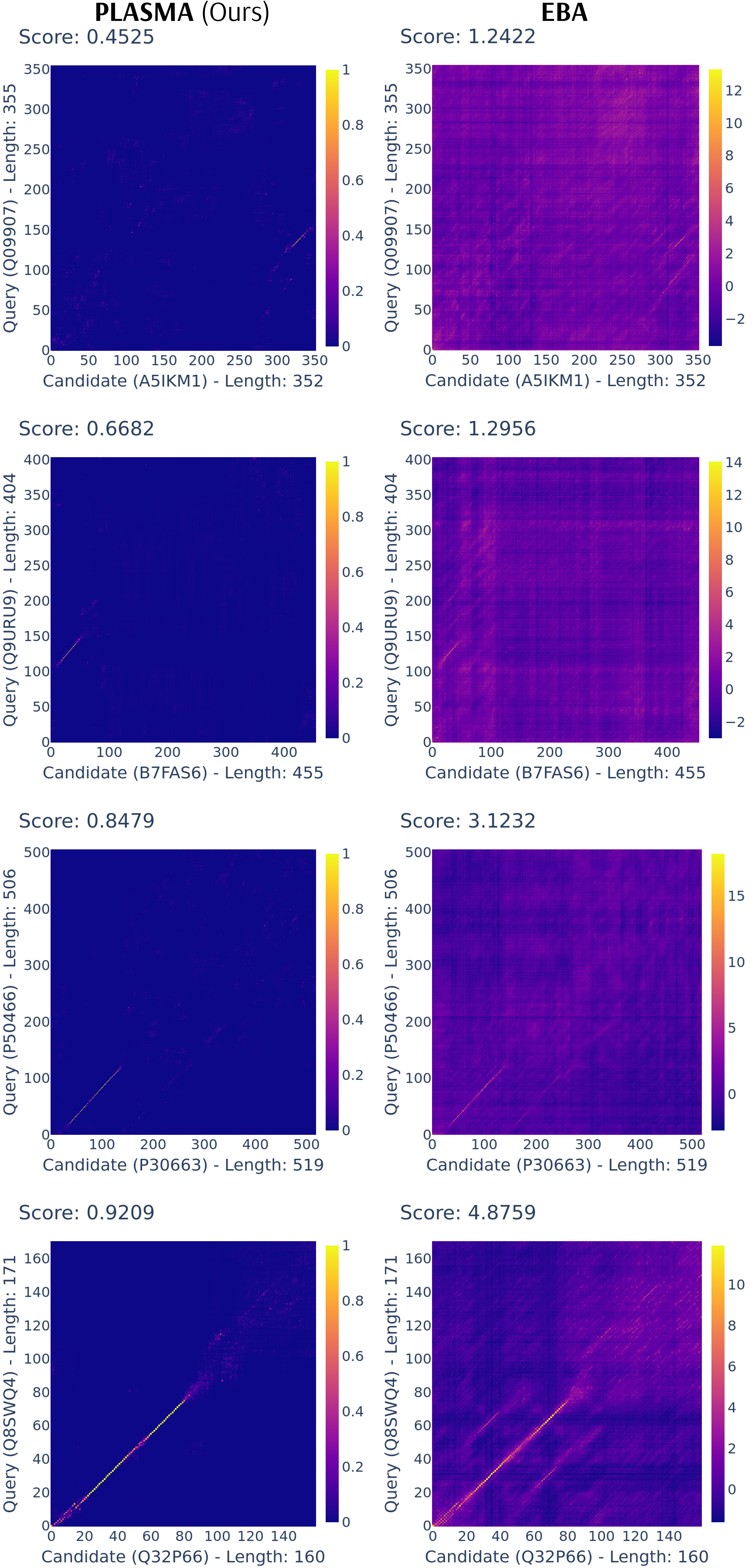}
    \caption{Alignment matrix visualizations of random positive pairs from \texttt{test\_inter}. (Part 2)}
    \label{fig:matrix-vis-2}
\end{figure}

\begin{figure}[p]
    \centering
    \includegraphics[width=0.7\textwidth]{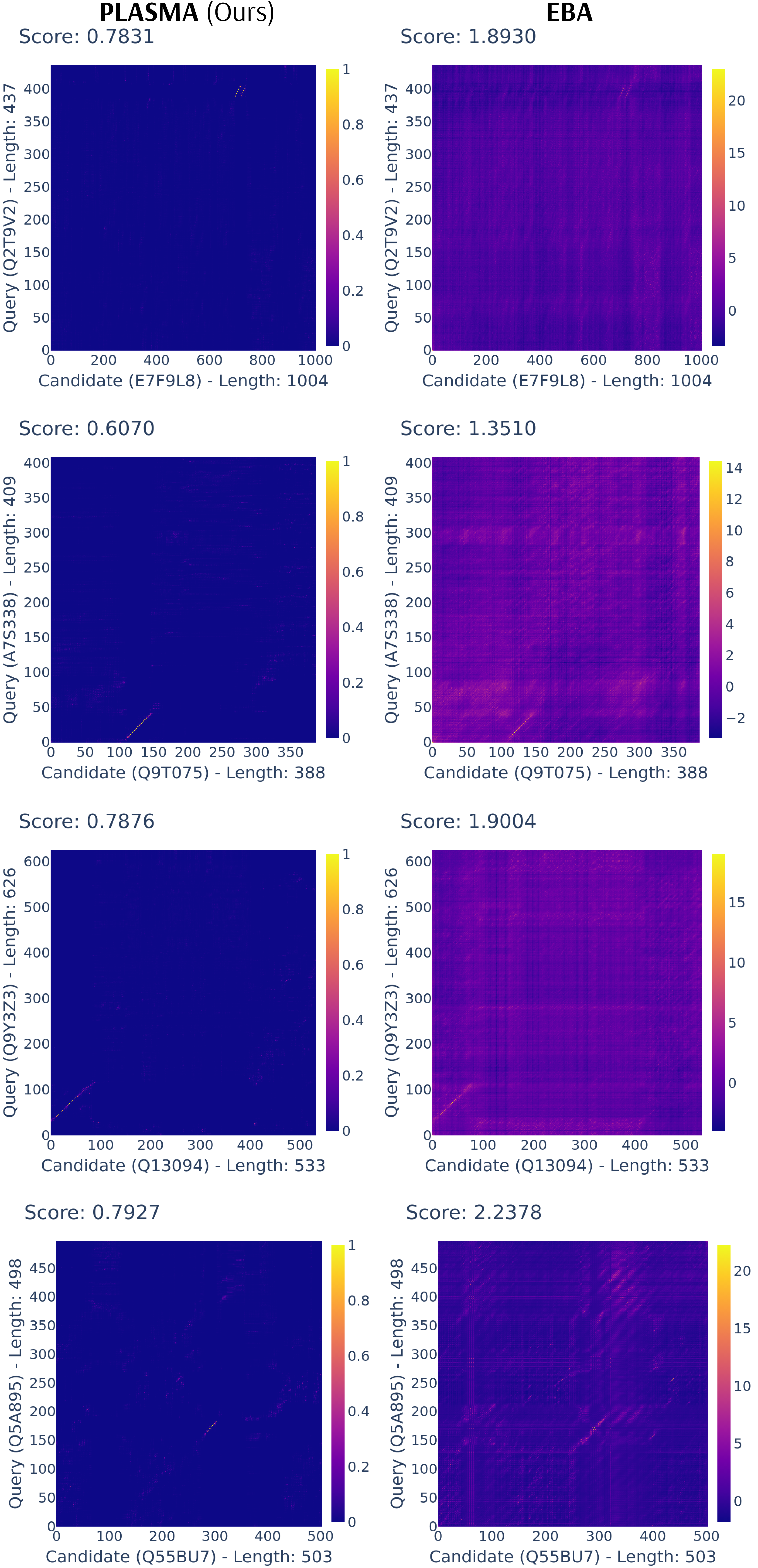}
    \caption{Alignment matrix visualizations of random positive pairs from \texttt{test\_inter}. (Part 3)}
    \label{fig:matrix-vis-3}
\end{figure}

\end{document}